\documentclass[traditabstract]{aa}

\usepackage{graphicx}
\usepackage{txfonts}
\usepackage{lipsum}
\usepackage{subcaption}         
\usepackage{lscape}             
\usepackage{placeins}           
                                
\usepackage{graphicx}
\usepackage{txfonts}
\usepackage{natbib}
\usepackage{enumerate}
\usepackage{bm}
\usepackage[breaklinks]{hyperref}
\hypersetup{
  colorlinks=true,
  linkcolor=blue,
  citecolor=blue,
  urlcolor=blue,
}

\bibpunct{(}{)}{;}{a}{}{,} 

\newcommand{\twelveco}{$^{12}$CO}

\newcommand{\cotwo}{$^{12}$CO(2--1)}
\newcommand{\cothree}{$^{12}$CO(3--2)}

\newcommand{\thirteencotwo}{$^{13}$CO(2--1)}

\newcommand{\rtwo}{$R_{21}$}

\newcommand{\lco}{$L^\prime_{\rm CO}$}

\newcommand{\lcotwo}{$L^\prime_{\rm CO(2-1)}$}

\newcommand{\ci}{C{\sc i}}
\newcommand{\cii}{C{\sc ii}}
\newcommand{\ha}{H$\alpha$}

\newcommand{\kms}{km\,s$^{-1}$}

\newcommand{\kkms}{K\,km\,s$^{-1}$}
\newcommand{\kkmspc}{K\,km\,s$^{-1}$\,pc$^{2}$}
\newcommand{\htwo}{H$_2$}
\newcommand{\Nhtwo}{$N_\mathrm{H2}$}

\newcommand{\nhtwo}{$n_{\rm H2}$}

\newcommand{\tb}{$T_{\rm B}$}

\newcommand{\aco}{$\alpha_{\rm CO}$}

\newcommand{\spit}{\textit{Spitzer}}
\newcommand{\hers}{\textit{Herschel}}

\newcommand{\logoh}{12$+$log(O/H)}

\newcommand{\Sigmalum}{$\Sigma_\mathrm{lum}$}

\newcommand{\avirial}{$\alpha_{\rm virial}$}
\newcommand{\mlum}{M$_\mathrm{lum}$}
\newcommand{\mvir}{M$_\mathrm{vir}$}

\newcommand{\cmtwo}{cm$^{-2}$}
\newcommand{\cmthree}{cm$^{-3}$}

\newcommand{\hi}{\rm H{\sc i}}

\newcommand{\micron}{$\mu$m}
\newcommand{\zzsun}{${\mathrm Z/Z}_\odot$}
\newcommand{\zsun}{$\mathrm{Z}_\odot$}
\newcommand{\msun}{$M_\odot$}
\newcommand{\msunpcsq}{$M_\odot$\,pc$^{-2}$}
\newcommand{\lsun}{$L_\odot$}
\newcommand{\msunyr}{$M_\odot$\,yr$^{-1}$}

\newcommand{\sigmav}{$\sigma_{\mathrm{V}}$}
\newcommand{\mdark}{M$_\mathrm{dark}$}
\newcommand{\pe}{P$_\mathrm{e}$}
\newcommand{\tff}{$t_\mathrm{ff}$}

\newcommand\T{\rule{0pt}{3ex}}       
\newcommand\B{\rule[-2.5ex]{0pt}{0pt}} 

\begin{document}

\title{Metal-poor molecular clouds in dwarf galaxies at parsec-scale resolution}

\author{L.~K. Hunt \inst{\ref{inst:hunt}}
\and
M. Heyer \inst{\ref{inst:heyer}}
\and
U. Lisenfeld \inst{\ref{inst:lisenfeld}}
\and
A. Wei\ss \inst{\ref{inst:weiss}}
\and
A. Bolatto \inst{\ref{inst:bolatto}}
\and
E. Brinks \inst{\ref{inst:brinks}}
\and
M. Rubio \inst{\ref{inst:rubio}}
\and 
D. Calzetti \inst{\ref{inst:heyer}}
\and
F. Combes \inst{\ref{inst:combes}}
\and
E. Corbelli \inst{\ref{inst:hunt}}
\and
B.~G. Elmegreen \inst{\ref{inst:elmegreen}}
\and
S. Garc{\'i}a-Burillo \inst{\ref{inst:santi}}
\and
C. Henkel \inst{\ref{inst:weiss}} 
\and
K.~M. Menten \inst{\ref{inst:weiss}}\thanks{In memory of Karl M. Menten, who passed away suddenly before
this work was completed. His inspiring advice and invaluable contributions
will be sincerely missed.}
\and
L.~J. Smith \inst{\ref{inst:smith}}
\and
A. Adamo \inst{\ref{inst:adamo}}
\and
B.~T. Draine \inst{\ref{inst:draine}}
\and
D.~A. Hunter \inst{\ref{inst:hunter}}
\and
N. Izumi \inst{\ref{inst:izumi1}}
\and
K.~E. Johnson \inst{\ref{inst:johnson}, \ref{inst:johnsonnrao}} 
}

\offprints{L. K. Hunt}
\institute{INAF - Osservatorio Astrofisico di Arcetri, Largo E. Fermi, 5, 50125, Firenze, Italy
\label{inst:hunt}
\email{leslie.hunt@inaf.it}
\and
Department of Astronomy, University of Massachusetts, Amherst, MA 01003, USA \label{inst:heyer}
\and
Departamento de Física Te{\'o}rica y del Cosmos, Universidad de Granada, 18071, Granada, Spain; 
Instituto Carlos I de Física Te{\'o}rica y Computacional, Facultad de Ciencias, 18071, Granada, Spain \label{inst:lisenfeld}
\and
Max-Planck-Institut f\"ur Radioastronomie, Auf dem H\"ugel 69, 53121 Bonn, Germany \label{inst:weiss}
\and
Department of Astronomy, University of Maryland, College Park, MD 20742, USA; 
Joint Space-Science Institute, University of Maryland, College Park, MD 20742, USA \label{inst:bolatto}
\and
Centre for Astrophysics Research, University of Hertfordshire, AL10 9AB, Hatfield, UK \label{inst:brinks}
\and
Departamento de Astronomía, Universidad de Chile, Casilla 36-D, Santiago, Chile \label{inst:rubio}
\and
Observatoire de Paris, LUX, College de France, CNRS, PSL, Sorbonne University, F-75014, Paris, France \label{inst:combes}
\and
Katonah, NY 10536, USA \label{inst:elmegreen}
\and
Observatorio Astron\'omico Nacional (OAN)-Observatorio de Madrid,
Alfonso XII, 3, 28014-Madrid, Spain \label{inst:santi}
\and
Space Telescope Science Institute, 3700 San Martin Drive, Baltimore, MD 21218, USA \label{inst:smith}
\and
Department of Astronomy, Oscar Klein Centre, Stockholm University, AlbaNova, SE-106 91 Stockholm, Sweden \label{inst:adamo}
\and
Princeton University Observatory, Peyton Hall, Princeton, NJ 08544-1001, USA \label{inst:draine}
\and
Flagstaff, AZ 86001,  USA \label{inst:hunter}
\and
Graduate School of Science, Nagoya University, Furo-cho, Chikusa-ku, Nagoya, Aichi 464-8602, Japan \label{inst:izumi1}
\and
Department of Astronomy, University of Virginia, Charlottesville, VA 22904-4325, USA \label{inst:johnson}
\and
Adjunct Astronomer, National Radio Astronomy Observatory, 520 Edgemont Road, Charlottesville, VA 22903, USA \label{inst:johnsonnrao}
}

   \date{}

\abstract{Understanding how stars form at low metallicity is challenging because carbon monoxide (CO) emission, 
which usually probes the molecular gas reservoir, is faint,
and tends to be concentrated in parsec-scale clumps.
To overcome this difficulty, in order to study the molecular clouds in metal-poor environments, 
we have acquired \cotwo\ ALMA maps at 3$-$4\,pc resolution 
of four nearby dwarf galaxies with nebular (O/H) metallicities 
ranging from $\sim 1/9$\,\zsun\ to $\sim 1/3$\,\zsun.
Significant CO signal in the cubes has been identified through two masks,
a ``strict'', high-confidence, mask and a ``broad'' one for better completeness. 
The former traces the cores of the clouds, while the latter includes fainter emission
in the vicinity of the cores.
Through a cloud-finding algorithm applied to the two sets of masked cubes, we have identified more than 200 molecular clouds across
the sample, with $\sim 50$ in NGC\,3109, at $\sim 1/9$\,\zsun, the lowest metallicity 
parsec-scale cloud population yet observed.
The best-fit size-linewidth relations for the resolved clouds are similar among all galaxies, with a slope like 
the Milky Way (MW), but with lower velocity dispersion at a given cloud size. 
For a fixed CO-to-\htwo\ MW conversion factor, the cloud 
virial ratios tend to be large, and show a weak increasing trend with increasing
cloud size, and decreasing apparent surface densities and metallicity.
We find evidence for faint but significant CO emission outside CO-bright cores,
spanning similar velocities to those observed in the cores. 
We attribute this to effective shielding by gas and dust,
despite the low metallicities of our sample,
since the CO tends to be cospatial with dust as traced by far-infrared emission.
}


   \maketitle

\nolinenumbers

\section{Introduction \label{sec:intro}}

The early Universe was populated by metal-poor, low-mass galaxies, and 
according to standard paradigms for galaxy formation, these dwarf galaxies resided in
low-mass dark-matter halos that were the sites of the first star formation
\citep[e.g.,][]{bullock17,behroozi19}.
Nearby low-metallicity dwarf galaxies are the closest, albeit imperfect, analogs to
this early population, and much effort has been devoted to understanding
how stars form in these tiny systems.
However, it is not yet clear how metal-poor dwarfs fit within the current framework for star formation,
and, by extension, how stars formed in the metal-poor early Universe.

The scenario for star formation that has emerged over the past decade 
envisions a balance between several opposing mechanisms acting on the gas: 
gravity collapses the clouds as they cool, producing high-density cores where stars form; feedback
from these stars disrupts the cores 
\citep[e.g.,][]{corbelli17,dobbs19}
stopping star formation and ejecting matter into the circumgalactic medium.
This balancing act is observed at the
macroscopic level as a low efficiency for the conversion of gas into stars.

Dwarf galaxies are at the extreme end of this balance where star formation
can barely occur: 
feedback effects are stronger because the interstellar pressure is low 
\citep[e.g.,][]{hunter18},
and gravitational potential wells are shallow. 
Cooling is slow because of the low metal and dust abundances. 
Although dark-matter halo abundance matching 
implies inefficient star formation (SF) in dwarfs 
\citep[e.g.,][]{behroozi19},
on a molecular cloud scale, stars can form in dwarf galaxies with an efficiency that may be comparable to 
that of giant spirals \citep[e.g.,][]{rubio15,elmegreen26b}. 

\begin{table}[t!]
\caption{Properties of observed galaxies}                 
\label{tab:sample}    
\renewcommand{\arraystretch}{1.3}
\begin{center}                        
\resizebox{\linewidth}{!}{
\begin{tabular}{cccccc}      
\hline\hline               
\T
Name & Distance & Dimensions$^a$ & \logoh & log$_{10}$ & log$_{10}$ \\
& & & & Stellar mass$^b$ & SFR$^b$ \\
& (Mpc) & (arcsec$\,\times\,$arcsec) &  & (\msun) & (\msunyr) \B \\
\hline                      
IC\,5152$^c$  & 1.96 & $290.4\times168.4$ & 7.92 & 7.19 & $-1.49$\\    
NGC\,2915$^d$ & 4.29 & $150.4\times85.7$  & 8.17 & 8.22 & $-1.62$\\    
NGC\,3109$^e$ & 1.33 & $187.0\times187.0$ & 7.73 & 7.89 & $-2.71$\\    
NGC\,5408$^d$ & 5.32 & $90.0\times90.0$   & 8.19 & 8.59 & $-1.35$\\    
\hline                                  
\end{tabular}
}
\end{center}
\noindent
{\small
$^a$~\citet{jarrett03}: total $K_s$ dimensions; \\
$^b$~Taken from \citet{hunt16}; \\
$^c$~\citet{tully13}; \citet{berg12}: TRGB distance, O/H, respectively;\\
$^d$~\citet{tully13}; \citet{aniano20}: TRGB distance, O/H, respectively; \\
$^e$~\citet{jacobs09}; \citet{lee03}: TRGB distance, O/H, respectively.  \\
}
\end{table}

The problem of probing the connection of star formation with molecular gas at low metallicity is
the inability 
to probe molecular clouds in a metal-poor environment.
The relatively high abundance of carbon monoxide (CO) in spiral galaxies makes it a common proxy of \htwo, 
because it can be easily excited within cold molecular clouds, 
thanks to low critical density and low excitation energy in the ground-state rotational transition.
However, the presence of CO emission as a tracer for the molecular gas
depends primarily on shielding.
While self-shielding of \htwo\ is highly efficient, occurring even at \htwo\ column densities
as low as \Nhtwo\,$\sim 10^{14}$\,\cmtwo\ \citep{dejong80},
self-shielding of CO and cross-shielding of CO by \htwo\ are inefficient processes.
Thus, additional shielding of CO from FUV radiation by dust is necessary to maintain significant CO content. 
At low metallicity, C and O abundances and the dust-to-gas ratios are reduced
\citep[e.g.,][]{aniano20}, leading to diminishing protection of CO against photodissociation.  

This means that the CO-emitting regions in low-metallicity environments are expected to be much smaller than
in more typical conditions in the interstellar medium \citep[ISM,][]{bolatto99,bolatto13}.
The CO retreats to the densest regions within the molecular cloud, the less dense regions being dominated by carbon in the form of \ci\ 
and \cii\ \citep{wolfire10}. 
Interestingly, this implies that in metal-poor galaxies, CO emission becomes a probe for the denser ISM, namely the component that is actively forming stars 
\citep[e.g.,][]{elmegreen26a}. 
Thus, 
in order to identify the structure of the ISM associated with star formation 
in metal-poor systems, 
parsec-scale resolution CO maps are needed.

However, at metal abundances lower than $\sim$1/3\,\zsun\ [\logoh\,$\sim$\,8.2]\footnote{Throughout this paper, 
we adopt a Solar abundance \logoh\,=\,8.69 \citep{asplund09}.},
CO historically has been exceedingly hard to detect
\citep[e.g.,][]{taylor98,schruba12,henkel22,hunter24}.
This is also approximately the metallicity threshold where dust-to-gas ratios
as a function of metallicity change slope \citep{remyruyer14},
and $q_\mathrm{PAH}$, the polycyclic aromatic hydrocarbon (PAH) abundance parameter,
approaches 0 \citep{aniano20}. 
At such metallicities, so far, only five metal-poor dwarf galaxies beyond the Magellanic Clouds
have resolved molecular cloud measurements with beams $\la$ 6-8\,pc
\citep{rubio15,oey17,schruba17,miura18,shi20}. 
Four of these studies \citep{rubio15,schruba17,miura18,shi20}
resolved the smallest clouds, thus
sampling the cloud population, but only over limited regions of the galaxy.
Moreover, of these, only WLM and Sextans\,B \citep{rubio15,shi20} 
have O/H $\,\leq\,$1/6 \zsun\ [\logoh$\,\la\,$7.9\,], with 
a total of 15 clouds found across both galaxies.
The small-number statistics 
complicate any joint analysis, thus
thwarting interpretation of these observations within a common theoretical framework.

To overcome this challenge, we have obtained observations with the Atacama Large Millimeter/submillimeter Array (ALMA) 
of four nearby star-forming dwarf galaxies beyond the Magellanic Clouds in \cotwo\ at $\sim 3-4$\,pc resolution:
IC\,5152, NGC\,2915, NGC\,3109, and NGC\,5408.
These galaxies have oxygen abundances \logoh\ ranging from 7.7 to 8.2
\citep{berg12,lee03,aniano20}, 
and were previously detected in \cothree\ with the single-dish Atacama Pathfinder EXperiment (APEX).
In what follows, we first describe the individual galaxies in Sect. \ref{sec:sample}.
The APEX observations are reported in Appendix \ref{app:apex}, and
the ALMA observations, and their reduction, in
Sect. \ref{sec:data}. 
We then apply cloud-finding algorithms and present the cloud properties and scaling relations in Sect. \ref{sec:clouds}.
Results are discussed in Sect. \ref{sec:discussion},
followed by our conclusions in Sect. \ref{sec:conclusions}.

\section{Observing sample \label{sec:sample}}

We targeted only galaxies with known dust emission, since dust signals the likely location
of CO emission, especially at low metallicity \citep[e.g.,][]{rubio15,schruba17,shi20}.
According to our estimates\footnote{These estimates are based on galaxies with far-infrared detections,
and the relation between CO luminosity, star-formation rate, and metallicity 
\citep{hunt20}.}, these are among the very few galaxies in which CO clouds could 
be detected at $\leq 4$\,pc resolution with ALMA.
Two targets, NGC\,2915 and NGC\,5408, were observed with \hers\ in the context of 
Key Insights into Nearby Galaxies: Far-Infrared Survey with \hers\ \citep[KINGFISH,][]{kennicutt11},
and the other two, IC\,5152 and NGC\,3109 both dwarf irregulars (dIrrs), by \spit\ within the Local Volume Legacy
\citep[LVL,][]{dale09}.
Table \ref{tab:sample} reports the properties of the sample galaxies,
including metallicity, stellar mass, and star-formation rate (SFR). 

The closest galaxy in our sample, NGC\,3109, at a distance of 1.3\,Mpc, is a member of
the Local Group \citep[LG,][]{vandenbergh99}, 
and is the primary part of a filamentary association of metal-poor dwarf galaxies that includes
Antlia, Sextans\,A, Sextans\,B, and Leo\,P \citep{bellazzini13}.
It is thought to be a ``backsplash'' galaxy,
namely one that was once within the virial radius of the Milky Way (MW) and was subsequently slingshot out of it
\citep[e.g.,][]{banik21}. 
At a distance of 2.0\,Mpc, IC\,5152 lies just beyond the LG \citep{zijlstra99}.
Despite the low metallicity of both these galaxies,
11\%\,\zzsun\ and 17\%\,\zzsun, respectively, both have emission
from PAHs \citep[][]{jackson06,haynes10}.

The two more distant galaxies, NGC\,2915 and NGC\,5408, are classified
optically as blue compact dwarfs \citep[BCDs,][]{meurer94,noeske03},
and both lie within \hi\ disks that extend well beyond the stellar distribution
\citep{meurer96,vaneymeren10}.
Because \hi\ kinematics suggest that dark matter dominates the rotation
curve in NGC\,2915, it has been proposed to be one of the darkest disk galaxies ever
measured \citep{meurer96}, even though its optical core is a BCD.
NGC\,5408 is the only galaxy of the four that hosts an ultra-luminous X-ray source
\citep{kaaret03}.

\section{APEX spectra and ALMA maps \label{sec:data}}

Prior to the ALMA observations, we tested the feasibility of our observing
program using the APEX telescope following the notion that 
``{\it Everything that can be detected with APEX, can be imaged by ALMA}'' (Menten, priv. comm.).
These observations are described in Appendix \ref{app:apex}.

\subsection{ALMA observations and processing \label{sec:alma}}

ALMA \cotwo\ 
maps were obtained using the band 6 receiver ALMA proposal \#2023.1.00227.S (PI Hunt) over Cycles 10 and 11.  
We required $\sim 3-4$\,pc beams and a Largest Angular Size (LAS) of $\sim 460-480$\,pc. 
This was achieved by observations with the 12-m array in two configurations for a total of 45.6\,hr estimated time,
in addition to Atacama Compact Array (ACA) 7-m maps (35.8\,hr), and Total Power (TP) observations 
(36.3\,hr) for the two closest galaxies.
These nearest targets, IC\,5152 and NGC\,3109, were covered by 13 and 22 pointings, respectively,
while the two more distant targets, NGC\,2915 and NGC\,5408 had 3 pointings each.
Like the APEX observations, these mosaics (see Fig. \ref{fig:apexpointings})
are either centered on the infrared peaks (NGC\,2915, NGC\,5408),
determined by \spit/MIPS or \hers/PACS imaging,
or centered to encompass the brightest infrared emission (IC\,5152, NGC\,3109).

The calibration of the data was carried out by the regional ALMA center. The subsequent data reduction and postprocessing was done using the 
pipeline created by the Physics at High Angular resolution in Nearby Galaxies (PHANGS) collaboration
\citep{leroy21}. 
The pipeline consists of three steps: 
\textit{(i) Staging of  the uv-data:} In this step, the 
{\em uv}-data associated with the target galaxy are extracted from the original measurement set, 
the continuum is subtracted, and a line-free continuum data set is produced. 
In order to spectroscopically resolve the small structures,
the channel width for all cubes is left at the intrinsic value of 0.32\,\kms.
\textit{(ii) Imaging and deconvolution of the data:} This step consists of repeated
calls to the \texttt{tclean} task, within CASA \citep[Common Software Applications,][]{casa}, 
which are interleaved with the creation of masks that guide the deconvolution and
checks for convergence. A mixture of multiscale and single-scale CLEAN calls are used during this process.
\textit{(iii) Postprocessing the data:} The pipeline applies
primary beam corrections, convolves the data to have a
circular synthesized beam, combines the interferometric
and total power data, converts the data to have units of Kelvin, and trims and
downsamples the cubes in order to save disk space.
Finally, the images are exported into FITS cubes. 

\begin{table}
\centering
\caption{Properties of the  \cotwo\ data cubes}
\label{tab:almadata}
\resizebox{\linewidth}{!}{
\renewcommand{\arraystretch}{1.3}
\begin{tabular}{lcccc}
\hline
\hline
\\
 &  \multicolumn{1}{c}{IC\,5152} & \multicolumn{1}{c}{NGC\,2915} 
& \multicolumn{1}{c}{NGC\,3109} & \multicolumn{1}{c}{NGC\,5408} \\ 
\\
\hline
\multicolumn{1}{l}{Resolution (beam FWHM)} \T \\
Angular & 0\farcs404 & 0\farcs134 & 0\farcs650 & 0\farcs150 \\ 
Physical (pc) & 3.8 & 2.8 & 4.2 & 3.9 \\
Pixel scale & 0\farcs116 & 0\farcs042 & 0\farcs200 & 0\farcs031 \B \\
\hline 
\multicolumn{1}{l}{1$\sigma$ noise$^\mathrm{a}$} \T \\
mJy\,beam$^{-1}$ & 2.82 & 1.33 & 3.17 & 1.54 \\
K & 0.397 & 1.701 & 0.173 & 1.580 \\
\msun & 31 & 71 & 16 & 126 \B \\
\hline 
\multicolumn{1}{l}{1$\sigma$ surface brightness noise$^\mathrm{a}$} \T \\
\kkms & 0.28 & 1.21 & 0.12 & 1.12 \\
\msunpcsq & 1.9 & 8.1 & 0.8 & 7.5 \B \\
\hline
Largest angular scale \T \\
Angular  & 49\arcsec\ & 22\arcsec\ & 74\arcsec\ & 18\arcsec \\
Physical (pc) & 466 & 458 & 477 & 464 \B \\
\hline
Area mapped $^\mathrm{b}$ \T \\
Angular (arcmin$^2$)& 0.69 & 0.23 & 1.14 & 0.23 \\
Physical (unprojected, kpc$^2$) & 0.23 & 0.35 & 0.17 & 0.54 \B \\
\hline
\hline
\end{tabular}
}
{\small
\begin{description}
\item[$^\mathrm{a}$]~Median noise over the unmasked data cube.
For the velocity-integrated surface brightness and mass sensitivities,
we assume a line width of 5 times the native channel width of 0.32\,\kms,
thus gaining an improvement of $\sqrt{5}$ in the sensitivities
\citep[see also][]{oakes25}.
\item[$^\mathrm{b}$]~Area coverage is defined by summing over the areas
in the coverage map with values $\ge 90$\% 
\citep[see][]{leroy21}.
\end{description}
}
\end{table}

We closely followed the individual steps of the PHANGS pipeline and also used most of the default parameters.  
For the imaging, the CASA gridding algorithm ``mosaic" is used and a Briggs weighting scheme is applied to the 
{\em uv-}data. 
We experimented with the robustness parameter $r$ in order to achieve an optimal combination of resolution, sensitivity and beam shape. 
In general, as $r$ decreases, the beam size decreases and the noise increases. PHANGS uses $r\,=\,0.5$ as a good compromise. 
However, for our data set, we found that $r\,=\,0$ gives a slightly better resolution (between 10\% and 20\%), 
only marginally higher noise ($\sim$ 15\%), and a better shape of the Point Spread Function (PSF), 
avoiding extended wings that were evident for $r=0.5$. 
Thus, in the present work, we use $r\,=\,0$. Some general properties of the final data cubes are given in Table~\ref{tab:almadata}.

After this, the pipeline enables the production of different data products; 
in particular the data cube can be smoothed to different resolutions and 
moment maps can be created. 
Following \citet[][PHANGS]{rosolowsky21},
to estimate signal completeness, we have constructed two masks of CO emission in the cube: 
a ``strict'' mask for high confidence, and a ``broad'' mask for high completeness.
The strict mask follows very closely the standard recipe defined for \texttt{cprops} \citep{rosolowsky06, rosolowsky21}. 
First, a core mask is created with all voxels with signal-to-noise (S/N) above 4 over two successive
velocity channels. Then, a lower S/N 
surrounding mask is made 
that includes all voxels with S/N above 2 in two adjacent 
velocity channels. The final strict mask consists
of all contiguous regions in the outer, surrounding, mask that contain any
pixels from the higher-significance core mask. 

In addition, we have created additional cubes at coarser resolutions than the native one ($\sim\,3-4$\,pc),
namely with beam full-width half-maxima (FWHMs) of 8\,pc, 16\,pc, and 32\,pc. 
We then produced a broad mask by taking the union of all strict masks
from the four different resolutions.
We also inspected the union mask including the 64\,pc and 128\,pc strict masks,
but found that they tend to collect features that can be very large,
up to a few hundred parsecs; 
these structures seem unrealistically large, 
especially compared to the small dimensions of the galaxies.
They also encompass a non-negligible velocity gradient due to the rotation
curve of the galaxy, rendering the measured velocity dispersions not representative
of the internal motions within the CO aggregations.
The aim of the strict and broad masks is to probe different hierarchical scales in the structure of the ISM
(see Sect. \ref{sec:clouds}).
Unlike the 32\,pc union mask, the higher resolutions, 8\,pc and 16\,pc, do not substantially
increase the range of scales probed by our data, relative to the native resolution.
Thus, for the broad mask, 
we adopt the 32\,pc union mask;
hereafter, unless otherwise specified, this mask will be referred to as the broad mask.

\begin{figure*}[!t]
\centering
\includegraphics[width=0.82\textwidth]{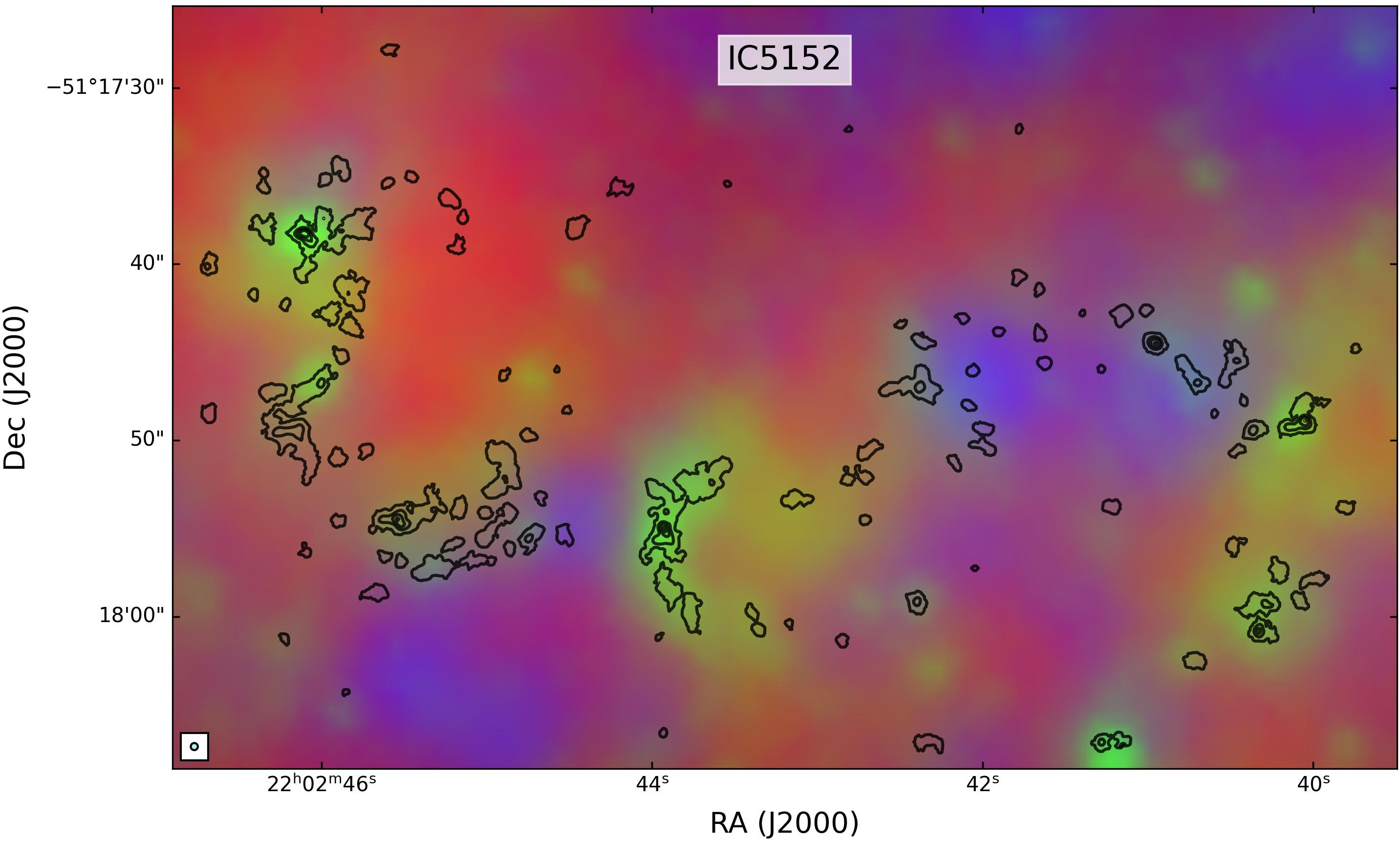} \\
\vspace{0.5\baselineskip}
\includegraphics[width=0.49\textwidth]{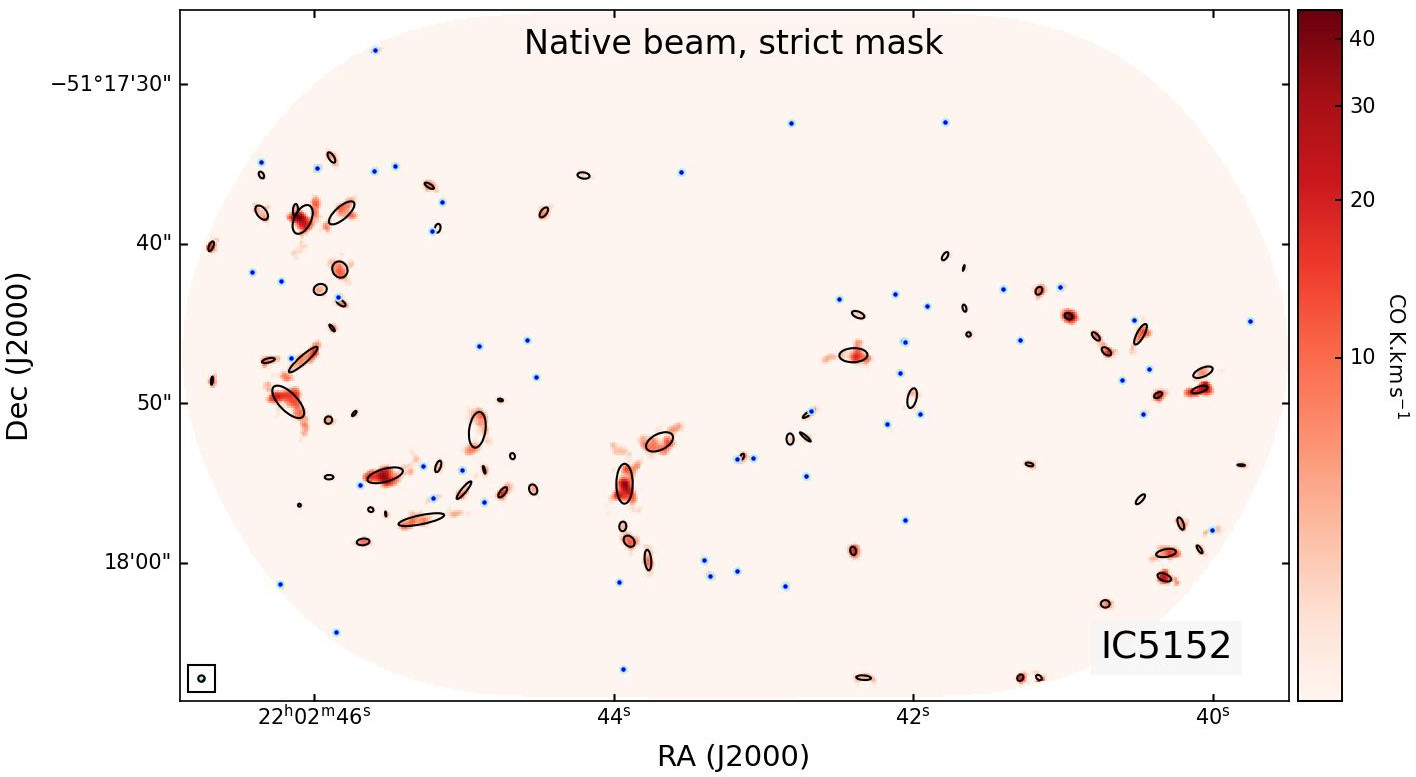} 
\includegraphics[width=0.49\textwidth]{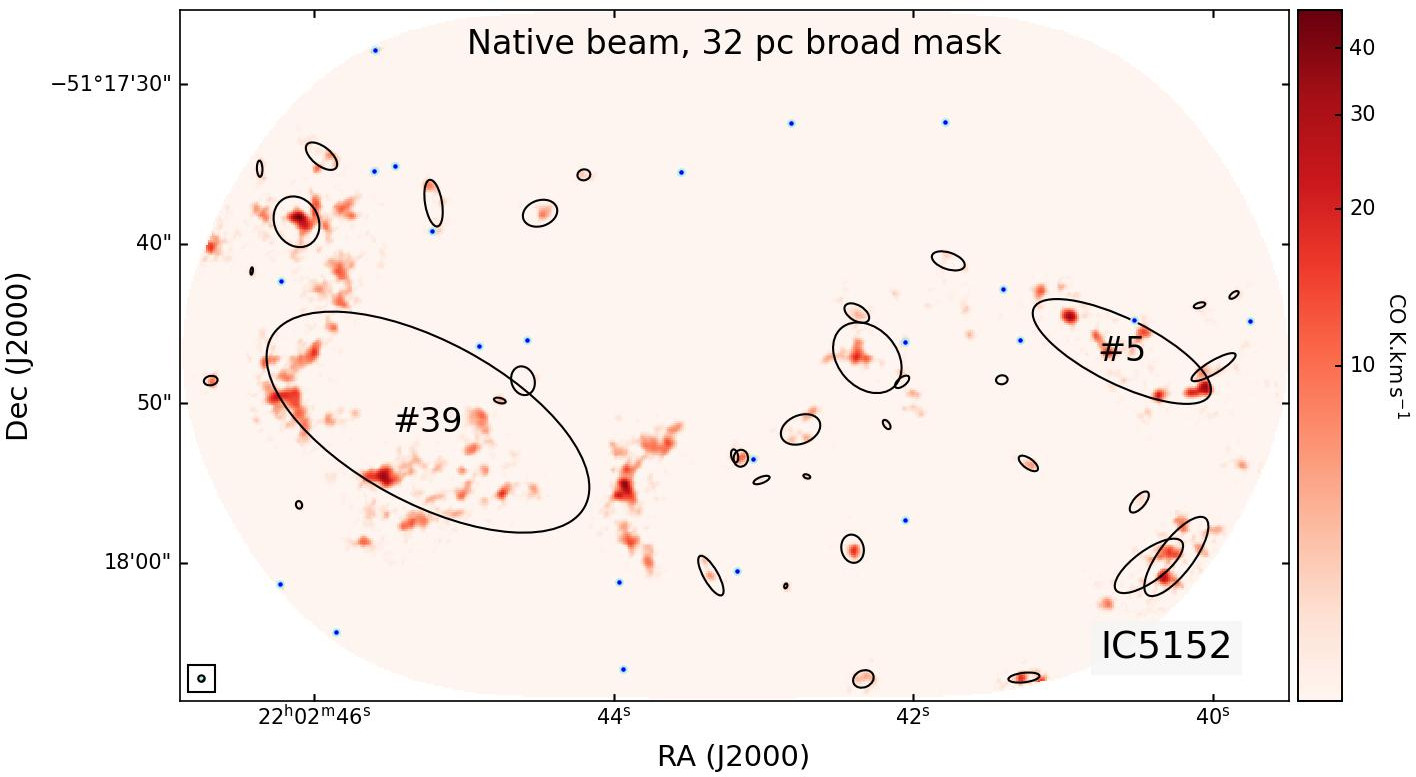} \\
\caption{Top panel: \cotwo\ integrated intensity contours (zero moment with strict mask) overlaid on the RGB image of IC\,5152
with red given by 70\,\micron\ (MIPS),
green by 8\,micron\ (\spit/IRAC channel 4),
and blue by NUV/GALEX \citep{gildepaz07,dale09}.
Bottom panels:
Integrated intensity zero-moment \cotwo\ map with an arcsinh color stretch,
and the minimum color corresponding to the 2$\sigma$ surface brightness noise
given in Table \ref{tab:almadata}.
The ALMA beam is shown in the lower left corner of each panel.
The locations of the selected clouds described in Sect. \ref{sec:cloudfinding} are overlaid as ellipses and filled 
blue circles;
the ellipses represent the deconvolved major/minor axes and position angles of the resolved
clouds, and the filled circles the unresolved ones.
The differences in the two bottom panels result from the two different masks applied before cloud selection and to the moment maps:
native strict mask (left) and broad 32\,pc (union) mask (right).
Broad-mask clouds \#5 and \#39 are labeled.
}
\label{fig:moment5152}
\end{figure*}

The integrated intensity \cotwo\ (zero-moment) map of IC\,5152 is shown in Fig. \ref{fig:moment5152}; 
the top panel shows CO contours overlaid on an RGB
rendering, where red is given by 70\,\micron\ (\spit/MIPS) 
to trace warm dust,
green by 8\,micron\ (\spit/IRAC channel 4) for PAH emission at 7.7\,\micron,
and blue by NUV/GALEX \citep{gildepaz07,dale09} for star formation.
The bottom panels show the zero-moment map as an arcsinh color stretch,
with the locations and shapes of the clouds identified in Sect. \ref{sec:cloudfinding}.
The remaining galaxies appear in Figs. \ref{fig:moment2915}, \ref{fig:moment3109}, and \ref{fig:moment5408}
(see Appendix \ref{app:moments}).
The ALMA coverage relative to the entire galaxy is illustrated in Fig. \ref{fig:apexpointings}.
The CO emission generally follows PAH emission (green tones), as also found by \citet{chown25} for WLM and NGC\,6822,
and tends to be associated with star formation (blue), and/or cool dust (red).
Similar trends are seen for the other galaxies.

Throughout this paper, we use a constant \aco\ to convert CO flux/luminosity to
\htwo\ mass, \aco$_{(1-0)}$\,=\,4.35\,\msun\,(K\,\kms\,pc$^2$)$^{-1}$
\citep{bolatto13}, including the contribution of helium,
together with a CO(2-1)-to-CO(1-0) line ratio \rtwo\ of 0.65
\citep{rosolowsky21,leroy22,schinnerer24}. 
These are meant to be only fiducial values, which we investigate
in Sect. \ref{sec:discussion}.

\begin{table}
\centering
\caption{Total CO luminosities and cloud statistics}
\label{tab:co}
\resizebox{\linewidth}{!}{
\renewcommand{\arraystretch}{1.3}
\begin{tabular}{lrrrr}
\hline
\hline
\\
 &  \multicolumn{1}{c}{IC\,5152} & \multicolumn{1}{c}{NGC\,2915} 
& \multicolumn{1}{c}{NGC\,3109} & \multicolumn{1}{c}{NGC\,5408} \\ 
\\
\hline
\multicolumn{5}{l}{Total log$_{10}$(\lcotwo/\kkmspc)} \T \\
Unmasked cube             & 4.93 & 4.03 & 4.26 & \multicolumn{1}{c}{$~~~~~-$} \\
Native strict masked cube & 4.62 & 3.96 & 3.89 & 3.72 \\
Broad (union) masked cube & 4.84 & 4.80 & 4.08 & 4.58\B \\
\hline
\multicolumn{5}{l}{Total log$_{10}$(\lcotwo/\kkmspc) in the clouds$^\mathrm{a}$} \T \\
Native strict mask      & 4.62 & 3.96 & 3.89 & 3.72 \\
Broad 32\,pc union mask & 4.83 & 4.79 & 4.08 & 4.58\B \\
\hline 
\hline 
\multicolumn{5}{l}{Total \lcotwo\ in the cores (native strict mask) with S/N $\geq 3^\mathrm{b}$ } \T \\
Number of cores              & 124 & 65 & 49 & 35 \\
Number of resolved cores     & 72  & 14 & 29 & 5 \\
log$_{10}$(\lcotwo/\kkmspc)$^\mathrm{a}$ & 4.62 & 3.96 & 3.89 & 3.72\B \\
\hline
\multicolumn{5}{l}{Total \lcotwo\ in the clouds (broad union mask) with S/N $\geq 3^\mathrm{b}$} \T \\
Number of clouds              & 58 & 69 & 45 & 44 \\
Number of resolved clouds     & 36 & 21 & 35 & 15 \\
log$_{10}$(\lcotwo/\kkmspc)$^\mathrm{a}$ & 4.83 & 4.79 & 4.08 & 4.57\B \\
\hline 
\hline 
\multicolumn{5}{l}{Total \lcotwo\ in the cores (native strict mask) with S/N $\geq 5^\mathrm{b}$ } \T \\
Number of cores              & 99 & 14 & 35 & 4 \\
Number of resolved cores     & 67  & 8 & 26 & 3 \\
log$_{10}$(\lcotwo/\kkmspc)$^\mathrm{a}$ & 4.61 & 3.79 & 3.88 & 3.52\B \\
\hline
\multicolumn{5}{l}{Total \lcotwo\ in the clouds (broad union mask) with S/N $\geq 5^\mathrm{b}$} \T \\
Number of clouds              & 36 & 6 & 26 & 3 \\
Number of resolved clouds     & 29 & 3 & 24 & 2 \\
log$_{10}$(\lcotwo/\kkmspc)$^\mathrm{a}$ & 4.83 & 4.75 & 4.07 & 4.32\B \\
\hline
\hline
\end{tabular}
}
{\small
\begin{description}
\item[$^\mathrm{a}$]~To be directly comparable with the total CO flux in the cube,
we have adopted the un-extrapolated values of \lcotwo\ for each cloud (see Sect. \ref{sec:cloudprops}).
\item[$^\mathrm{b}$]~The cloud tallies include only those with S/N thresholds as specified on the peak \tb\
within the structure. As described in the main text (see Sect. \ref{sec:cloudprops}), we call clouds defined with the strict mask
``cores'', and those with the broad union mask ``clouds''.
\end{description}
}
\end{table}

\subsection{Total CO luminosities \label{sec:coflux}}

Following \citet{leroy21}, 
we assessed the noise properties of the cubes by examining the distribution of the noise and the S/N in three dimensions.
The noise does not strongly depend on frequency within the spectral window.
Like for the PHANGS galaxies,
the S/N is dominated by a Gaussian-like distribution centered around zero with unit width, and a skewness
toward larger \tb\ values
\citep[see e.g., Fig. 16:][]{leroy21}. 
Thus, given that the noise distribution is well behaved, like \cite{leroy21}, 
we simply sum over the cube to obtain the total luminosity, \lco.
Specifically, we calculate:
\begin{equation}
\mathrm{L^\prime_{CO,\,total}}\ \mathrm{(K\,km\,s^{-1}\,pc^2})\,=\,\mathrm{D_{pc}^2}\ \sum_i \mathrm{T}_{\mathrm{B},i}\ \delta \varv\ \Omega_\mathrm{pixel} \quad ,
\label{eqn:lco}
\end{equation}
\noindent
where $i$ is the sum over all voxels, of channel width $\delta \varv$, 
with D$_\mathrm{pc}$ the distance to the galaxy in parsecs,
and $\Omega_\mathrm{pixel}$ the pixel solid angle. 
This value can be compared to the analogous sums of CO emission at native resolution within the strict mask,
and within the broad mask (as the union of all strict masks up to a 32\,pc beam size).
The results are reported in Table \ref{tab:co}.
Throughout the paper, we rely only on these two masks, strict and broad (32\,pc), applied to the cube at native resolution.

The broad masks for IC\,5152 and NGC\,3109 capture between 
81\% and 66\%
of the total CO flux, respectively;
these are also the galaxies for which it was necessary to obtain TP in order to ensure the LAS of $\sim 460-480$\,pc. 
For NGC\,2915, instead, the broad mask contains $\sim 6$ times more CO flux than the sum
obtained from the unmasked cube itself.
This can occur when there is little signal, because the masked regions do not include the negative noise fluctuations
that are contained in the sum of the cube \citep[see][]{leroy21}. 
The problem is even more severe for NGC\,5408, where the total sum of the unmasked cube is slightly negative.
These two galaxies were not observed in TP, since the LAS was achieved by just combining the 12-m and 7-m arrays.
Because of their larger distance, and our wish to maintain a constant physical resolution,
NGC\,2915 and NGC\,5408 have the smallest beams,
as well as the lowest noise per beam (see Table \ref{tab:almadata}).
It is not clear how to otherwise recover the total CO flux for these two galaxies except through masking.

\vspace{-\baselineskip}
\section{Molecular clouds at low metallicity \label{sec:clouds}}

Given the paucity in the literature of statistical samples of molecular clouds 
at high resolution in low-metallicity galaxies,
our observations 
are designed to detect the small molecular clouds expected to reside in such metal-poor environments.
Here we describe the search for molecular entities traced by CO in our sample galaxies.

\subsection{Cloud finding \label{sec:cloudfinding}}

The primary goal of this study is to evaluate the dynamical state of molecular clouds in each galaxy,
and the  key properties believed to impact star formation and feedback processes. 
The exquisite angular resolution of the ALMA observations reveals distinct, compact CO structures that are closely 
clustered both spatially and kinematically. 
The aggregation of clustered clumps likely comprises a larger molecular cloud.  To account for this hierarchical structure, 
we apply 
both the strict- and the broad 32\,pc union masks, described in Sect. \ref{sec:alma},
to the native-resolution cube.
The value of this 
comparison
is to connect the distinct features identified at the native resolution to form larger structures while also collecting 
faint emission, if present, in the areas between the compact clumps. 
The masks define the regions where there is CO signal for cloud definition. 

Localized regions of \cotwo\ emission 
are identified in position-position-velocity (ppv) space using 
the native resolution cube with the two different masks. 
These are the two extremes of the masking regimes, from strict at the highest resolution to
broad encompassing the largest scales, and their comparison should provide insight into the
nature of the CO clouds at low metallicity.
The {\tt pycprops} module, a \texttt{python} implementation of the {\tt cprops} package \citep{rosolowsky06,rosolowsky21},
finds the set of connected voxels within the mask and assigns a cloud label. 
As we are interested in the global properties of the 
clouds, we do not further decompose the clouds 
nor do we construct dendrogram charts \citep{rosolowsky08}.

\begin{figure*}[t!]
\centering
\includegraphics[height=0.24\linewidth]{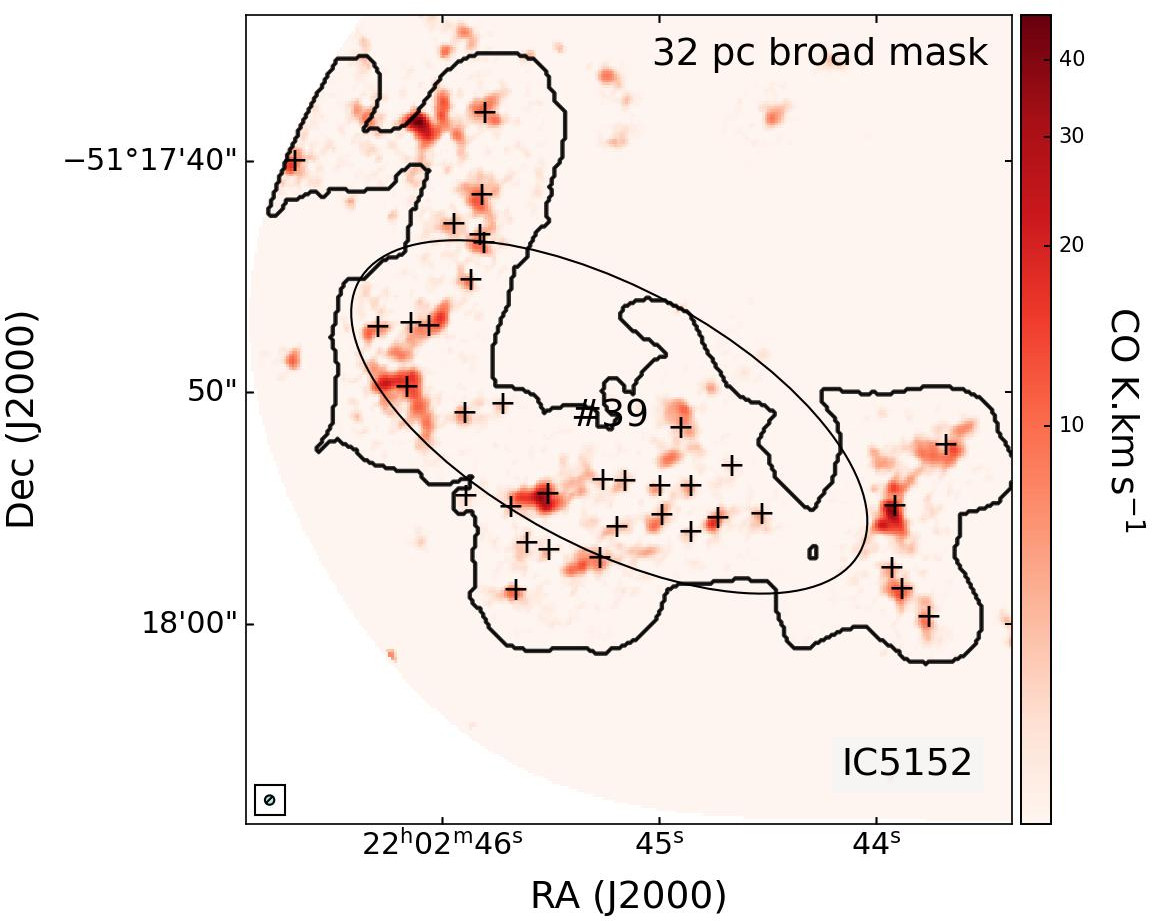}
\hspace{0.01\linewidth}
\includegraphics[height=0.23\linewidth]{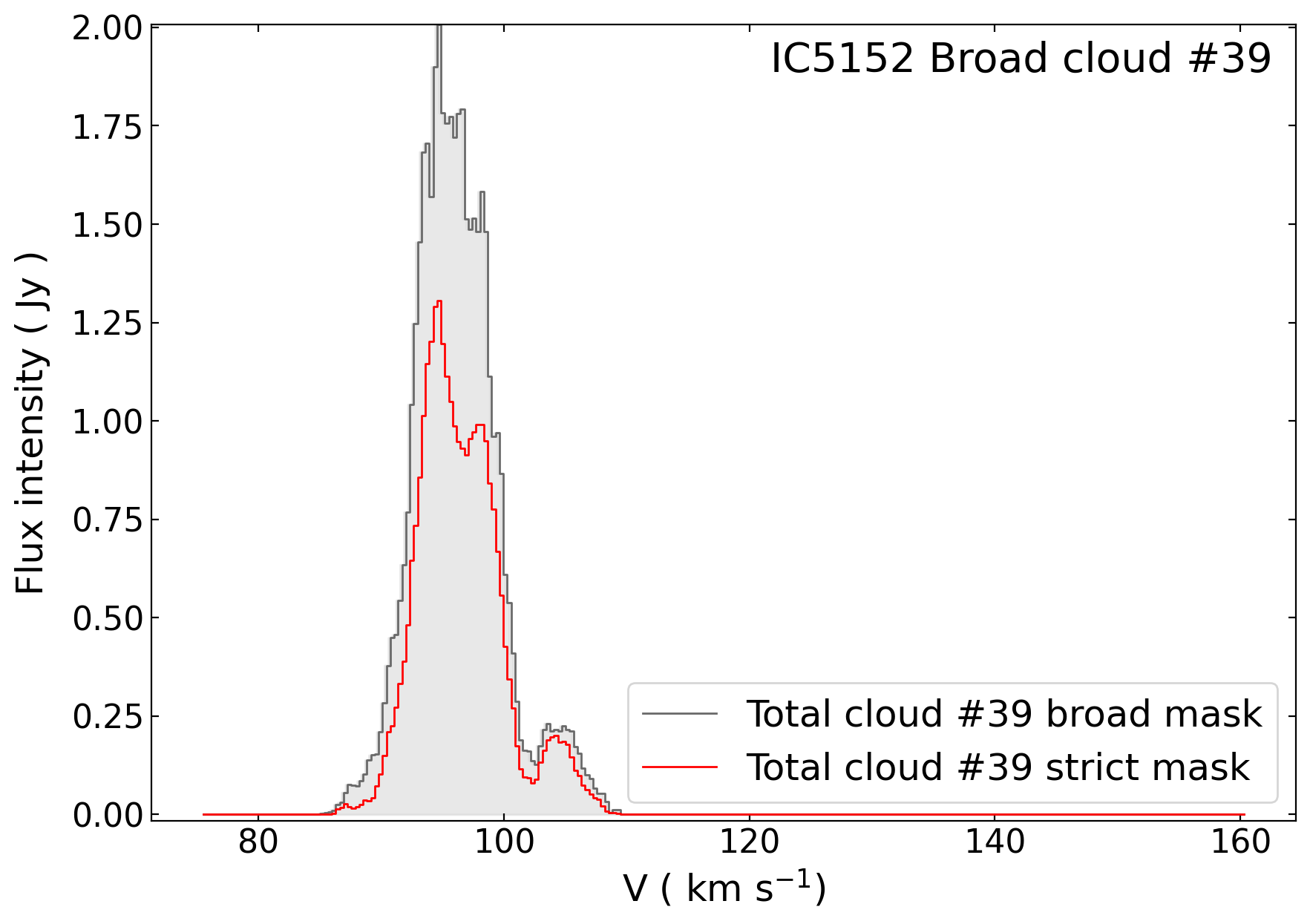}
\hspace{0.005\linewidth}
\includegraphics[height=0.23\linewidth]{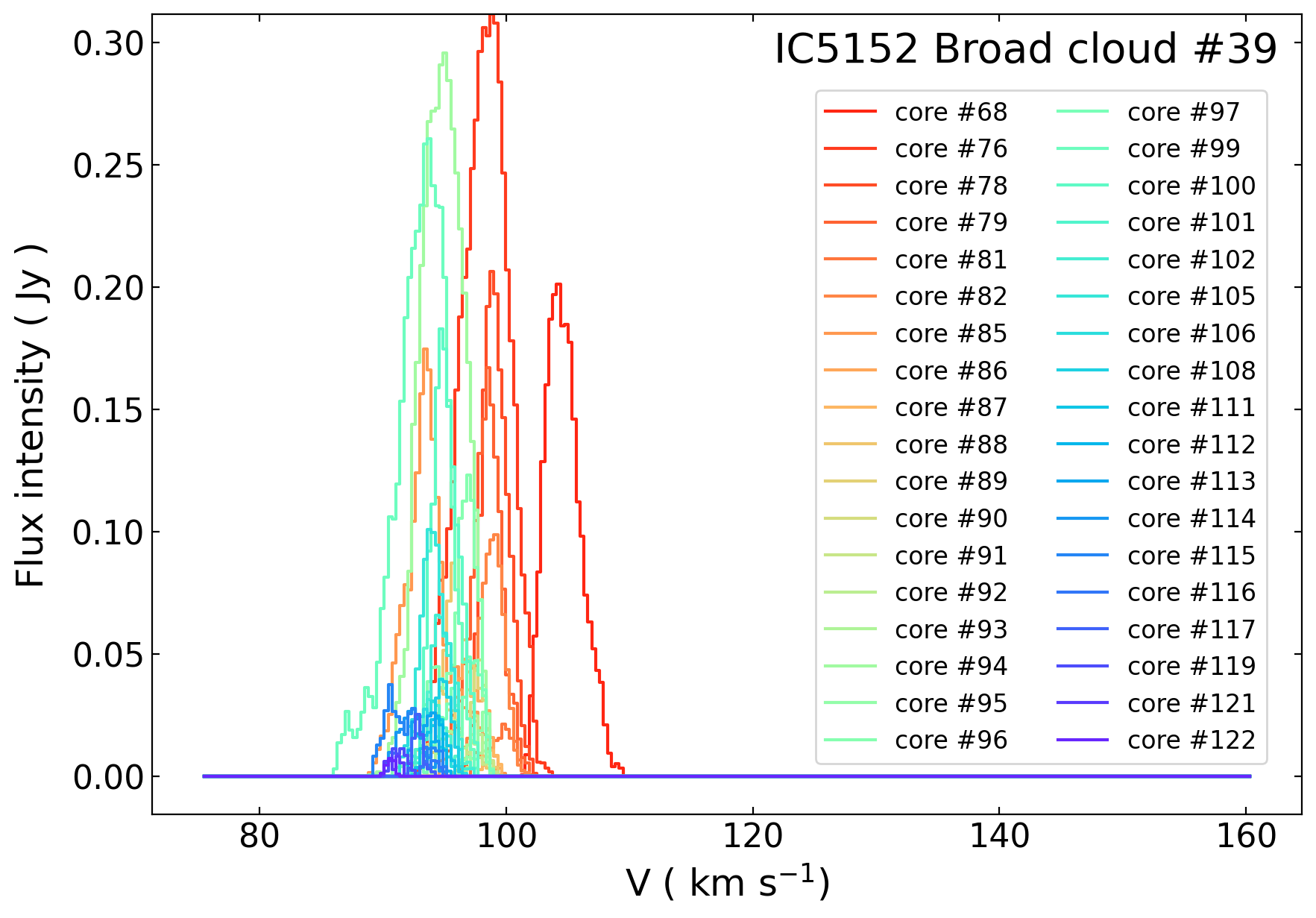} \\
\hspace{0.03\linewidth}
\includegraphics[height=0.24\linewidth]{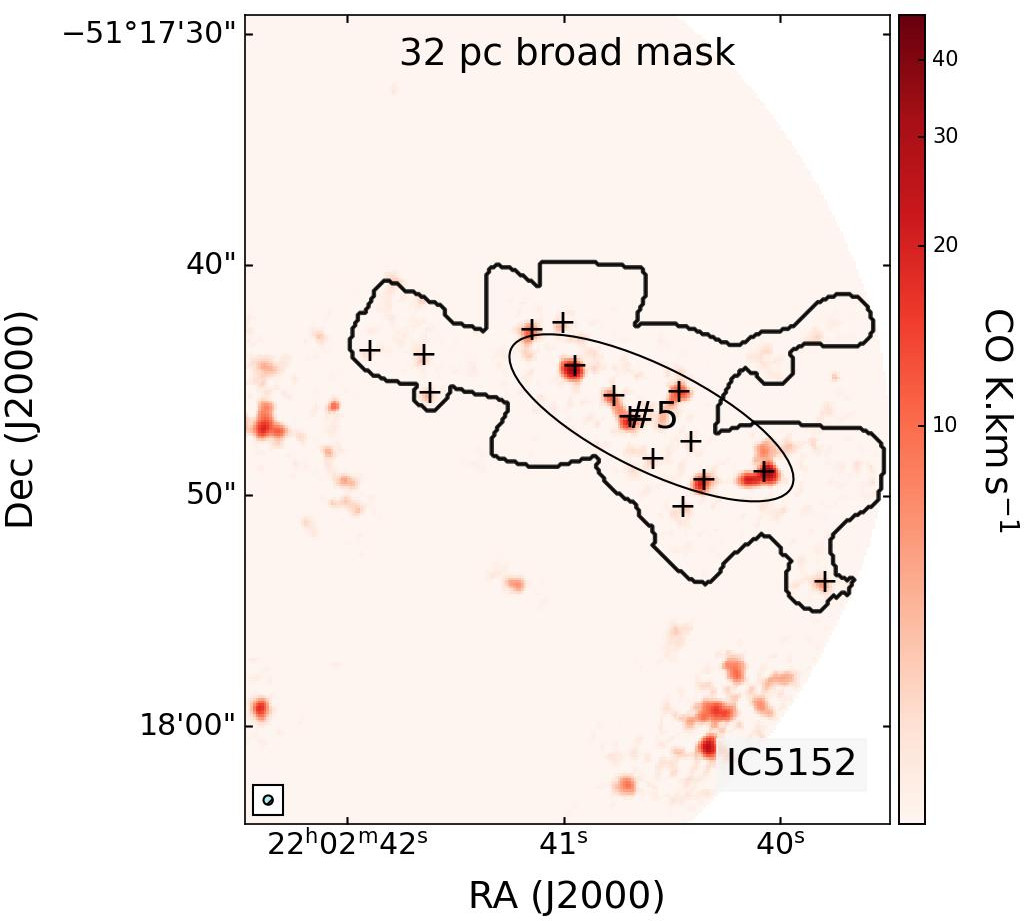}
\hspace{0.015\linewidth}
\includegraphics[height=0.23\linewidth]{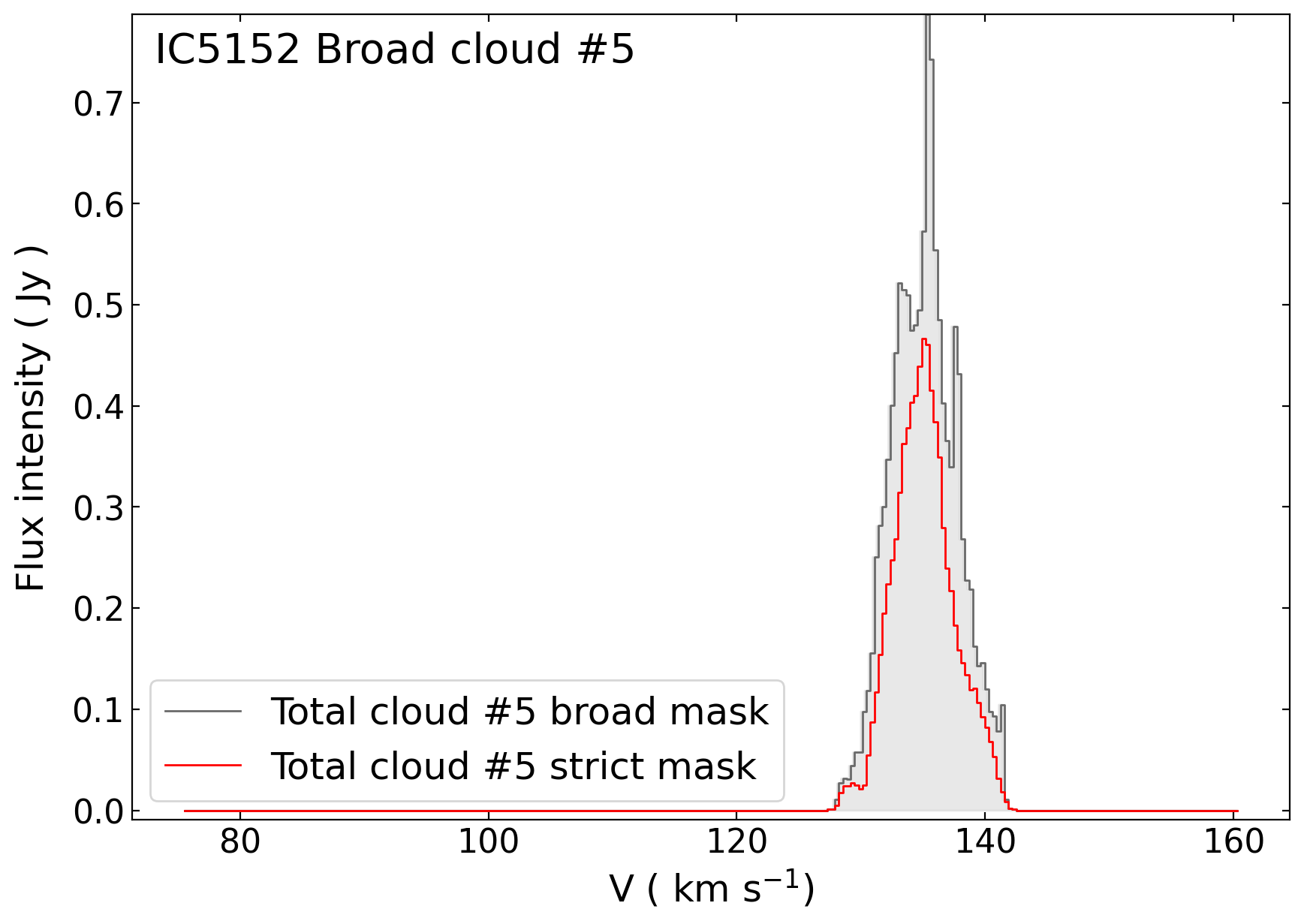}
\hspace{0.005\linewidth}
\includegraphics[height=0.23\linewidth]{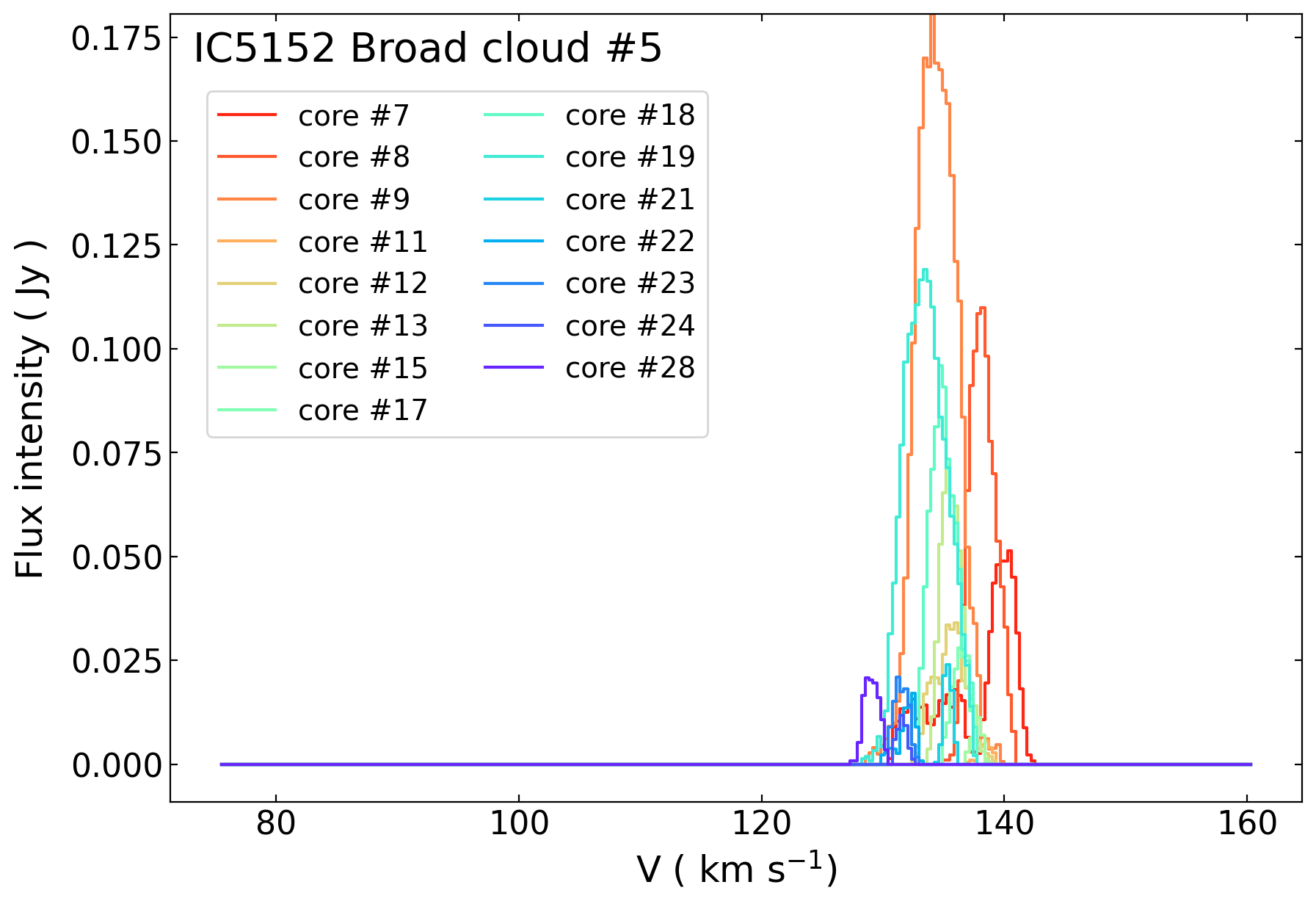}
\caption{A close-up view of the labeled broad-mask Clouds \#39 (top row) and \#5 in IC\,5152 (bottom) overlaid on the \cotwo\ moment zero map.
Left panels: the broad mask (heavy black) contours of the extent of the emission associated with the clouds, together with the elliptical
approximation of its dimensions.
The 36 small cores within Cloud \#39 (top) and the 15 cores within Cloud \#5 (bottom) are shown as $+$ signs.
Middle panels: the spectra of the broad-masks Cloud \#39 (top) and \#5 (bottom) are shown as gray filled histograms,
and the total spectra of the sum of the emission from the cores is given by the red curves.
Right panels: the individual spectra of the 36 (15) cores found in the strict masks within 
Cloud \#39 (\#5) identified by the broad masks.
}
\label{fig:ic5152_cloud5_39}
\end{figure*}

\begin{figure*}[h!]
\centering
\includegraphics[width=0.49\linewidth]{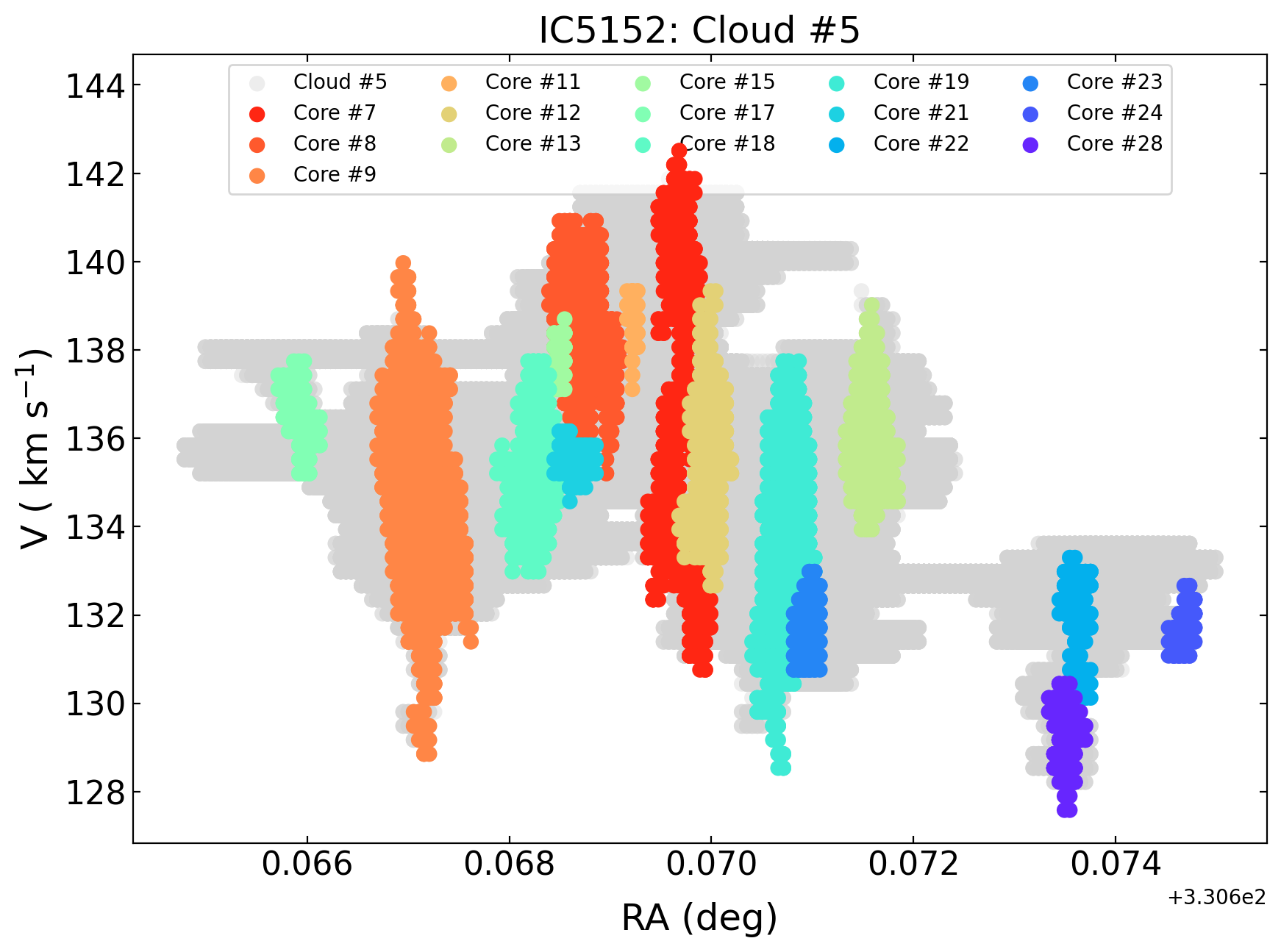}
 \hspace{0.01\linewidth}
\includegraphics[width=0.49\linewidth]{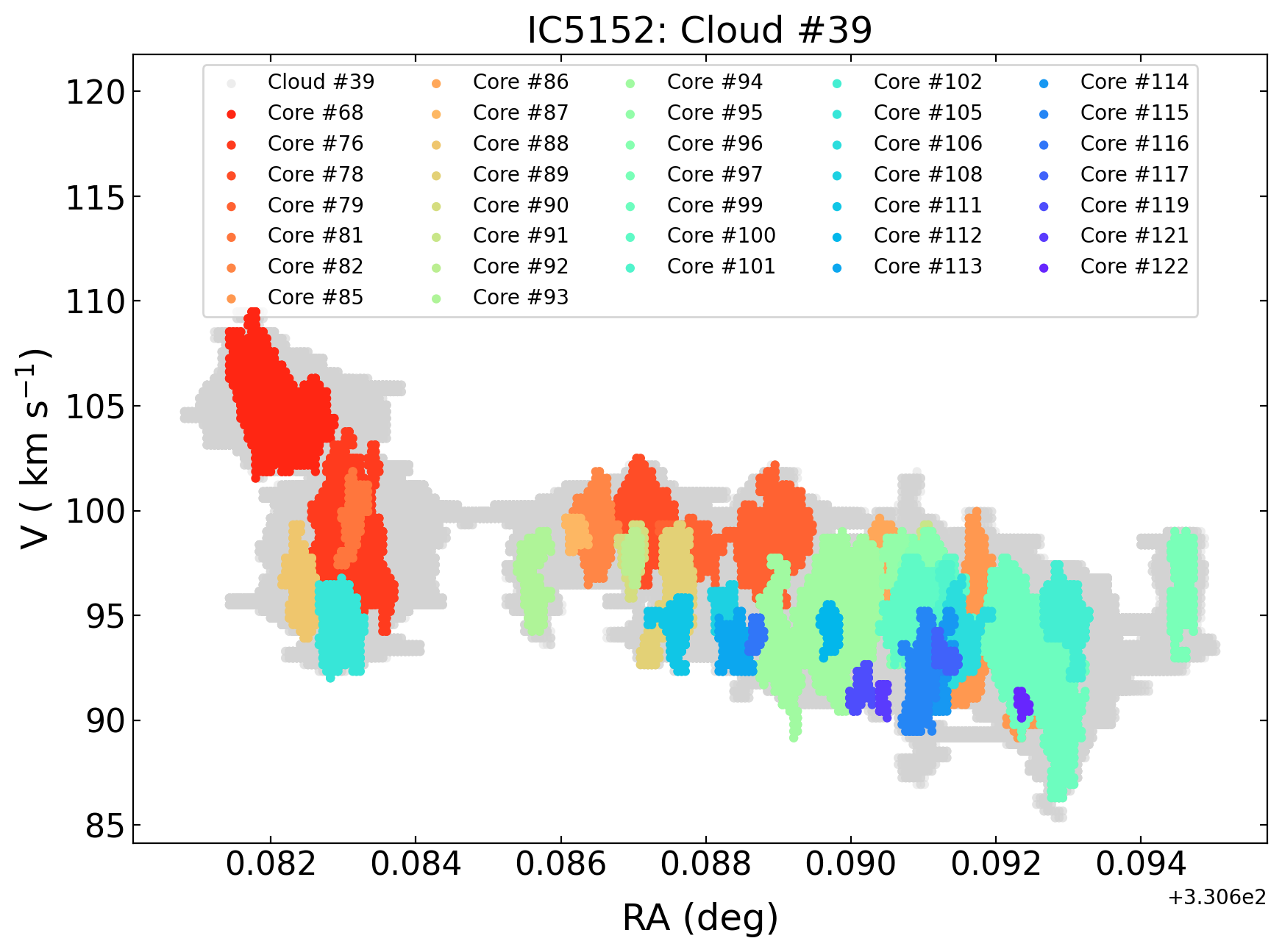}
\caption{The velocity structure of the voxels within the labeled broad-mask Clouds \#5 (left panel) and \#39 (right) in IC\,5152 plotted against RA.
The rotation axis for IC\,5152 is roughly oriented north-south.
The overall broad-mask cloud emission is shown as light gray regions, while the 
individual cores are color coded as in the legend 
(see also Fig. \ref{fig:ic5152_cloud5_39}).
}
\label{fig:ic5152_cloud5_39_velRA}
\end{figure*}

\subsection{Cloud properties \label{sec:cloudprops}}

In addition to cloud labeling, {\tt pycprops} carries out several other critical tasks.  
Cloud sizes, velocity dispersions, and fluxes are initially calculated based on the voxels defined by the mask.  
These are determined by calculating various moments of the
emission along the spatial and spectral axes of the emission within the mask:
the line widths come from the intensity-weighted variance along the spectral direction,
and the sizes from the intensity-weighted second moments over the two spatial axes.
Specifically, the spatial variances in the x and y directions of the cube
are calculated within the isosurface that defines the boundary of the cloud through the mask, together with a covariance term. 
These are placed in a covariance matrix that is then diagonalized to determine
the major and minor axes of the distribution of the emission, together with its position angle.
The fluxes rely on the zeroth moment, namely the sum of the intensity within the mask.
For more details, see \citet{rosolowsky06} and \citet{rosolowsky21}.

These quantities are calculated within the mask, so they only consider emission above the threshold defined by the mask.
However, a more useful boundary of the clouds is the isophote where the brightness temperature is zero.  
Accordingly, {\tt pycprops} extrapolates these properties to those that would be inferred using the ideal \tb\,=\,0\,K threshold.  
Subsequently, these extrapolated spatial and velocity (square root, rms) variances are deconvolved with respect to the beam size, and to the spectral resolution
set in the PHANGS pipeline, respectively. 
The velocity dispersion \sigmav\ is defined by the deconvolved velocity rms variation.
The deconvolved cloud size R is calculated as the geometric mean of the deconvolved rms variations 
along the major and minor axes of the ellipse defined by the diagonalized covariance matrix, 
$\sigma_\mathrm{maj,d}$, $\sigma_\mathrm{min,d}$:
\begin{equation}
\mathrm{R}\,=\,\eta \sqrt{\sigma_\mathrm{maj,d}\,\sigma_\mathrm{min,d}}
\label{eqn:radius}
\end{equation} 
Following \citet{rosolowsky21}, we adopt $\eta\,=\,\sqrt{2\,\ln{2}}\,=\,1.177$,
which corresponds to the conversion of the spatial rms variation $\sigma$ to the half-width at half-maximum (HWHM),
assuming a two-dimensional Gaussian distribution of the surface brightness of the cloud.

For clouds with initial angular sizes less than or commensurate with the beam size, {\tt pycprops} assigns the
cloud radius to the not-a-number (nan) value.  
{\tt pycprops} also calculates uncertainties of several key properties using the bootstrap method.  
Our application used 256 iterations of bootstrapping.
Details of these calculations are described in \citet{rosolowsky06} and \citet{rosolowsky21}.

Although for structural properties we generally adopt the extrapolations to \tb\,=\,0\,K,
when we compare flux in the clouds with flux in the cube, we use the un-extrapolated values
(see Table \ref{tab:co}).
This is to ensure that additional, extrapolated, flux does not confound the comparison.

Applying the methods described in Sect. \ref{sec:cloudfinding} to our data processed with 
Briggs weighting and a robust parameter $r$ of 0, 
we identify two sets 
of molecular clouds for each galaxy 
at the native resolution: 
one using the strict mask, and another using the broad 32\,pc union mask.  
The numbers of identified clouds, resolved clouds, and the total CO luminosity in the clouds within each mask 
are reported in Table \ref{tab:co}.
For each galaxy, the total CO luminosity in the clouds, 
\lco$_\mathrm{,clouds}$,
is just 
$\sum_i \mathrm{L}^\prime_\mathrm{CO,cloud,i}$ 
summed over the clouds [see also Eq.\,\eqref{eqn:lco}],
with $\mathrm{L^\prime_{CO,cloud,i}}$ as derived by \texttt{pycprops}. 
The basic properties of the cores/clouds for each galaxy are compiled in Tables \ref{tab:IC5152-props-strictmask}-\ref{tab:NGC5408-props-Broad32} 
(see Appendix \ref{app:proptables}).

Considering a S/N$\geq 3$ (S/N$\geq 5$) on the peak \tb\ within the individual masked structures (see Table \ref{tab:co}),
in total across the sample, we find 273 (152) 
clouds with the strict mask,
of which 120 (104) 
are resolved at $\sim 3-4$\,pc resolution;
the broad mask gives fewer clouds, 216 (71), 
of which 107 (58) 
are resolved.
To ensure robust cloud properties (Sect. \ref{sec:cloudscaling}),
we consider only the resolved clouds with S/N $\geq 5$ on the peak brightness temperature
within the cloud, \tb\ (see Table \ref{tab:co}),
as output by \texttt{pycprops}. 
Hereafter, the clouds found within the strict mask will be designated as cores,
and those with the broad mask as clouds.

It can be seen from Table \ref{tab:co} that the CO emission contained in the masked cube is the same to within a few percent
as the CO contained in the cores and clouds identified with the masks.
This is a feature of 
\texttt{pycprops}, and the procedure used to generate the masks,
in which a seeded watershed algorithm associates pixels 
around a local maximum, thus allocating virtually all of the available emission into clouds \citep{rosolowsky21}.

\subsection{Cores, clouds, and masks \label{sec:beamdilution}}

The difficulty in defining clouds is broadly illustrated in the bottom panels of Fig. \ref{fig:moment5152},
where the locations and sizes of the cores and clouds identified within the strict and broad masks 
are shown for IC\,5152 (see Appendix \ref{app:moments} for the remaining galaxies).
A more detailed example is given in Fig. \ref{fig:ic5152_cloud5_39} for Clouds \#5 and \#39 identified within the broad mask.
Within these larger clouds in IC\,5152, there are 15 and 36 smaller cores, respectively,
found with the strict masks. 
The spectra of the individual 
cores are shown in the right panels, and their sum, as red histograms in the middle panels.
The sum of the cores comprises 
63\% (64\%) 
of the emission of the Cloud \#39 (\#5) spectrum, shown in gray in the middle panels.
These fractions are close to the overall fraction of CO emission found in the strict mask relative to the broad mask
(see below).

Other broad-mask clouds also contain multiple cores, such as Cloud 
\#13 in IC\,5152 with 7 cores; 
Clouds \#33 and \#43 in NGC\,2915 have 6 and 5 cores, respectively. 
There are also other clouds with 4 cores or fewer in NGC\,2915 and NGC\,3109
(see Fig. 
\ref{fig:ngc3109_cloud25});
NGC\,5408 has only clouds with 1 or 2 cores.
In NGC\,3109, at a metallicity of $\sim 11$\%\,\zzsun, the compact emission in the 
two clouds with 4 cores (\#17, \#25) 
comprises 
between 60\% and 70\% of the 
emission found with the broad mask.
IC\,5152, slightly more metal rich with $\sim 17$\%\,\zzsun, shows similar fractions, 
with the sum of the cores equivalent to 63-64\% of the cloud emission. 
Even at the low metallicities of these galaxies,
the CO is not strictly confined to small, compact cores,
but rather there is 
faint emission beyond them,
possibly as connecting filaments or bundles,
or a multitude of faint unresolved sources.

To assess the total amount of faint CO emission outside the cores, identified by the strict mask,
we can compare the overall CO content in the strict and broad 
masks (Table \ref{tab:co}). 
For IC\,5152 and NGC\,3109 (with S/N$\geq$3), we find that 
the strict mask contains
$\sim 62-65$\%
of the flux identified in the broad mask, 
and for NGC\,2915 and NGC\,5408, 
the fraction is $\sim 15$\% for both galaxies.
Given the relatively large numbers of clouds (broad mask) and cores (strict) 
(see Table \ref{tab:co}),
this could be considered a fairly robust comparison. 
Considering clouds with a S/N on the peak \tb\ of 5 rather than 3 (see Table \ref{tab:co}),
the fractions do not change significantly (although the fraction for NGC\,2915 drops to 11\%).
It appears that roughly half or more of the CO emission in our sample galaxies is found outside the dense cores
identified by the strict masks.
The brightness temperatures from these outer envelopes are too faint to be included
in the strict mask, but detectable CO flux can accumulate above the noise when summing over the
larger areas of the broad-mask clouds.

As will be investigated in Sect. \ref{sec:cloudscaling},
the difference in the properties of cores (found by strict masks) and clouds (broad masks)
is a consequence of the hierarchical structure of the ISM
\citep[e.g.,][]{stanimirovic01,vazquez19,hacar23},
and the nature of the watershed algorithms (here \texttt{pycprops}) and masks used to identify
molecular aggregations.
Because the masking is based on singling out local maxima in the spectral cube
(see Sect. \ref{sec:alma}),
clouds separated by less than the beam size tend to be blended together into
larger structures.
Thus, the derived properties of cores and clouds directly depend 
on the applied mask and the data resolution, since watershed algorithms
with a strict mask
tend to aggregate emission into regions comparable to the beam size
\citep{pineda09,leroy16,rosolowsky21}.
Therefore, the broad union mask enables larger structures to be identified.

Figure \ref{fig:ic5152_cloud5_39_velRA} illustrates the velocity structure for all the voxels
within Clouds \#5 and \#39 in IC\,5152; the overall broad-mask cloud emission is shown as light gray, and the individual cores
are color coded (arbitrarily) as reported in the legend.
These are the largest clouds in IC\,5152, and also
contain systematic velocity gradients that are present at various levels inside the dispersions measured by \texttt{pycprops}. 
Consequently, the clouds identified with the broad mask will show a general trend 
of increasing internal kinematic energy, which is a combination of core-to-core dispersions or turbulence,
and systematic motions possibly including rotation or shear.
However, it is interesting that the cores show roughly an equal spread in velocity
as the overall cloud, implying some degree of velocity coherence even in the broad-mask clouds.

\begin{figure*}[t!]
\centering
\includegraphics[width=\linewidth]{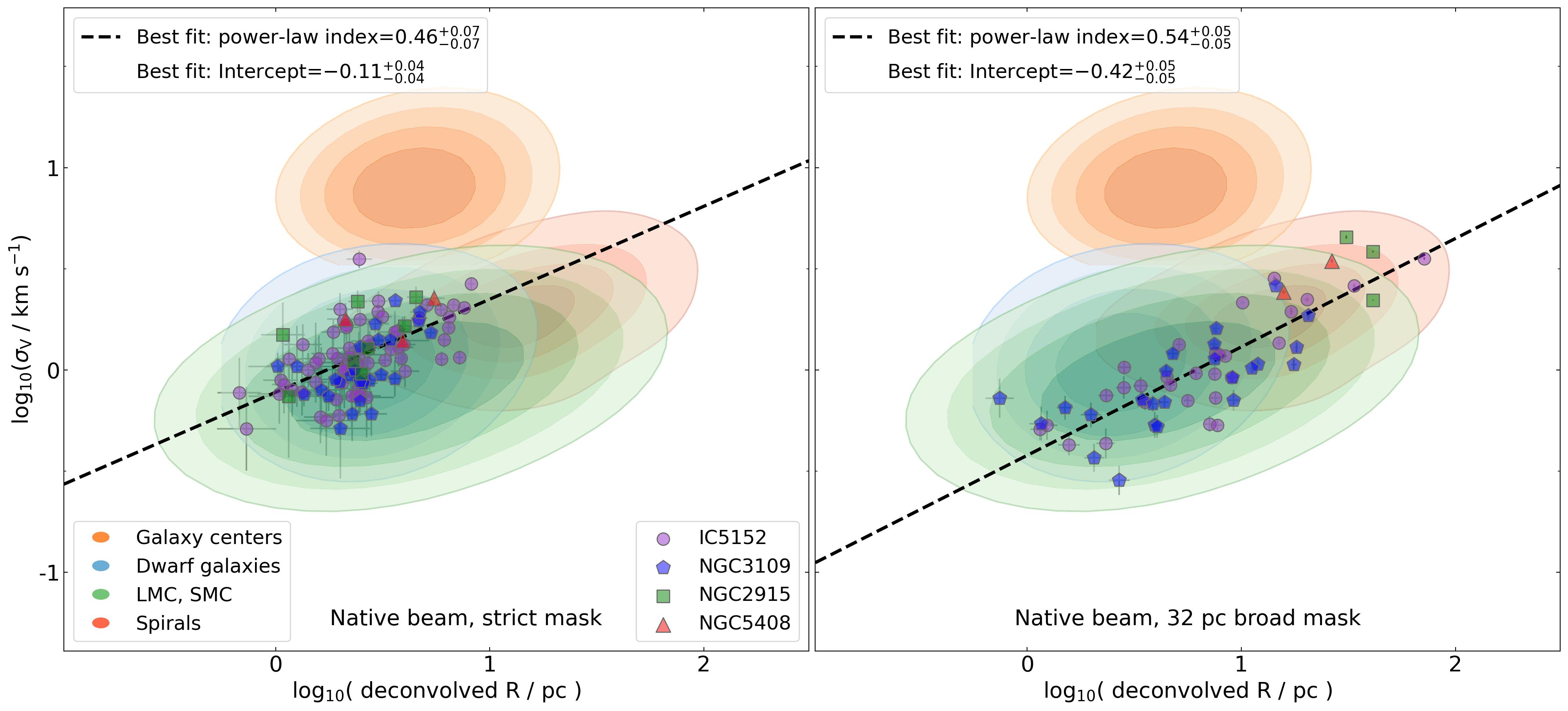}
\caption{Line-width \sigmav\ vs. cloud radius (deconvolved R) for the clouds in our sample selected with two masks:
the native-resolution strict mask (left panel), and the 32-pc broad mask (right).
The groupings for giant molecular clouds from other nearby galaxies, including dwarf galaxies and the Milky Way, 
are represented as colored contours that describe their 2D distributions;
see text for more details.
The best-fit \texttt{linmix} regression is shown as a dashed line.
The left panel, selected using the native-resolution strict mask, captures the cloud cores,
while the right panel with the broad mask encompasses also larger aggregations.
Only cores/clouds with S/N$\geq$5 on the peak temperature are considered.
}
\label{fig:cloudprops_1}
\end{figure*}

\subsection{Resolved cloud scaling relations \label{sec:cloudscaling}}

The correlation of molecular cloud properties such as line width, size, and mass give important insight
into their hierarchical structure.
However, as described in Sect. \ref{sec:beamdilution},
the concept of cloud as an aggregation of molecular emission is ill defined.
Dense clouds, and their denser cores, are generally found as emission peaks within a larger
more extended 
structure, but the definition of a ``cloud'' or ``core'' is highly data dependent,
as the beam size, noise, and dynamical range of the observations strongly influence the resulting cloud catalogs.

The left panels of Fig. \ref{fig:ic5152_cloud5_39} clearly show that clouds are neither circular nor
elliptical in cross section; the assumptions of spherical clouds or ellipsoidal shapes
are merely a ``distillation'' of the cloud properties.
Thus, the relations of line width \sigmav\ and mass inferred from CO luminosity \mlum\ to cloud size (deconvolved radius, R,
the geometric mean of the major and minor ellipse axes)
are, of necessity, approximations. 

In what follows, we determine the best-fit regression in log-log space with \texttt{linmix}\footnote{\url{https://linmix.readthedocs.io/en/latest/index.html}}.
With a Bayesian approach, \texttt{linmix} \citep{kelly07} takes into account heteroscedastic data, possibly correlated measurement errors, and intrinsic scatter. 
As mentioned above, we include in the analysis only those cores/clouds with S/N$\,\geq\,5$ on the maximum \tb\ within the structure
(see Table \ref{tab:co}).
This is to ensure that only the clouds with the most robust property estimations
are considered.
The results of the \texttt{linmix} fits for the size-linewidth relation and the CO luminosity \mlum-size  relations are
given in Table \ref{tab:linmix}.

\begin{table}[!t]
\centering
\caption{\texttt{linmix} regressions for scaling relations: $\log_{10}(y) = \gamma\,\log_{10}(\mathrm{R/pc}) + b$}
\label{tab:linmix}
\renewcommand{\arraystretch}{1.3}
\small{
\begin{tabular}{lrr}
\hline
\hline
\T 
 &  \multicolumn{1}{c}{Strict mask} & \multicolumn{1}{c}{Broad mask} \B \\
\hline
\multicolumn{3}{c}{ $y\,=\,\sigma_\mathrm{V}\ /\ \mathrm{km s}^{-1}$ }  \T \\
\hline
$\gamma$ & $0.46^{+0.07}_{-0.07}$ & $0.54^{+0.05}_{-0.05}$  \\
$b$ & $-0.11^{+0.04}_{-0.04}$ & $-0.42^{+0.05}_{-0.05}$ \\
$\sigma$ (dex) & \multicolumn{1}{c}{0.12} &  \multicolumn{1}{c}{0.15} \\
Number points & \multicolumn{1}{c}{104} & \multicolumn{1}{c}{58} \B\\
\hline
\multicolumn{3}{c}{ $y\,=\,\mathrm{M}_\mathrm{lum}\ /\ $\msun }  \T \\
\hline
$\gamma$ & $1.77^{+0.16}_{-0.16}$ & $1.78^{+0.10}_{-0.11}$ \\
$b$ & $2.75^{+0.08}_{-0.08}$ & $2.16^{+0.10}_{-0.10}$ \\
$\sigma$ (dex) & \multicolumn{1}{c}{$0.27$} &  \multicolumn{1}{c}{$0.34$} \\
Number points & \multicolumn{1}{c}{104} & \multicolumn{1}{c}{58} \B\\
\hline 
\hline
\end{tabular}
}
\end{table}

Figure \ref{fig:cloudprops_1} shows the correlation between line width, velocity dispersion \sigmav,
and deconvolved radius R;
what we call cores, identified with the strict mask, appear in the left panel, and the clouds
with the broad mask in the right.
The \texttt{linmix} fit in log space is given by a dashed regression line.
Also shown are two-dimensional (2D) distributions of samples from the literature,
including spiral galaxies
\citep[MW, NGC\,300:][]{heyer09,faesi18},
galaxy centers
\citep[MW and NGC\,253 centers:][]{oka01,leroy15},
the Magellanic Clouds
\citep[LMC, SMC:][]{wong11,ohno23},
and other nearby dwarf galaxies
\citep[NGC\,625, NGC\,6822, Sextans\,B, WLM:][]{imara20,schruba17,shi20,rubio15}.
We have not included the dwarf galaxies NGC\,5253 \citep{miura18} or II\,Zw\,40 \citep{kepley16} in the comparison,
because their starburst nature skews their molecular clouds to higher \sigmav, more akin to galaxy
centers than to the more quiescent dwarfs studied here.
Each distribution has been smoothed with a 2D Gaussian kernel for visualization.

For both the cores (strict mask) and clouds (broad mask), we find a power-law index of $\sim$\,0.5,
similar to the canonical value for the MW \citep{dame86,solomon87,heyer15,rice16},
but shallower than the index of $\sim 0.6-0.7$ found for clouds beyond the MW \citep{bolatto08,oakes25},
or from simulations \citep{colman24}. 
This could imply that cloud self-gravity is dominating the relation, rather than
shear, as formulated by \citet{xie25}; 
because dIrrs typically have low shear and relatively thick disks, the clouds in the sample galaxies 
do not reach \citet{xie25}'s proposed shear transition scale at the disk thickness.

The \sigmav\ fitted intercept for the cores 
($b=-0.11$, corresponding to $0.78$\,\kms) is similar to  
the MW value of $\sim 0.7-1.2$\,\kms\ \citep{dame86,solomon87}\footnote{\citet{solomon87} use $\eta\,=\,1.91$
rather than the lower $\eta$ used here; if we use $\eta\,=\,1.91$ to estimate the deconvolved radius R, we find smaller
intercepts: $b=-0.21$, for the cores, and $b=-0.54$ for the clouds.},
and within the range of the \sigmav\ intercept values for dendrogram trunks and leaves
in the SMC \citep{ohno23} and the LMC \citep{wong19}.
The \sigmav\ intercept ($b\,=\,-0.42$) for the clouds is two times ($0.3$ dex) lower than for the cores,
possibly implying a decreasing role of core self-gravity and core-level feedback, and
a lesser degree of turbulent support for the larger aggregations.

However, the low \sigmav\ intercept for the clouds could also be due to a problem with the \texttt{pycprops}
estimates of the velocity dispersions with the broad mask.
Such a problem could arise because unlike the strict mask applied to the native resolution cube,
the broad union mask applied to the native resolution cube leaves negative voxels.
Since the cloud sizes and line widths are estimated using the second moments of the
distribution of the emission along the spatial and spectral axes of the cube,
the negative voxels could skew the inferred sizes and velocity dispersions to 
lower values.
We have checked that this is not the case by comparing \sigmav\ of the individual
cores in a given cloud, with \sigmav\ inferred for the broad-mask clouds containing those cores.
Considering cores and clouds with S/N$\geq$5 as shown in Fig. \ref{fig:cloudprops_1},
the mean ratio of the core \sigmav\ with the overall cloud \sigmav\ is $\sim 1.05\,\pm\,0.25$,
thus not the factor of two expected from the difference in the \sigmav\ intercepts for cores 
(strict mask) and clouds (broad).
Moreover, if the larger clouds were to have an erroneously lower \sigmav\
than the smaller ones, because of the effect of more negative voxels in the larger clouds,
we would expect a shallower slope for the clouds;
instead, the power-law slopes of the core and the cloud relations are the same,
to within the uncertainties.
In essence, the lower \sigmav\ intercept is due to a shift to the right,
to a larger de-convolved size for the clouds,
in which \sigmav\ remains roughly constant even though the clouds are larger.
This point will be discussed further 
in Sect. \ref{sec:discussion}.

The mass \mlum\ inferred from \lcotwo\ is given as:
\begin{equation}
\mathrm{M_{lum}}\,=\,\alpha_\mathrm{CO(2-1)}\,L^\prime_\mathrm{CO(2-1)}\,=\, \frac{\alpha_\mathrm{CO(1-0)}}{R_{21}}\,L^\prime_\mathrm{CO(2-1)} \quad ,
\label{eqn:mlum}
\end{equation}
\noindent
with \aco$_{(1-0)}$\,=\,4.35\,\msun\,(K\,\kms\,pc$^2$)$^{-1}$
\citep{bolatto13}, including the contribution of helium,
together with a CO(2-1)-to-CO(1-0) line ratio \rtwo\ of 0.65
\citep[][see also Sect. \ref{sec:alma}]{rosolowsky21}. 
Since \rtwo\ is known to depend on physical conditions
within the galaxy such as level of SF activity \citep[e.g.,][]{leroy22,schinnerer24,lee26}, a single \rtwo\ value
may introduce additional scatter, both within the galaxies and across the sample.
The global specific SFR of the four galaxies varies almost two orders of magnitude (see Table \ref{tab:sample}),
with IC\,5152 having the highest and NGC\,3109 the lowest,
despite their similar metallicities.
Thus, applying a constant \rtwo\ for all systems
could be overestimating \mlum\ where SF surface density is high,
and underestimating it where low \citep[e.g.,][]{leroy22}. 

\begin{figure*}[h!]
\centering
\includegraphics[width=\linewidth]{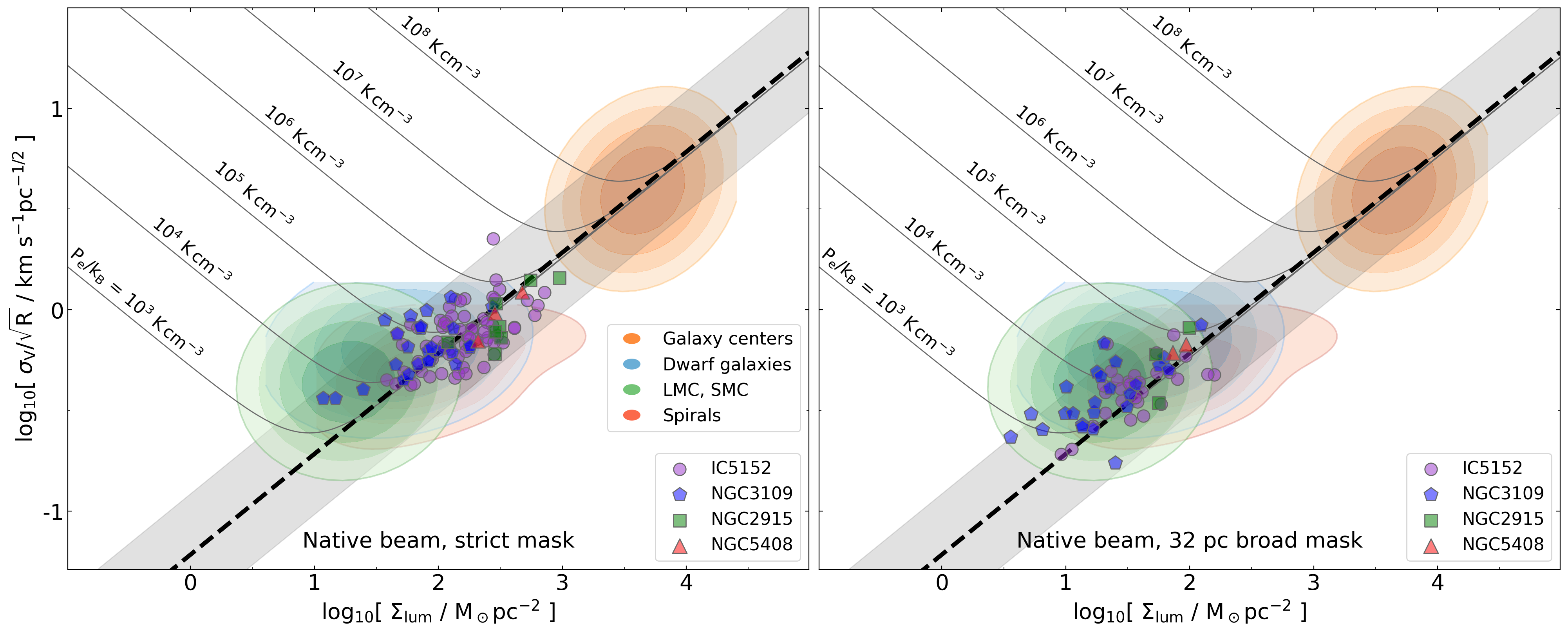}
\caption{Virial relation of \sigmav/$\sqrt{\mathrm{R}}$ plotted against molecular surface density 
\Sigmalum\ for our sample cores from the strict masks (left panel) and clouds from the broad masks (right).
Also shown are the groupings for other cloud samples as in Figs. \ref{fig:cloudprops_1} and \ref{fig:cloudprops_2}.
The black dashed line corresponds to virialized clouds with external pressure \pe\,=\,0 (\avirial\,=\,1), 
while the gray shaded area above and below virialization corresponds to a factor of two excursion on either side.
The gray curves show the contribution of \pe\ for pressure-bounded virial equilibrium
according to the formulation by \citet{field11}.
Only cores/clouds with S/N$\geq$5 on the peak temperature are shown.
}
\label{fig:pressure}
\end{figure*}

The correlation between luminous mass \mlum\ and deconvolved cloud radius R is shown in Fig. \ref{fig:cloudprops_2}.
As in Fig. \ref{fig:cloudprops_1}, other galaxies from the literature are shown as 2D distributions
outlined by contours.
The surface density \Sigmalum\ is calculated as the inferred luminous mass \mlum\ divided by the projected area:
\begin{equation}
\mathrm{\Sigma_{lum}}\,=\,\frac{\mathrm{M_{lum}}}{\pi\,\mathrm{R}^2}  \quad .
\label{eqn:sigmalum}
\end{equation}
\noindent
A power-law index of 2 would be expected were the clouds (or cores) to have a constant surface density \Sigmalum;
however, we find shallower indices (see Table \ref{tab:linmix}), implying that
the cloud surface densities are lower for larger radii.
Not surprisingly, for a radius of 1\,pc, the surface densities \Sigmalum\ inferred from the best fit are higher
for the cores ($\sim$\,560\,\msunpcsq) than the clouds ($\sim$\,140\,\msunpcsq).
These values\footnote{If we had
adopted $\eta\,=\,1.91$ to estimate R, the best-fit surface densities \Sigmalum\ 
for a radius of 1\,pc would 
be lower, $\sim 240$\,\msunpcsq\ and $\sim 60$\,\msunpcsq, respectively.}
are consistent with the low metallicity of our sample, 
and the requirement that the presence of CO requires about 1 magnitude
of visual extinction \citep{pineda08}, as found also for the CO cores in WLM \citep{rubio15},
and discussed further by \citet{elmegreen26a}.

As illustrated by \citet{heyer09} and several other studies,
the Larson \citep{larson81} relations among cloud radius R, velocity dispersion \sigmav, and luminous mass \mlum\
are mathematically coupled.
Assuming gravitational equilibrium and a spherical geometry, the cloud's virial mass, \mvir, can be expressed as:
\begin{equation}
\mathrm{M_{vir}}\,=\,5 \sigma_v^2 \mathrm{R}/f\,G \quad ,
\label{eqn:mvir}
\end{equation}
\noindent
where $f$ is a geometrical constant that describes the assumed
mass density profile $\rho(r)$;
for $\rho(r)\,\propto\,r^{-k}$, \texttt{pycprops} assumes
$k\,=\,1$ such that $f = 10/9$
\citep{oakes25,rosolowsky06}.
The factor of 5 is for a sphere of constant density \citep[e.g.,][]{field11}.
Under the assumption of virialization,
by substituting \mlum\ for \mvir, and converting to 
surface density \Sigmalum\ (Eq. \ref{eqn:sigmalum}), it can be seen that
the ratio of the velocity dispersion \sigmav\ and square root of the radius depend on the surface density
\citep{keto86,heyer09}:
\begin{equation}
\frac{\sigma_v}{\sqrt{\mathrm{R}}}\,=\, \left( \frac{f\,\pi\,G}{5} \right)^{1/2} \Sigma_\mathrm{lum}^{1/2} \quad .
\label{eqn:heyer}
\end{equation}
\noindent
Equation \eqref{eqn:heyer} can be understood as a 
description of the relative importance
of self-gravity and external pressure as a confining force
\citep{keto86,heyer09,field11}.

Figure \ref{fig:pressure} shows \sigmav/$\sqrt{\mathrm{R}}$ for the clouds in our sample and,
as in Figs. \ref{fig:cloudprops_1} and \ref{fig:cloudprops_2},
various samples from the literature;
the cores from the strict masks appear in the left panel, and from the 32\,pc broad union mask in the right.
Also illustrated are contours 
for various values of pressure-bounded virial equilibrium as formulated by \citet{field11},
as well as \avirial\,=\,1, shown by the black dashed line, and a factor of two variation
around it indicated by the gray shaded area.
It can be seen from Fig. \ref{fig:pressure} that the cores with the highest surface densities
appear to be more 
gravitationally bound,
in need of less external pressure, than the lower density ones.
A similar feature was also noted by \citet{keto86}. 

\begin{figure*}[t!]
\centering
\includegraphics[width=\linewidth]{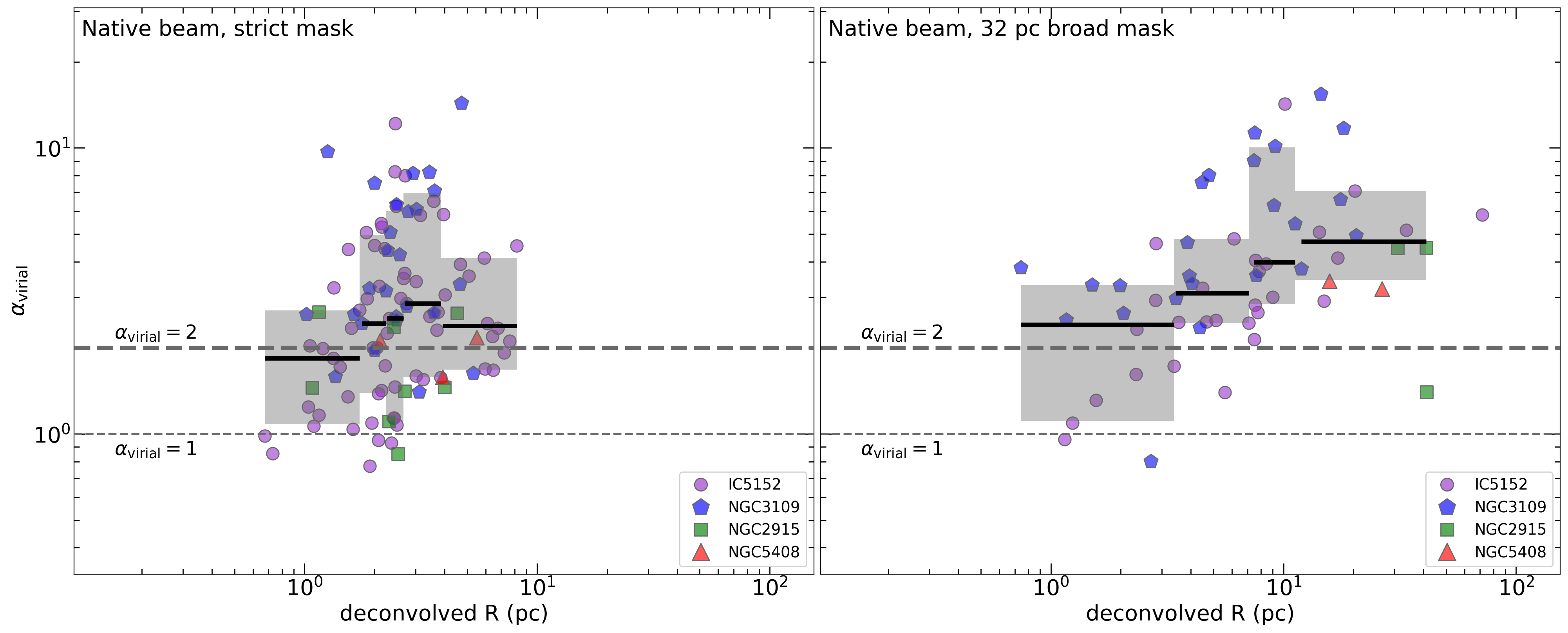} \\
\centering
\includegraphics[width=\linewidth]{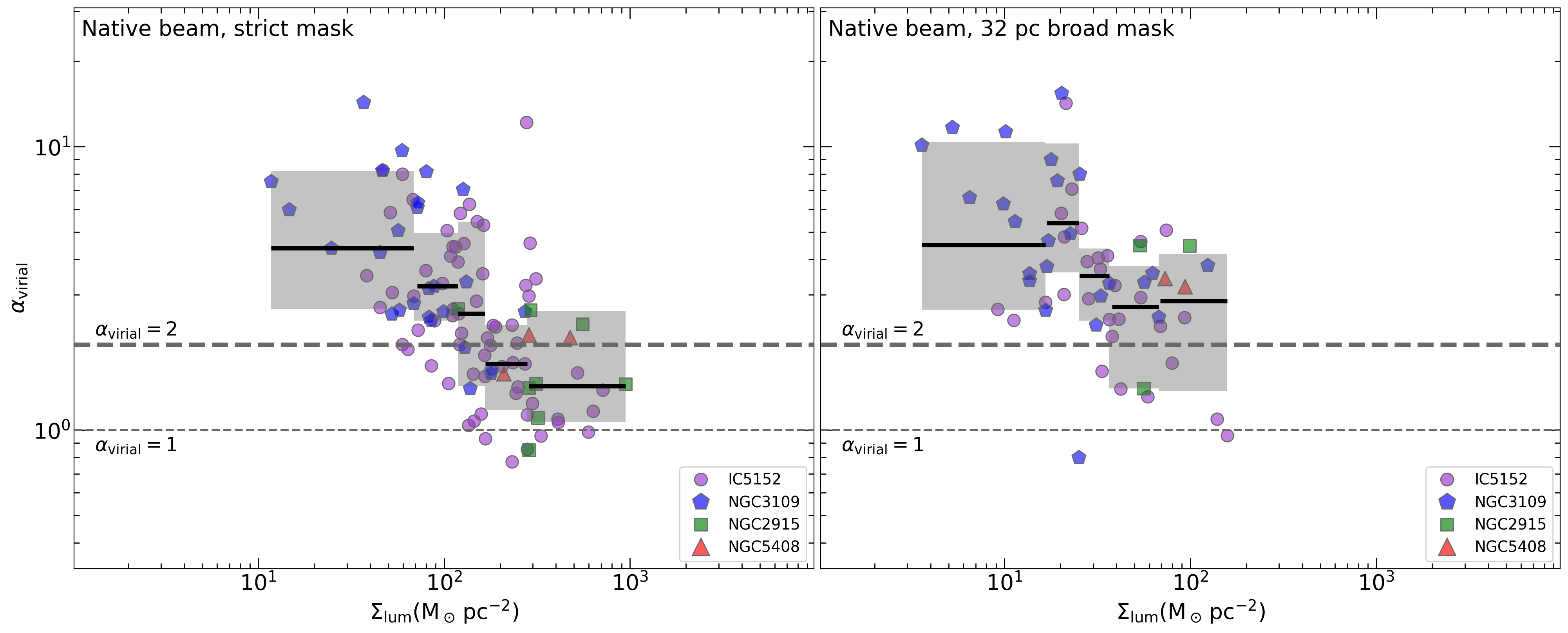}
\caption{Top panels: Virial ratio, \avirial\ vs. deconvolved cloud radius R for the clouds in our sample,
identified with  the native-resolution strict mask (left panel), and the 32-pc broad (union) mask (right).
Bottom panels: Virial ratio, \avirial\ vs. \Sigmalum\ for the cores identified with  the native-resolution strict mask 
(left panel), and the 32-pc broad (union) mask (right).
The gray regions show $\pm\,1\sigma$ in bins with the same number of data points,
and the  solid black segments correspond to the bin medians.
Virial ratios, \avirial, of 1 and 2 are marked by horizontal dashed lines.
}
\label{fig:virial}
\end{figure*}

Figure \ref{fig:pressure} also reveals a degeneracy,
namely between external pressure confinement
and the \aco\ conversion used to calculate \mlum\ (Eq. \ref{eqn:mlum}).
It could be that the more tenuous cores/clouds, that is those with lower surface densities,
require a larger \aco\ than the ``standard'' MW one that we 
have adopted here. 
A larger \aco\ would move points to the right, thus closer to virial equilibrium
as illustrated by the black dashed line.

\subsection{Virial parameter \label{sec:virial}}

We define the virial ratio \avirial\ as
\citep[e.g.,][]{bertoldi92,rosolowsky21,oakes25}:
\begin{equation}
\alpha_\mathrm{virial}\,=\,2\,M_\mathrm{vir}/M_\mathrm{lum} \quad, 
\label{eqn:virial}
\end{equation}
where \mvir\ is given by Eq. \eqref{eqn:mvir} and \mlum\ by Eq. \eqref{eqn:mlum}.
The factor of 2 in the numerator ensures a proper accounting of the CO luminosity since
half the mass is contained inside the FWHM by definition; thus the virial mass is an estimate
of the dynamical mass within the FWHM, and should be compared with \mlum/2.
The variation of \avirial\ as a function of deconvolved radius is reported in Fig. \ref{fig:virial} (top panel),
where the cores found in the strict mask are given in the left panel, and the larger aggregations
identified with the 
broad 
mask in the right.
Figure \ref{fig:virial} (bottom panel) also shows \avirial\ plotted against 
\Sigmalum, with cores in the left panel, and the larger aggregations of clouds in the right.

The virial parameter \avirial\ covers a large range of values. 
The strict-mask cores show variations in \avirial\ from 0.8 to 14; 
the broad-mask clouds show similar variations with \avirial\ ranging from 0.8 to 15. 
The mean (median) values of \avirial\ across all strict-mask cores and broad-mask clouds are 3.2 (2.5) and 
4.4 (3.5),
respectively.
Figure \ref{fig:virial} shows that the binned, median values of \avirial\ 
increase slightly with increasing cloud size for the strict-mask cores, 
and monotonically for the broad-mask clouds;
similar trends are seen for the variations of \avirial\ and \mlum\ (not shown).
The strict-mask cores in Fig. \ref{fig:virial} show a clear trend for \avirial\
to decrease with increasing surface density, \Sigmalum,
and the trend is also present, if less pronounced, for the broad-mask clouds (right panel).

\section{Discussion \label{sec:discussion}}

The ALMA data presented here 
provide an opportunity to investigate the cold, molecular ISM in sub-solar metallicity environments.
The scaling relationships described in
Sect. \ref{sec:cloudscaling} offer insights to the gas pressure in regions of molecular gas,
the role of CO-dark gas in low metallicity molecular clouds, and
the dynamical
state of the overall molecular cloud population.

\subsection{Turbulence and pressure \label{sec:turbulence}}

Several different approaches can be used to construct a relationship between the measured velocity dispersion and
size.  \citet{goodman98} define four distinct methods based on how size-line width relationships are assembled.
Type~1 relationships are based on 
multiple gas tracers and multiple clouds or cores \citep{larson81,
fuller92, caselli95}.  Type~3 is similar to Type~1 using multiple gas tracers but considers
only a single cloud or core
\citep{larson81, caselli95}.
A Type~2 relationship uses a single gas tracer on multiple clouds to compile velocity dispersions and sizes
\citep{solomon87}.
Type~4 relationships are constructed from a single gas tracer on a single cloud using methods
such as structure functions \citep{miesch94},
Principal Component Analysis \citep{heyer04}, and
$\Delta$-variance \citep{ossenkopf08}.
Figures~\ref{fig:cloudprops_1} and \ref{fig:cloudprops_2} are 
Type~2 relationships in which the points are collected from multiple clouds using a
single molecular gas tracer [in this study, \cotwo]. 
In our case, it is an extended Type~2 relationship as it includes
multiple clouds from four different galaxies. With only a single tracer, we can not assemble Type~1 and Type~3
relationships from our data.   A Type~4 relationship requires high spatial dynamic range observations
of the single cloud of interest, 
which cannot be achieved here.

A size-line width relationship is characterized by a power law $\sigma_v=\sigma_v(1\,\mathrm{pc})\ R^{\gamma}$,
where $\gamma$ is the power law index and $\sigma_v(1\,\mathrm{pc})$ is a normalization coefficient (denoted as $b$ in Table \ref{tab:linmix}). 
It is important to note that Type~2 relationships are not directly connected to widely applied functions/measures
of interstellar turbulence such as the velocity structure function or power spectrum unless the
true structure function for each cloud within the multiple cloud sample has similar values of both
$\gamma$ and $\sigma_v(1\,\mathrm{pc})$ \citep{larson81, heyer04}.  If this is not the case, then the scatter
in the Type~2 relationship would be large and possibly not even recognized as a correlation.
Conversely, if the points in a Type~2 relationship 
are strongly correlated, then this necessarily implies the individual clouds that
comprise this relationship have similar values of the power-law index and the coefficient.  
Based on Monte Carlo modeling of the scatter in Type~2 relationships derived from Type~4 relationships for a sample of 27 clouds in the Milky Way, 
\citet{heyer04} found a strict upper limit of the index and coefficient fractional variances to be $0.08-0.12$.
This tight correlation in a Type~2 relationship allows one to infer the mean values of the size-line width parameters
in Type~4 relationships over an ensemble of clouds.
The Type~2 points are basically the endpoints at the largest spatial scale of the true velocity structure function or Type~4 relationship
of each individual cloud
 \citep{heyer04}.   The overlay of these endpoints from each cloud forms the Type~2 relationship.

The Type~2 size-line width relationships in Figure~\ref{fig:cloudprops_1} for both
strict-mask cores and broad-mask clouds exhibit correlations.
This result is surprising  given the differences in metallicity/environment
between the four galaxies,
although the derived power-law parameters are biased by IC\,5152 and NGC\,3109 that
contribute the most points.   
Yet, \sigmav\ and R values from NGC\,2915 and NGC\,5408 fall within the
scatter. 
The tight correlations indicate similar values of $\gamma$ and
$\sigma_v(1\,\mathrm{pc})$ for each cloud in each
galaxy.  A value of $\gamma\sim$ 0.5 is consistent with the index of the first-order structure function
for Burgers turbulence \citep{burgers48}  in which the velocity field is dominated by hydrodynamic shocks
from converging gas flows.
Nevertheless, self-gravitating clouds with a constant pressure boundary condition give a slope of 0.5
regardless of what causes the motion \citep{elmegreen89}.
Hence, the nature of the turbulence within the cores/clouds cannot be definitively established from our data.

The coefficient, \sigmav(1\,pc), provides an approximate measure of turbulent gas motions and
pressure at a fixed scale within the cold ISM of the clouds.
For the strict-mask cores, \sigmav(1\,pc)\,=\,0.78$\pm$0.07\,\kms, 
comparable to the value of 
$0.72$\,\kms\
found by \citet{solomon87} for giant molecular clouds in the Milky Way\footnote{This value is corrected from 1\,\kms\ quoted in \citet{solomon87} 
to account for the difference in the values adopted for the galactocentric radius of the Sun (10\,kpc vs. 8.5\,kpc) that affects the cloud size.}.
The \sigmav(1\,pc) value
for the set of broad-mask clouds is a factor of two lower, 0.38$\pm$0.04\,\kms.  

The mass density of a cloud, $\rho_{cl}$, is estimated by
3$M_\mathrm{lum}/(4\pi R^3$) and the corresponding pressure of the cloud 
$P_{cl}\,=\,\rho_{cl}\,\sigma_\mathrm{V}^2/k_B$ 
in units of \cmthree\,K, where $k_B$ is the Boltzmann constant.  
The median values of $P_{cl}$ 
for the strict-mask core and the broad-mask cloud samples 
are: 2.8$\times10^5$\,\cmthree\,K and 1.9$\times10^4$\,\cmthree\,K, respectively.
Assuming a volume filling factor of unity, 
we find median number densities 
\nhtwo\ of
769\,\cmthree\ for the strict-mask cores,
and 
64\,\cmthree\ for the 
clouds.

The median gravitational free-fall times \tff\ of the cores/clouds, estimated as
$t_\mathrm{ff}=\sqrt{3\,\pi/32\,G\,\rho_{cl}}$, 
are similar to those of larger, more metal-rich, spiral galaxies.
For the strict cores, 
the mean \tff$\,\approx1.4$\,Myr, and for the clouds,
\tff$\,\approx5.5$\,Myr,
roughly half of $\sim11$\,Myr, 
the mass-weighted free-fall time within 1.5\,kpc apertures
on cloud scales of $\sim 150$\,pc  
estimated for the PHANGS galaxies by \citet{sun22,schinnerer24}. 
At $\sim 90$\,pc resolution, \citet{rosolowsky21} find for the PHANGS
galaxies \tff$\,\sim6.4$\,Myr,
comparable to the \tff\ of 5.5\,Myr found for the clouds in our sample.
Larger clouds with an equivalent \Sigmalum\ would be expected to have longer \tff.

It is important to remember that the derived pressures and densities in these CO cores/clouds
are degenerate with \aco.
Here we have used a MW value, but a larger 
value
would increase the densities and pressures significantly,
and diminish even further the inferred free-fall times.

\subsection{Dynamical state or CO-dark molecular gas? \label{sec:codark}}

The dynamical state of a cloud is evaluated by the virial ratio, \avirial,  previously defined in 
Sect. \ref{sec:virial}.
Equipartition between kinetic and gravitational
energy densities corresponds to \avirial\ $\sim$ 2;
we take \avirial\,=\,2 as the approximate threshold of bound and unbound states.
An \avirial\ of unity would imply a strongly self-gravitating cloud or core. 
In the absence of magnetic fields,
a value of \avirial\,$\ll 1$ indicates dominant self-gravity and potentially, the gravitational collapse of a cloud.
\citet{ballesteros11} argue that values of \avirial $<$ 2 can occur if a cloud is in a state of
hierarchical and chaotic gravitational collapse.  In this case, gravity, not turbulence,
is responsible for the non-thermal motions of the cloud.
While the condition \avirial\,$>2$ implies an internally, overpressured cloud or core with respect 
to its self-gravity, both random 
and systematic uncertainties for \avirial\ can blur this interpretation. A more robust sample 
of overpressured clouds can be identified with the condition \avirial\,$>5$. 
In this case, the cloud could be
expanding until it reaches an equilibrium with the ISM environment
determined by thermal, turbulent, magnetic and cosmic ray pressures pushing it up vertically and the
weight of the stars, gas, and dark matter pulling it down 
 \citep{elmegreen89}.

In the context of dynamical equilibrium, 
high \avirial\ values (see Sect. \ref{sec:virial}) imply that most of the strict or broad mask
clouds are overpressured with respect to self-gravity.  
However, there is a degeneracy with the conversion of \lco\ to \mlum:
\avirial\ could be overestimated
if our applied \aco\ value is too low owing to the environmental conditions of the four galaxies.  
At low metallicity, it is well known that larger values  
of \aco\ are expected relative to the Milky Way mean value 
\citep[e.g.,][]{israel97,schruba12,narayanan12,bolatto13,hunt15,accurso17,hunt20,hunt23}.

Because of this, in the low-metallicity environment of dIrr galaxies and BCDs, large
values of \avirial\ may occur by underestimating the amount
of luminous, molecular mass. This is because the reduced
metallicity implies a lower CO column density for the same
\htwo\ column compared to the MW. This will reduce the CO line
excitation temperature (via a reduced line trapping, leading
to strongly sub-thermal excitation) and thereby diminishing the CO
luminosity \citep[e.g.,][]{hu22}. 
In extreme cases the CO emission may even become optically thin,
even further reducing the CO luminosity for a given \htwo\ column.
In addition, the reduced shielding against UV radiation is
expected to 
further decrease the CO abundance in
the diffuse gas \citep[$A_V \sim 0.2-1.5$\,mag,][]{seifried20},
due to the  different self-shielding requirements of column density for
\htwo\ and \twelveco\ molecules \citep[e.g.,][]{vandishoeck88,bolatto99,wolfire10}.  
Both effects combined result in a CO-dark
\htwo\ component specifically in the low column-density regions of molecular clouds. 
In the Milky Way, the \htwo\ gas not associated with CO comprises on average $\sim$30\% of the
total molecular gas mass \citep{pineda13}.  This effect is especially significant in regions/galaxies with
low metallicities where there are low abundances of C and O to form CO and build up the self-shielding layer.
This CO-dark \htwo\ component has also been recognized in several previous studies of CO emission in dIrr galaxies
\citep[e.g.,][]{rubio15, archer22, elmegreen26a}.

To 
constrain the amount of CO-dark \htwo\ gas in these galaxies,
we equate the core/cloud 
mass to \mvir. 
This is equivalent to ascribing variations of \avirial\ 
to variations of the true \aco\ relative to the MW \aco\ value
used here to convert CO luminosity to \mlum.
Thus, the total CO-dark \htwo\ mass is obtained by summing the difference of \mvir\ and \mlum\ over the clouds:
\begin{equation}
\Sigma_i\ \mathrm{M_{dark,i}} = \Sigma_i\ (\mathrm{M_{vir,i}-M_{lum,i}}/2) \quad ,
\label{eqn:codark}
\end{equation}
\noindent
where index $i$ corresponds to the core/cloud number.
The fraction of 2 in the denominator of \mlum$_\mathrm{,i}$ is for consistency with
Eq. \eqref{eqn:virial},
because of the necessity to compare \mvir, calculated within the FWHM, with half of the total luminosity, \mlum.
Considering only the resolved 
broad-mask clouds with S/N$\geq 5$,
the fractions of the total \htwo\ mass contributed by the CO-dark component,
namely \mdark/\mvir,
range from 0.69 (NGC\,5408) to 0.86 (NGC\,3109).
The resolved strict-mask cores with S/N$\geq 5$ have slightly lower CO-dark fractions:
0.64 (IC\,5152), 0.46 (NGC\,2915), 0.75 (NGC\,3109), and 0.51 (NGC\,5408).
If we set to 0 the few cases where \mvir\,$<$\,\mlum/2 
(1/29 in IC\,5152 broad clouds, 1/24 in NGC\,3109 broad clouds, 5/67 in IC\,5152 strict cores, 1/8 in NGC\,2915 strict cores,),
these fractions change very little, $\la 0.01$,
 implying that these structures do not contribute very much to the overall molecular gas mass budget.

It is important to remember that these results depend on several assumptions.
Our adopted virial mass [Eq. \eqref{eqn:mvir}] is based on a spherical geometry and a specified density
profile through the geometrical constant $f$. 
Because the clouds identified with the broad masks, and perhaps even the cores with the strict ones,
may be filamentary rather than spherical, we have also
calculated for each structure the virial mass per unit length that would be obtained for virial equilibrium in a cylinder 
\citep[e.g.,][]{ostriker64,fiege00,freundlich14}: 
\begin{equation}
\mathrm{M_{vir,cyl}}\,=\,2 \sigma_v^2\,/\,G \quad .
\label{eqn:mvir_cyl}
\end{equation}
\noindent
For an ideal filament, the virial mass/length is independent of longitudinal extent \citep{fiege00}.
Using $\mathrm{M_{vir,cyl}}$ in Eq. \eqref{eqn:codark}, and comparing with \mlum\ per unit length (assumed to be $2R$), we have 
re-estimated the CO-dark \htwo\ fractions, \mdark/$\mathrm{M_{vir,cyl}}$.
The resulting fractions for the resolved broad-mask clouds with S/N$\geq$5 span
0.58 (NGC\,5408) to 0.77 (NGC\,3109),
slightly smaller than the analogous fractions assuming a spherical shape for the clouds.
The resolved strict-mask cores with S/N$\geq 5$ have similarly lower CO-dark fractions:
0.48 (IC\,5152), 0.21 (NGC\,2915), 0.65 (NGC\,3109), and 0.31 (NGC\,5408).
Thus, the inferred dark-CO fractions depend somewhat on the assumed geometry, at least for the strict cores,
which are nevertheless probably more spherical than filamentary.

These \mdark/\mvir\ fractions assuming a spherical geometry for the broad clouds are not far from 
the $\ga$90\% fractions found for
the tiny clouds in WLM, using a methodology that is independent of a CO-to-\htwo\ conversion factor
\citep{archer22}.
However, we do not directly measure CO-dark \htwo\ gas.
Our calculations implicitly
assume that the CO-dark \htwo\ is co-spatial with the CO-bright component,
and thus contribute to the gravitational potential on the same scale as the resolved clouds.
The larger \mdark/\mvir\ fractions of the clouds relative to the cores
suggest that the broad mask is revealing more CO-dark gas. 
Nevertheless, if there is CO-dark gas that is not spatially coincident on cloud scales with the CO, namely, beyond the confines of the CO mask,
we have no way of measuring its dynamical effects. 

There is an another important caveat to this exercise:
other ISM components besides CO-dark \htwo\ in the clouds may contribute to 
the self-gravity of the regions and therefore to the CO velocity dispersions,
which appear in \avirial.
These components include \hi, ubiquitous in dwarf galaxies, 
as well as disk stars and dark matter. 
The column density of \hi\ rarely exceeds 
$\sim 10$\,\msunpcsq\ in spirals
\citep[e.g.,][]{bigiel08}, and \hi\ clouds in dwarf galaxies
have roughly this same column density \citep{hunter19}.
However, in the metal-poor nearby dwarf WLM ($\sim 13$\% \zzsun),
\hi\ columns can be twice this
\citep{archer22}, and in I\,Zw\,18, at $\sim 3$\% \zzsun, 
as much as ten times higher \citep{lelli12}.
Thus, particularly for the more metal-poor galaxies in our sample,
IC\,5152 and NGC\,3109, atomic gas could be a contaminant in our virial mass estimates.
Disk stars and dark matter could also contribute on the larger scales of ($\sim 500$\,pc) disk thickness for cloud confinement. 
All things considered, the \avirial\ values estimated here are expected to be upper limits for
the true amount of CO-dark \htwo\ gas
\citep[see also][]{ballesteros06,johnson15}.

\begin{figure}[t!]
\centering
\includegraphics[width=\linewidth]{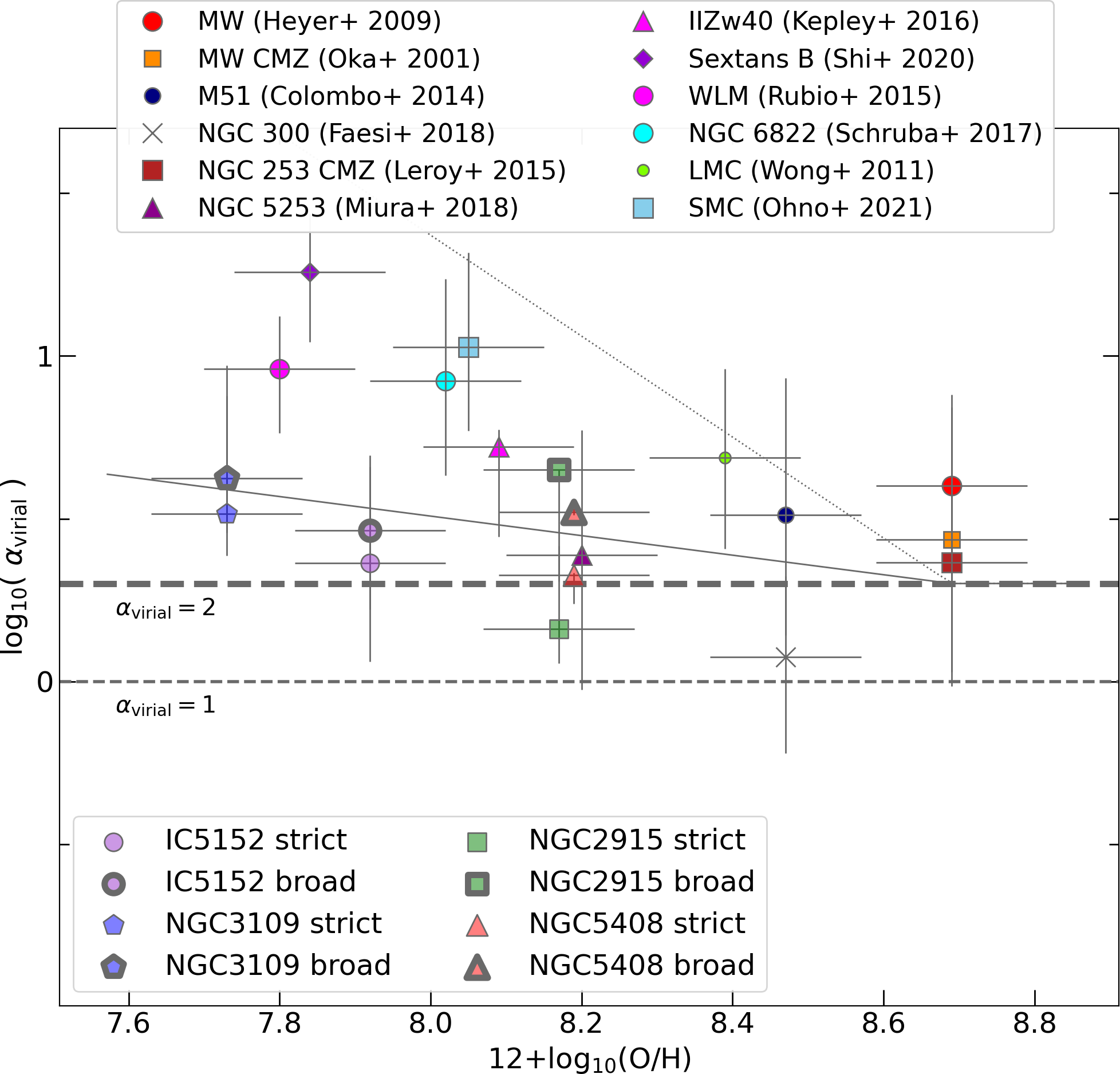}\\
\caption{Virial factor medians \avirial\ vs. nebular oxygen abundance, \logoh.
As in the lower-left legend, clouds from the broad masks are distinguished
from the cores identified in the strict masks by their heavy symbol border.
The error bars correspond to $\pm\,1\sigma$, or the 16th and 84th percentiles for each galaxy,
and for $\pm\,1$\,dex in O/H;
as in previous figures, only cores/clouds with S/N\,$\geq$\,5 are considered.
Samples from the literature are also shown
including those from Figs. \ref{fig:cloudprops_1} and \ref{fig:cloudprops_2},
as well as the dwarf starbursts
II\,Zw\,40 \citep{kepley16}, 
NGC\,5253 \citep{miura18}.
The literature \avirial\ values have been calculated in the same way as those here.
The global trend found by \citet{hunt20}
with \aco\,$\propto$\,\zzsun$^\beta$ metallicity below \zsun\ (and constant for more metal-rich environments) with a power-law index $\beta$ of
$-1.55$ is shown as a dotted line,
and $\beta\,=\,-0.3$ as a light solid line.
The horizontal heavy dashed lines correspond to the MW \aco\ used here as a fiducial value.
}
\label{fig:virialratio_oh}
\end{figure}

Figure \ref{fig:virialratio_oh} shows the dependence across galaxies of median \avirial\
 on global oxygen abundance \logoh.
In our sample, NGC\,3109 with a metallicity \logoh\,=\,7.73 ($\sim$\,11\% \zsun),
is the most metal-poor galaxy yet detected hosting a well-defined resolved CO cloud population.
Also included in the figure are the galaxies used in Figs. \ref{fig:cloudprops_1}
and \ref{fig:cloudprops_2},
as well as the dwarf starbursts, NGC\,5253 \citep{miura18} and
II\,Zw\,40 \citep{kepley16}.
The literature \avirial\ values have been calculated in the same way as those here.
The vertical error bars correspond to 1$\sigma$ (16th, 84th percentile) deviations from the medians of \avirial\
of the clouds in each galaxy.
Also shown are the expected global trends for a metallicity dependence
of \aco, namely \aco\,$\propto$\,\zzsun$^\beta$, where $\beta\,=\,-1.6$
\citep{accurso17,hunt20},
and a much shallower trend with $\beta\,=\,-0.3$,
perhaps more consistent with the resolved cloud data.
The much steeper metallicity dependence of \aco\ with metallicity found by \citet[][$\beta\,=\,-3.4$]{madden20}
would be inappropriate for the 
cloud populations in Fig. \ref{fig:virialratio_oh}.

\begin{figure*}[t!]
\centering
\includegraphics[height=0.33\linewidth]{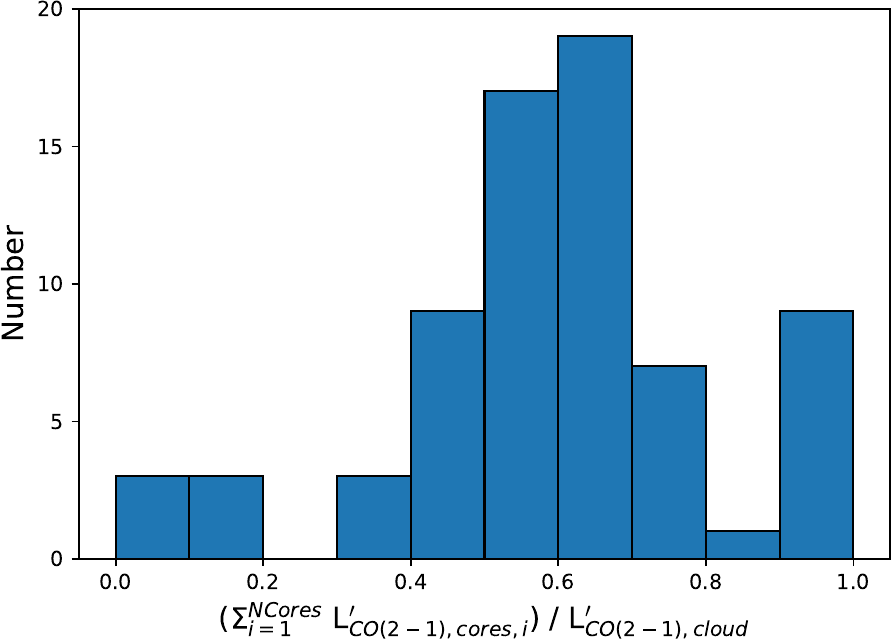}
\hspace{0.05\linewidth}
\includegraphics[height=0.33\linewidth]{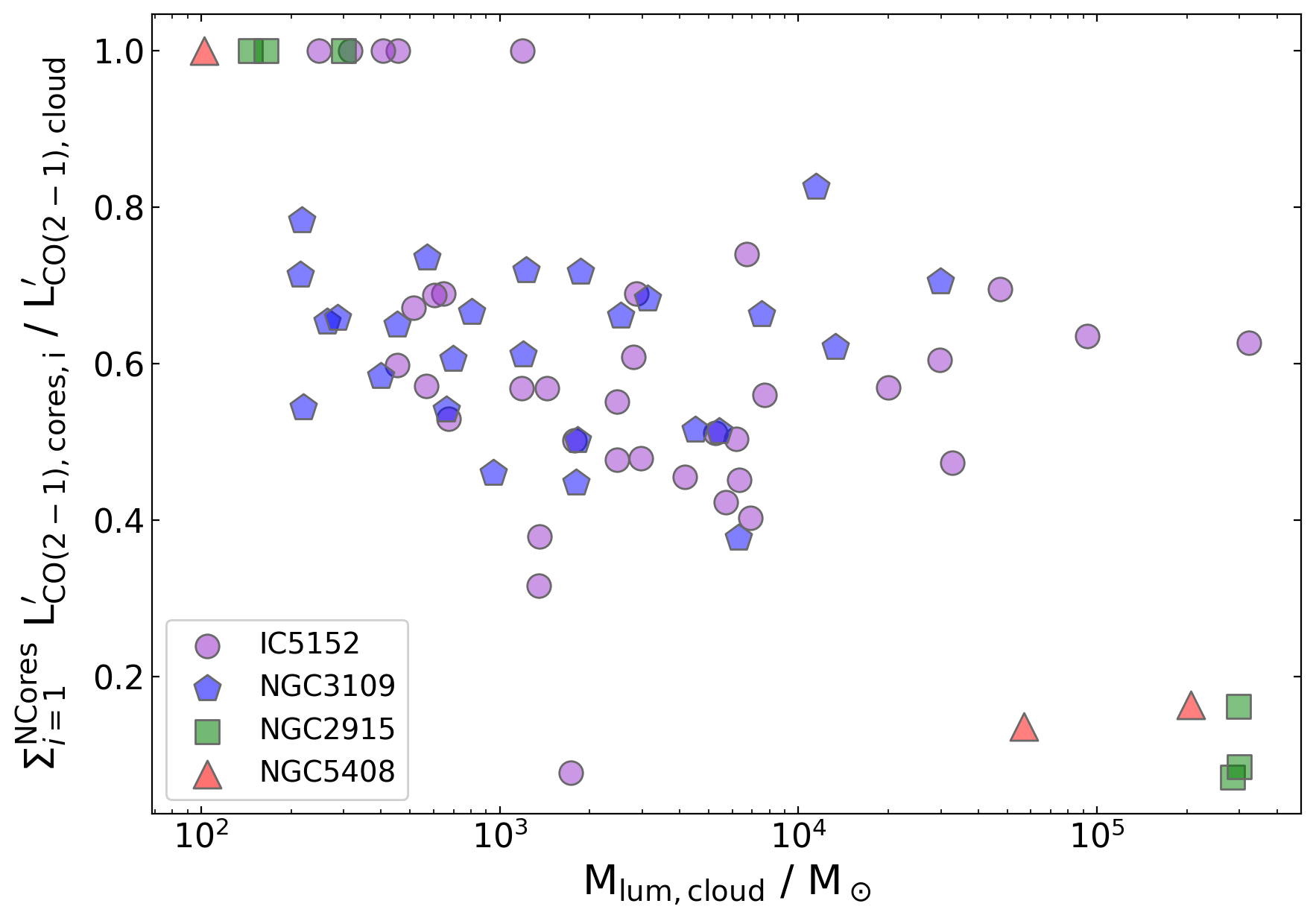}
\caption{Left panel: Histogram of the fractions of \cotwo\ luminosity found
in the sum of the cores over the broad-mask cloud within which they reside;
Right panel: the same fractions plotted against \mlum.
As in Table \ref{tab:co}, we consider only the un-extrapolated values
for both cores and clouds, and include in the fractions only those
cores and clouds with S/N$\geq 5$.
}
\label{fig:coresclouds}
\end{figure*}

If we make the assumption that \avirial\ is related to \aco,
Fig. \ref{fig:virialratio_oh} shows that 
the core/cloud populations of our sample galaxies and those from
the literature do not require as steep a metallicity dependence for \aco\
as the global CO values.
Such a result is not new, and was first noted by \citet{bolatto08}
who found that in low-metallicity galaxies, the detectable, bright, CO emission 
is ``relegated to the density peaks''.
Furthermore, the properties of these clouds, something like the ``tips of the icebergs'',
are similar to the CO clouds in the MW and other more metal-rich galaxies.
Thus, the \aco\ metallicity dependence is diluted, with the implication that we are not able to estimate the 
amount of CO-dark \htwo\ gas using \avirial\ of resolved clouds.

The observed trend of \avirial\ with metallicity shown in Fig. \ref{fig:virialratio_oh}
is commonly found when using optically thick CO tracers of \htwo, and scaling with a standard \aco.
However, an opposite trend, namely with \avirial\ decreasing with decreasing metallicity was found by \citet{lin25}
using $^{13}$CO.
To estimate \htwo\ mass, rather than the canonical conversion factor,
\citet{lin25} employed assumptions for the \htwo/$^{12}$CO and 
$^{12}$CO/$^{13}$CO abundance ratios, and their variation with metallicity.
They concluded that low-metallicity molecular clouds 
are in a sub-virial state.
We find the converse, namely that \avirial\ for our sample galaxies indicates 
mostly supervirial molecular clouds with, however, the caveat of the unknown \aco\ necessary to convert our
\cotwo\ measurements to \htwo\ mass.
Future work is needed to resolve this ambiguity. 

If we go further with the equivalence of \avirial\ and \aco,
we see that \avirial\ also varies with core/cloud size with the largest clouds,
and those with the lowest surface brightness \Sigmalum\ or column density, 
showing the largest \avirial (Fig. \ref{fig:virial}).
Such behavior would be consistent with the predictions
of \citet{hu22} and the observations of \citet{hunt23}
who found that in addition to metallicity, \aco\ depends on CO intensity (surface brightness, column density) and beam size
\citep[see also][]{hunter24}.

Ultimately, the virial mass, the virial parameter \avirial, and the amount of CO-dark \htwo\ gas
depend on the coefficient $\eta$ that we have used to convert the rms moments
of \texttt{pycprops} to cloud radii [see Eq. \eqref{eqn:radius}].
If we were to adopt the larger $\eta\,=\,1.91$ used by \citet{solomon87} and others,
all parameters based on virial mass would be significantly larger.
As also noted by \citet{rosolowsky21}, the definition of cloud size should be taken into
account when comparing with the results presented here.

\subsection{Cores versus clouds: faint intracloud CO? \label{sec:diffuse}}

In galaxies like the MW, M\,33, or M\,51, as much as $\sim 50$\% of the CO emission 
is diffuse, rather than residing in dense CO molecular clouds
\citep[e.g.,][]{duval16,muraoka23,pety13}.
For our sample galaxies,
comparing the CO emission contained in the cores relative to the clouds 
shows that at least 40\% of the CO emission in the broad-mask clouds is found outside
the dense cores identified by the strict mask 
(see Table \ref{tab:co}).
Additional CO emission is also evident in Figures \ref{fig:ic5152_cloud5_39} 
and \ref{fig:ngc3109_cloud25} for clouds in which 
the summed spectra 
of broad mask voxels are greater than the summed spectra summed from voxels of all cores within 
each cloud.
Figure \ref{fig:coresclouds} (left panel) shows the histogram of the fraction of summed core luminosities
with respect to the cloud luminosity for all clouds with S/N$\geq$5.
The luminosities have not been extrapolated to the 0\,K isophote\footnote{We have repeated this calculation 
using luminosities extrapolated to the 0\,K isophote.  The distribution is similar to that shown 
in Fig. \ref{fig:coresclouds}, but with some fractions exceeding unity because of the
up to 30\% uncertainties in the extrapolated values \citep{rosolowsky21}.}.
If there were no 
additional emission beyond 
the strict mask boundaries, then this fraction should be equal to $\sim$1.  The histogram 
demonstrates that most of the luminosity fractions fall within the range of 0.4 to 0.8.  
Clouds with fractions of unity are very compact clouds identified in the strict mask only,
but are also picked out in the broad 32\,pc mask in the union process.
These structures do not contribute much mass, as shown in the right
panel of Fig. \ref{fig:coresclouds}.
The more massive clouds, particularly in the more distant galaxies,
NGC\,2915 and NGC\,5408, tend to have much more faint 
emission, not recovered in the strict mask.

With our available data, we cannot distinguish whether this emission comprising 
1 minus the fractions shown in Fig. \ref{fig:coresclouds}
originates from a collection of faint, compact, unresolved cores or diffuse, filamentary emission distributed beyond the strict mask boundaries.
A diffuse component would be more exposed to the far-ultraviolet radiation field such that the CO 
abundance would be lower relative to a strongly, self-shielded core, making any emission very faint. 
In addition, the median molecular volume density (assuming the MW \aco)
in the broad-mask clouds is 
$\sim 60$\,\cmthree\ such that the CO excitation is 
sub-thermal, causing lower surface brightness.
This emission is detected by ALMA because of the reduced noise of the smoothed data and summing faint CO signal over 
many positions within a given cloud mask. 

At face value, this is surprising because in a low-metallicity environment,
the tiny cores are thought to be the only surviving remnant of photodissociation
\citep[e.g.,][]{vandishoeck88,bolatto99,wolfire10,bolatto13}.  
We surmise that the 
CO with this faint, diffuse, emission survives within the cloud because it is sheltered by 
shielding from dust (and gas). 
This is suggested by the distribution of the cores shown in Figs.
\ref{fig:moment5152}, \ref{fig:moment3109}, \ref{fig:moment2915}, and \ref{fig:moment5408},
which show significant PAH (8\,\micron) and dust (70\,\micron) dust emission
at the location of the CO detections.
The emission may also be caused by a sea of fainter, well-shielded cores so that the
shielding issue may be less severe.

It is obvious that this result depends on the way we have defined cores and clouds,
and on the spatial resolution and dynamic range of the data.
However, tests we have conducted suggest that our finding is relatively robust,
since it remains qualitatively unchanged with the various strict vs. broad union masks we have considered.
It is intriguing that even at $\sim 11$\% \zzsun\ (NGC\,3109),
we find evidence for the survival of a significant amount of CO with very faint, diffuse, emission.
This component tends to fill in the gaps between the CO peaks,
shifting the cloud to the right in the size-linewidth plot,
while the velocity \sigmav\ remains roughly the same
(see Fig. \ref{fig:cloudprops_1}).
We find that 
the velocity dispersion of the entire cloud is dominated by the velocity dispersion internal to the individual CO-bright cores, 
while the inter-core velocity dispersion is small. 
This suggests that in these entities the cores move together because of large-scale bulk velocity fields, or perhaps that the small inter-core velocity
dispersion 
is related to the low surface-mass density of the
diffuse inter-core gas 
\citep{wong19,muraoka23}. 

\section{Summary and conclusions \label{sec:conclusions}}

We have presented \cotwo\ ALMA data at $\sim 3-4$\,pc resolution (see Table \ref{tab:almadata})
of four nearby low-metallicity dwarf galaxies:
IC\,5152, NGC\,2915, NGC\,3109, and NGC\,5408 (see Table \ref{tab:sample}),
with nebular (O/H) metallicities ranging from 
$\sim 1/9$\,\zsun\ to $\sim 1/3$\,\zsun.
Processing with the PHANGS pipeline gives strict and broad masks,
together with the moment zero maps; 
the strict and broad masks (obtained by the union of strict masks up to 32\,pc beams)
enable the identification of molecular cores (with the strict masks), and clouds (with the broad),
as described in Sect. \ref{sec:cloudfinding}.
With S/N$\geq 5$, we find a total of 152 strict-mask cores
and 71 broad-mask clouds, 70-80\% of which are resolved.
NGC\,3109, at 
$\sim 1/9$\,\zsun, is the lowest metallicity galaxy yet
observed with a numerous population of molecular cores/clouds detected in CO
at parsec-scale resolution.
The analysis of the subset of this cloud population resolved by our
observations gives the following key results.

(i)~The slopes 
of the size-linewidth relations are
comparable to those found in the MW,
and consistent with the index of the first-order structure function for
Burgers turbulence (see Sect. \ref{sec:turbulence}).
The fitted \sigmav\ intercept for the cores is comparable to
that of the MW ($\sim 0.8$\,\kms\ vs. $\sim 0.7-1.2$\,\kms),
but that of the clouds is a factor of two lower than the cores
(see Sect. \ref{sec:cloudscaling}, Fig. \ref{fig:cloudprops_1}),
perhaps implying less need for turbulent support. 

(ii)~The estimated surface densities, mass densities, pressures, and free-fall times
of the cores and clouds (Sect. \ref{sec:turbulence}, Fig. \ref{fig:pressure})
are comparable to those in larger, more metal-rich spiral galaxies,
although the inferred density of $\sim 60$\,\cmthree\ for
the clouds would imply sub-thermal excitation.
However, this result depends on our use of a standard \aco, and the application
of a larger value would increase the inferred surface and mass volume densities.

(iii)~Assuming a fixed Galactic value of 
\aco$_{(2-1)}$ 
(Sect. \ref{sec:alma}),
the virial parameter \avirial\ 
for the cores and clouds in our sample ranges from roughly unity to 
$\sim 15$.
(Sect. \ref{sec:virial}, Fig. \ref{fig:virial}). 
\avirial\ derived in this way 
increases for increasing cloud size and decreases for increasing surface density, \Sigmalum.

(iv)~If we attribute the variation of the virial parameter \avirial\
to the presence of CO-dark gas
(Sect. \ref{sec:codark}),
a metallicity dependence for \aco,
\aco\,$\propto$\,\zzsun$^\beta$
with $\beta\,\approx\,-0.3$, would be consistent with the data.
This agrees with earlier work \citep[e.g.,][]{bolatto08}, 
confirming that the cores/clouds traced by CO at low metallicity
resemble the molecular clouds in more metal-rich, more massive galaxies
(Fig. \ref{fig:virialratio_oh}).

(v)~We find a faint, diffuse CO emission component beyond the CO-bright cores defined by the strict mask (Sect. \ref{sec:diffuse}). 
The fraction of this component is typically 40\% of the emission in the overall cloud,
although it can range from zero to $\sim$90\% (Fig. \ref{fig:coresclouds}).
We cannot determine if it arises from filamentary structures or from a plethora of faint, unresolved cores. 
The clouds with no diffuse component tend to be small and do not contribute much to the overall molecular emission. 
The velocity dispersion of these structures is dominated by the velocity dispersion internal to the CO-bright cores themselves 
(Fig. \ref{fig:ic5152_cloud5_39_velRA}). 
The diffuse CO-emitting gas must be shielded by dust, since the regions where it is observed are bright at 8 and 70\,\micron. 

Future work will focus on the \thirteencotwo\ isotopologue detected in our sample
galaxies, the kinematics of the molecular gas,
and a quantitative comparison of CO emission with tracers of star formation and other components of the ISM.
The exploration of diffuse CO emission at low metallicities is possible only with ALMA; 
to better understand the nature of star formation in a metal-poor ISM is fundamental for
our view of the early Universe.

\begin{acknowledgements}
This paper makes use of the following ALMA data: ADS/JAO.ALMA\#2023.1.00227.S. ALMA is a partnership of ESO (representing its member states), NSF (USA) and NINS (Japan), together with NRC (Canada), MOST and ASIAA (Taiwan), and KASI (Republic of Korea), in cooperation with the Republic of Chile. The Joint ALMA Observatory is operated by ESO, AUI/NRAO and NAOJ. 
We thank the anonymous referee for insightful comments that helped clarify and improve the paper.
We are grateful to the PHANGS collaboration for making publicly available a vast collection of 
extremely useful software and documentation for the analysis of ALMA data
\citep{leroy21,rosolowsky21}.
EC acknowledges financial support from INAF-Mini Grant RF-2023-SHAPES.
UL acknowledges support by the research grant
PID2023-150178NB-I00, financed by MCIU/AEI/10.13039/501100011033 and
from the Junta de Andalucía (Spain) grant FQM108.
MR wishes to acknowledge partial support from ANID(CHILE) through Basal FB210003.
\end{acknowledgements}

\bibliographystyle{aa} 
\bibliography{phylo}

\begin{appendix}

\section{APEX observations \label{app:apex}}

\begin{figure*}[t!]
\includegraphics[height=0.38\textwidth]{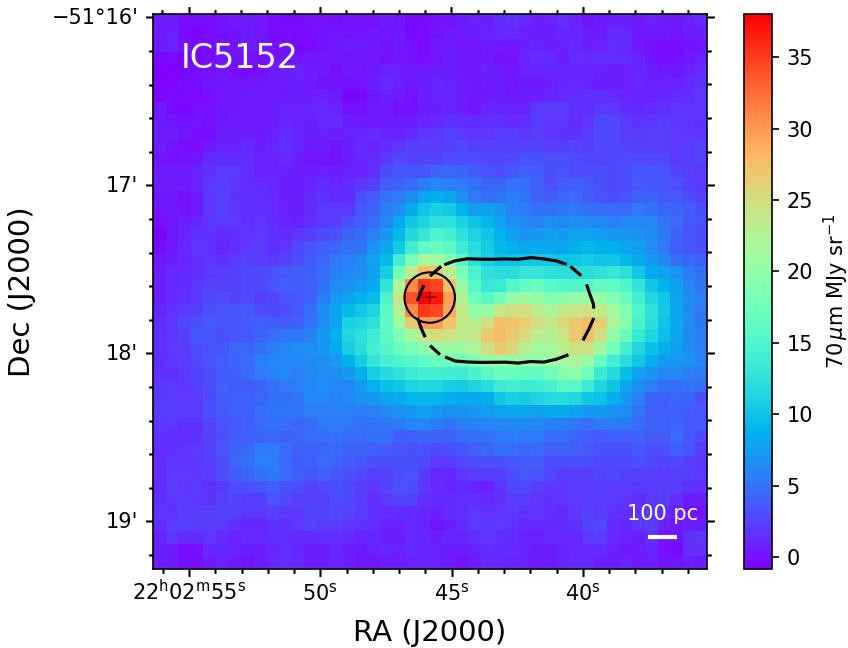}
\includegraphics[height=0.38\textwidth]{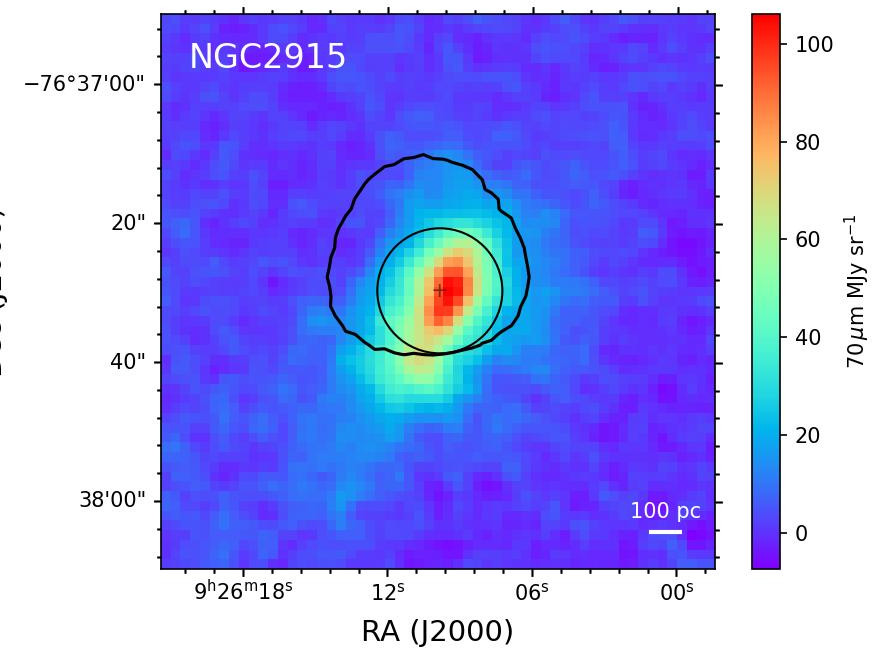} \\
\\
\includegraphics[height=0.3325\textwidth]{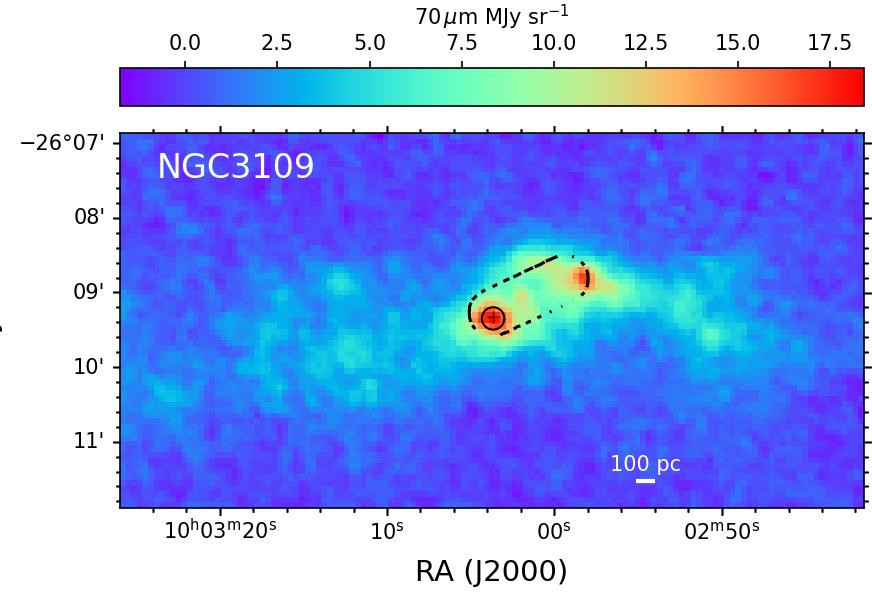}
\includegraphics[height=0.38\textwidth]{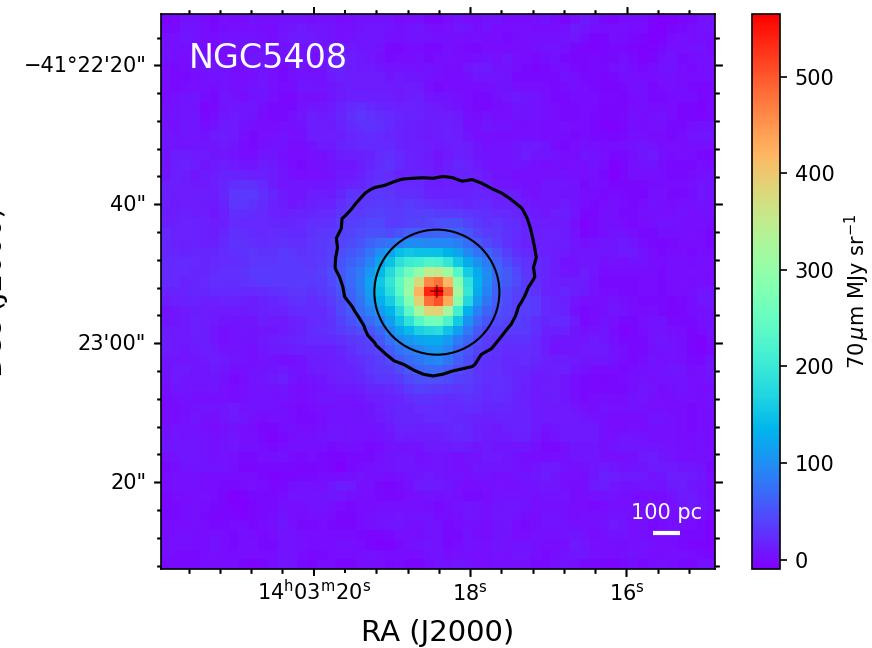}\\
\vspace{-\baselineskip}
\caption{Herschel/PACS 70\,\micron\ (Spitzer/MIPS for IC\,5152, NGC\,3109) images of the target galaxies.
The APEX pointing center is shown as a black cross, and the APEX beam of $\sim$18\arcsec\ as a black circle. 
Rough outlines of the ALMA coverage are given as heavy (in some cases dashed) black curves (see also Figs. \ref{fig:moment5152},
\ref{fig:moment2915}, \ref{fig:moment3109}, and \ref{fig:moment5408}).
}
\label{fig:apexpointings}
\end{figure*}

APEX observations\footnote{Project IDs: m-0109.f-9525a-2022 (PI: Wei\ss); m-0110.f-9503c-2022 (PI: Henkel)} were carried out using the SEPIA\,345 receiver on APEX
between July 30th and Sep. 27th 2022 under good weather conditions (PWV=0.8-1.0\,mm).
SEPIA\,345 is a dual polarization, sideband separating receiver with an instantaneous IF (intermediate frequency) bandwidth of 8\,GHz
for each sideband operating between 283–365\,GHz \citep{2022A&A...668A...2M}.
The receiver was tuned to the redshifted $^{12}$CO(3-2) frequency in the USB (upper sideband) for each galaxy.
APEX's full-width half-maximum (FWHM) beam size at the target frequencies is $\approx18$\arcsec. Observations were carried out
towards the peak of the MIPS or PACS 70\,$\mu$m emission in each galaxy (see Fig\,\ref{fig:apexpointings})
using the symmetric wobbler switch mode with a wobbler throw of 60\arcsec\ and a wobbler rate of 1.5\,Hz.
Data were recorded using eight 4-GHz-wide fast-Fourier transform spectrometer (FFTS) backends
\citep{2012A&A...542L...3K} configured to a spectral resolution of 0.24\,KHz ($\approx0.2$\,\kms).
The APEX pointing centers are illustrated in Fig. \ref{fig:apexpointings}, 
where they are overlaid on the IR 70\,\micron\ images. 

The total on-source integration times are listed in Table\,\ref{tab:apexresults}.
Pointing was checked every $\sim$1h on nearby CO emitting stars and found to be
stable within 3\arcsec\ (rms). Signal calibration was performed using the standard
hot-cold-sky temperature load scheme at a time interval of $\sim 7$\,min. For the receiver coupling to the
antennae we adopt 35\,Jy\,K$^{-1}$ based on regular measurements of planets conducted by the APEX
staff\footnote{\href{https://www.apex-telescope.org/telescope/efficiency/index.php}{https://www.apex-telescope.org/telescope/efficiency/index.php}}.

\begin{figure*}[h!]
\includegraphics[width=0.245\textwidth]{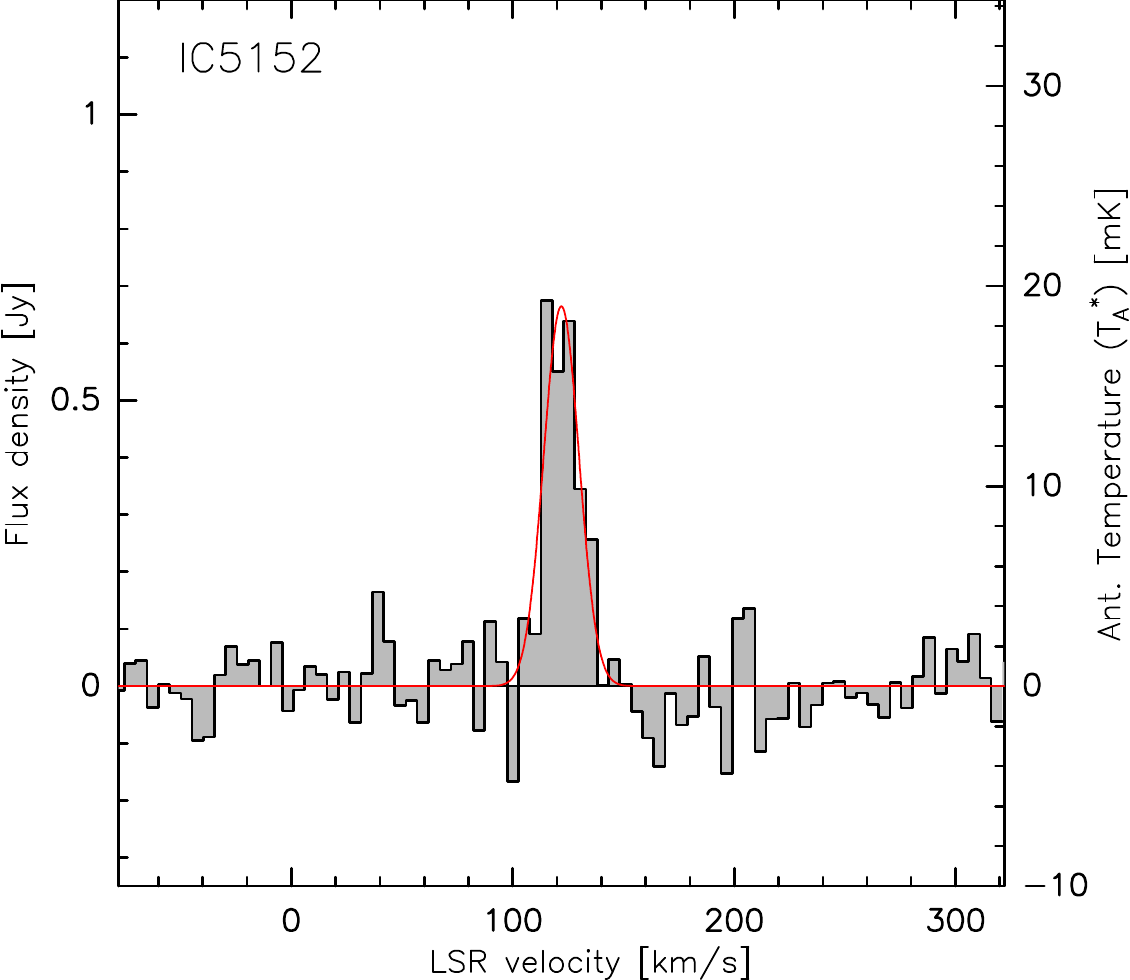}
\includegraphics[width=0.245\textwidth]{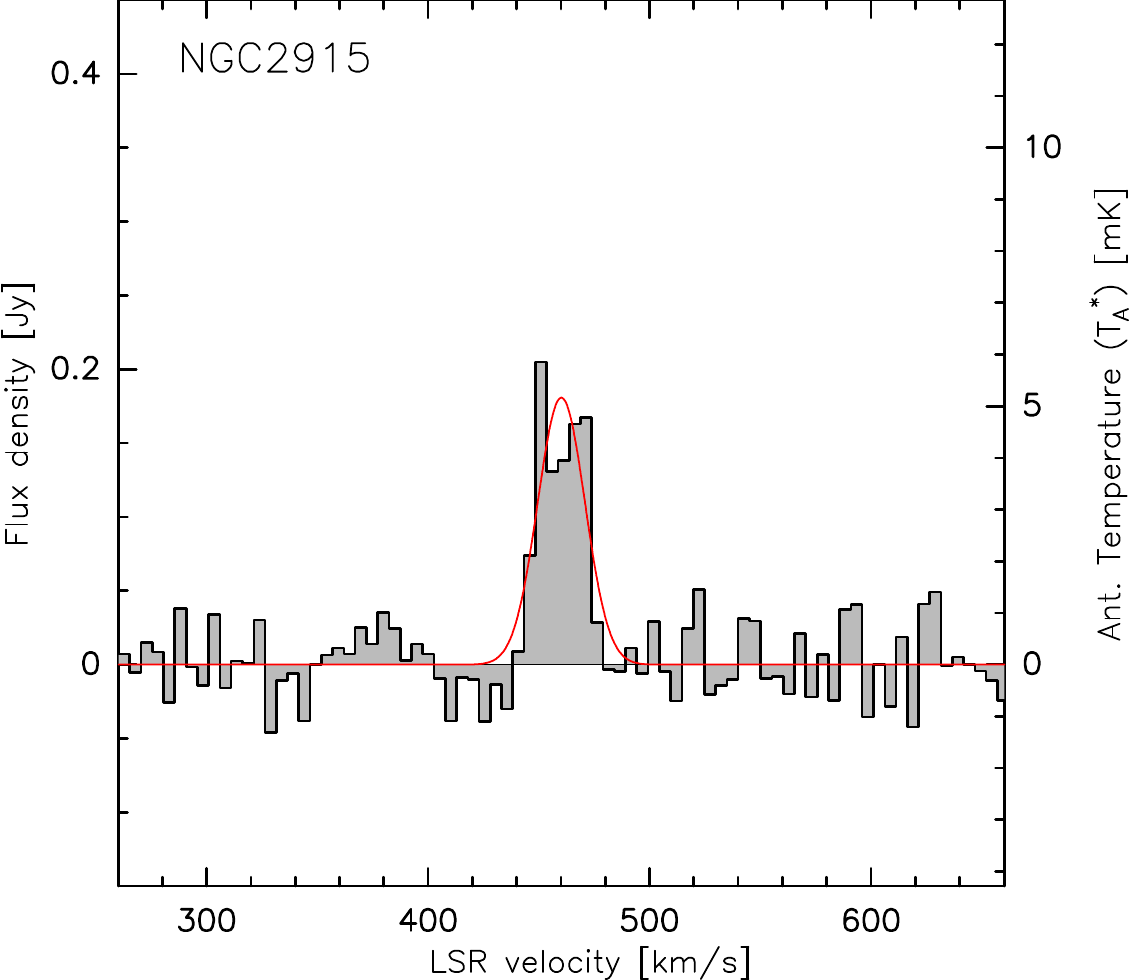}
\includegraphics[width=0.245\textwidth]{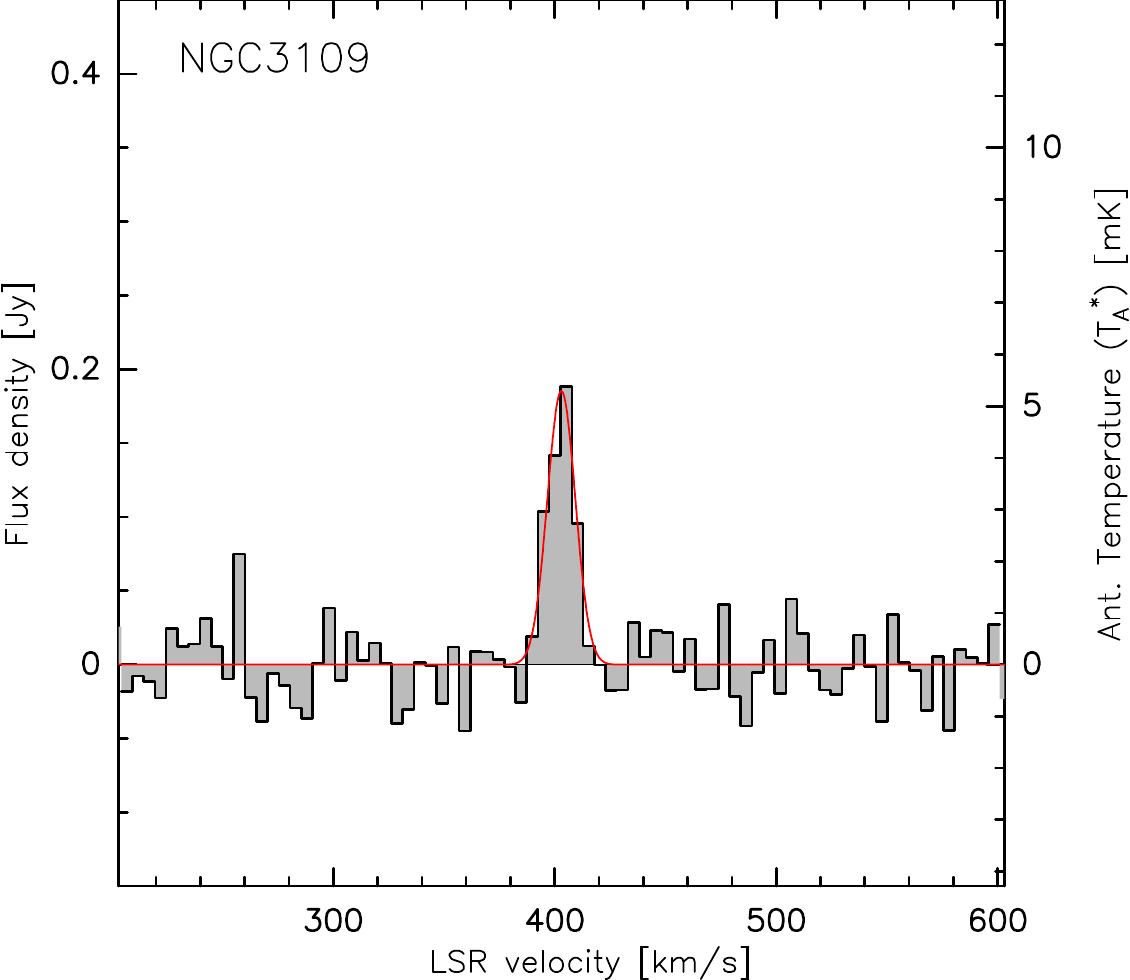}
\includegraphics[width=0.245\textwidth]{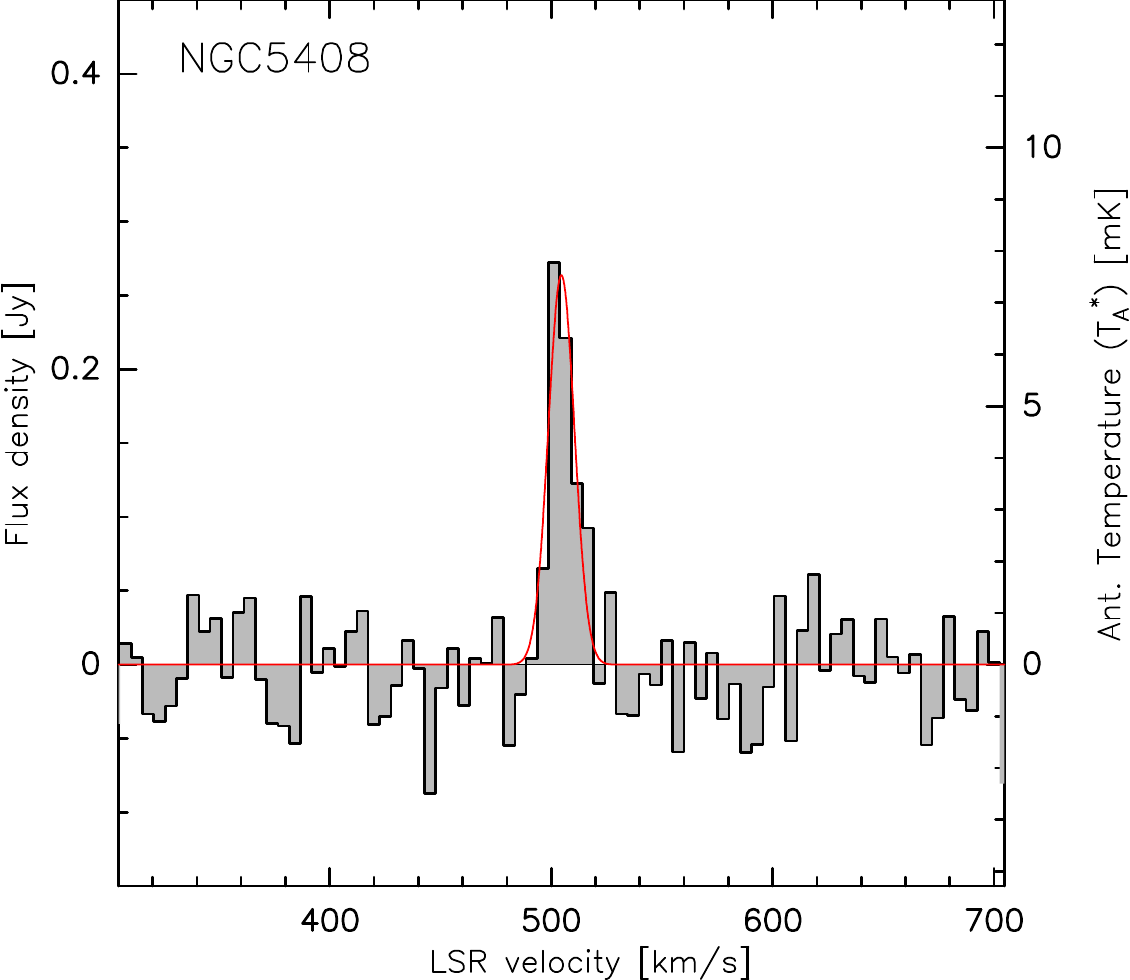}
\caption{\cothree\ APEX spectra centered on the apertures illustrated in Fig. \ref{fig:apexpointings}
The red curve is a Gaussian $+$ continuum fit to the spectrum.
}
\vspace{-\baselineskip}
\label{fig:apexspectra}
\end{figure*}

Data were processed using the GILDAS/CLASS\footnote{\href{https://www.iram.fr/IRAMFR/GILDAS}{https://www.iram.fr/IRAMFR/GILDAS}}
software package. For each of the spectra we extracted a 800\,\kms\ wide range around the $^{12}$CO(3-2) line,
resampled the velocity resolution to 5.0\,\kms\ and subtracted a linear baseline. All individual spectra were finally
co-added in a noise-weighted manner. The resulting spectra are shown in Fig.\,\ref{fig:apexspectra}. Line parameters are summarized in
Table\,\ref{tab:apexresults}, together with 
the integrated fluxes found by APEX within an 18\arcsec\ aperture.

\begin{table*}[h!]
\caption{APEX observations and results}                 
\label{tab:apexresults}              
\vspace{-1.8\baselineskip}
\begin{center}                        
\resizebox{\linewidth}{!}{
\begin{tabular}{cccccccccc}      
\hline\hline               
Name & \multicolumn{2}{c}{Position} & T$_{\rm int}$ & Noise$^{a}$    & I$_{\rm CO(3-2)}$  & FWHM & v$_{\rm LSRK}$ & L$'$ & L  \\
     & \multicolumn{2}{c}{J2000}    & [min]       & [mJy\,b$^{-1}$] & [Jy\,b$^{-1}$\,km\,s$^{-1}$] & [km\,s$^{-1}$] & [km\,s$^{-1}$] & [10$^{3}$K\,km\,s$^{-1}$\,pc$^2$] & [\lsun] \\
\hline                      
IC\,5152   & 22:02:45.83 & -51:17:40.1 & 121 & 45.6 & 13.3$\pm$ 0.9 & 18.8$\pm$1.5 & 98.9$\pm$0.7  &  13.8$\pm$1.0 & 18.4$\pm$1.3\\    
NGC\,2915  & 09:26:10.50 & -76:37:27.0 & 830 & 24.0 &  4.8$\pm$ 0.4 & 25.1$\pm$1.8 & 461.7$\pm$1.0 &  24.2$\pm$1.8 & 32.0$\pm$2.4\\    
NGC\,3109  & 10:03:03.70 & -26:09:22.6 & 620 & 25.5 &  2.9$\pm$ 0.3 & 14.9$\pm$1.7 & 381.3$\pm$0.8 &   1.4$\pm$0.1 &  1.9$\pm$0.2\\        
NGC\,5408  & 14:03:18.50 & -41:22:52.0 & 445 & 30.9 &  3.9$\pm$ 0.4 & 13.9$\pm$1.9 & 504.6$\pm$0.8 &  30.0$\pm$3.2 & 39.7$\pm$4.3\\     
\hline                                  
\end{tabular}
}
\end{center}
\vspace{-\baselineskip}
\tablefoot{$^{a}$ Noise levels are given for a spectral resolution of 5.0 km s$^{-1}$}
\end{table*}

\section{ALMA integrated intensity maps and overlays \label{app:moments}}

Here we illustrate the ALMA integrated intensity \cotwo\ zero-moment maps,
obtained with the strict (left panels) and broad masks (right) with CO
contours overlaid on RGB images, and with the cloud shapes shown
for the two different masks used for identification overlaid on the moment map.
These are the equivalent of Fig. \ref{fig:moment5152}, but for the remaining
three galaxies in the sample.

\begin{figure*}[t!]
\centering
\includegraphics[width=0.8\textwidth]{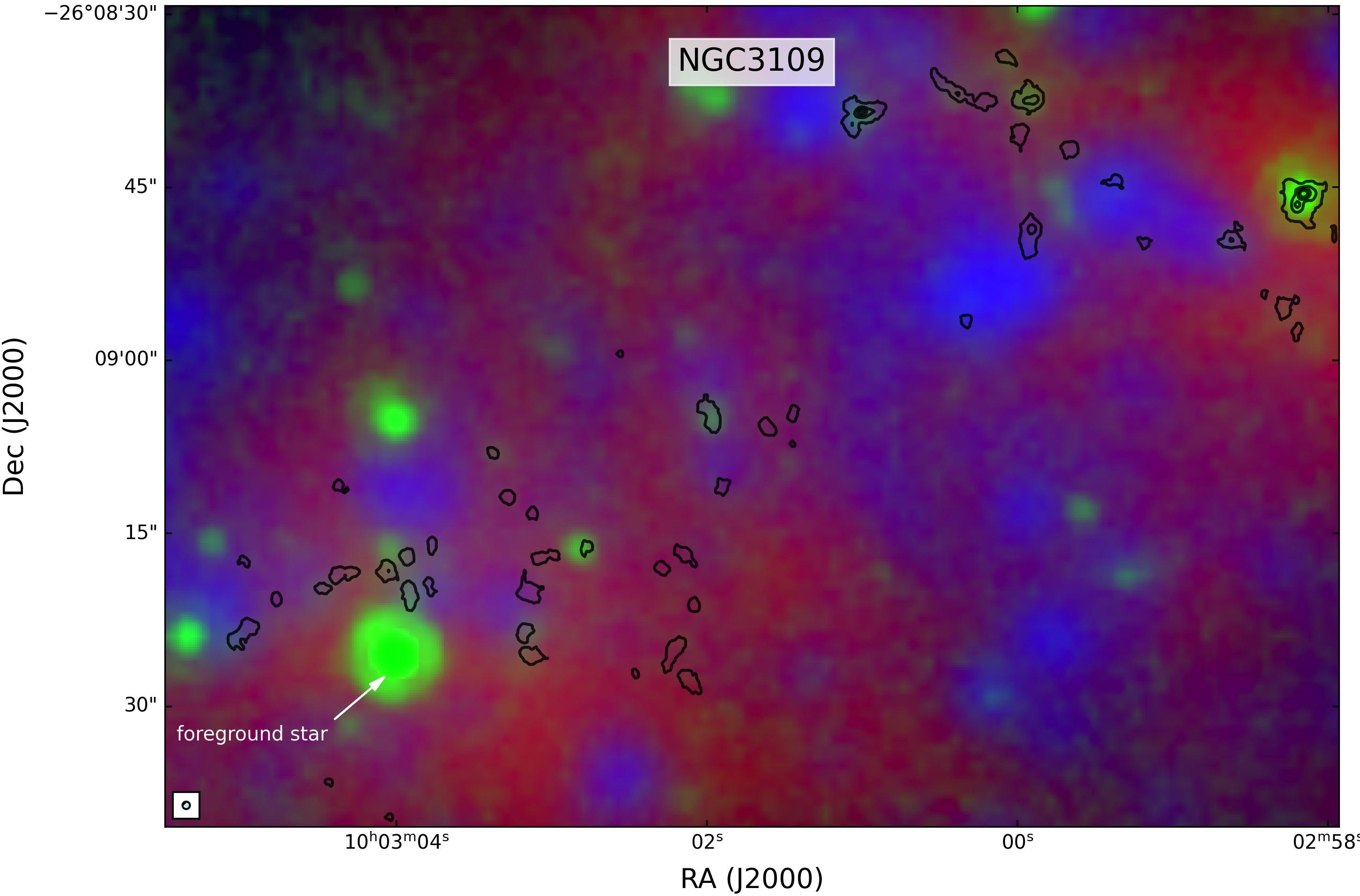} \\
\vspace{0.5\baselineskip}
\includegraphics[width=0.49\textwidth]{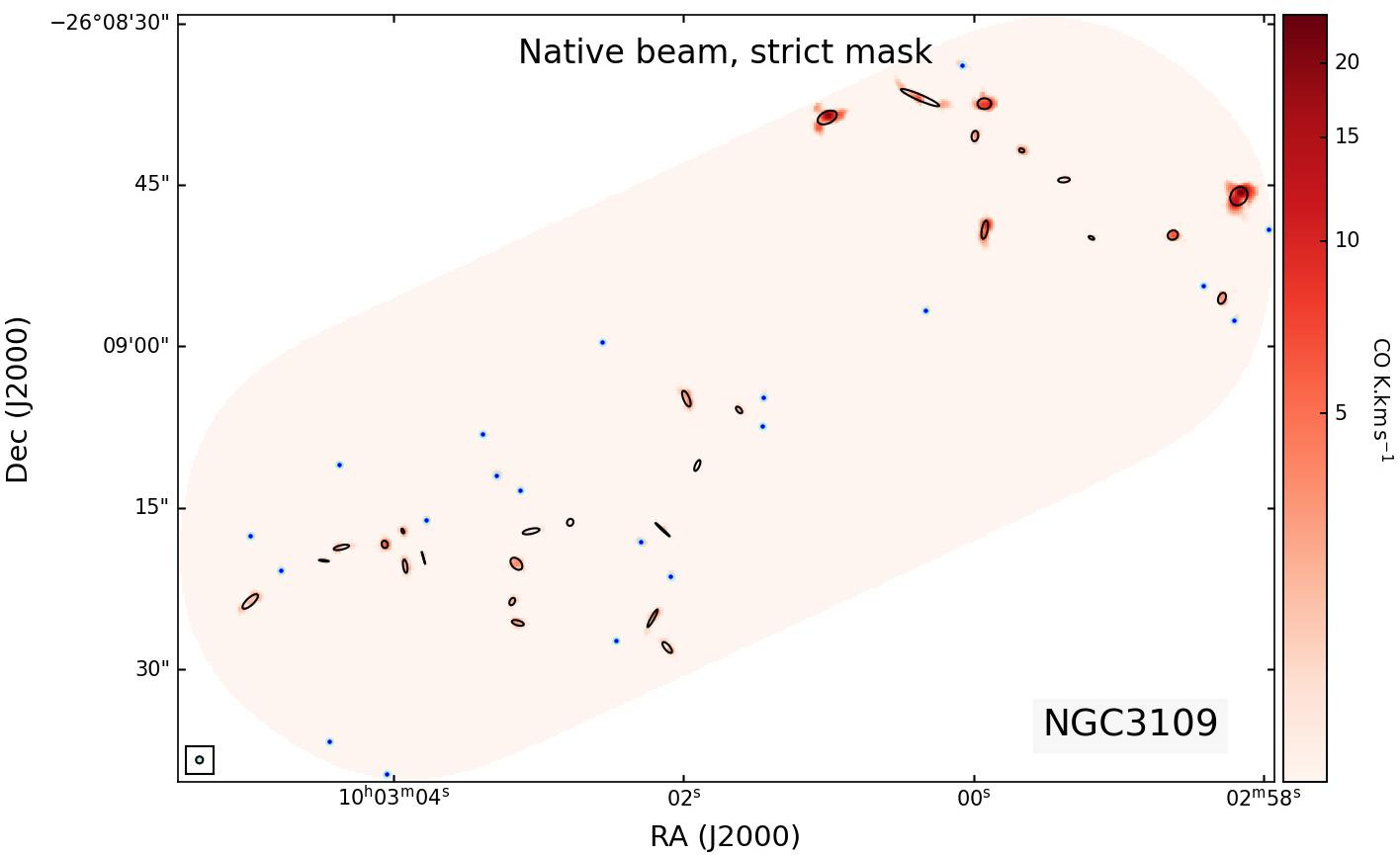} 
\vspace{0.5\baselineskip}
\includegraphics[width=0.49\textwidth]{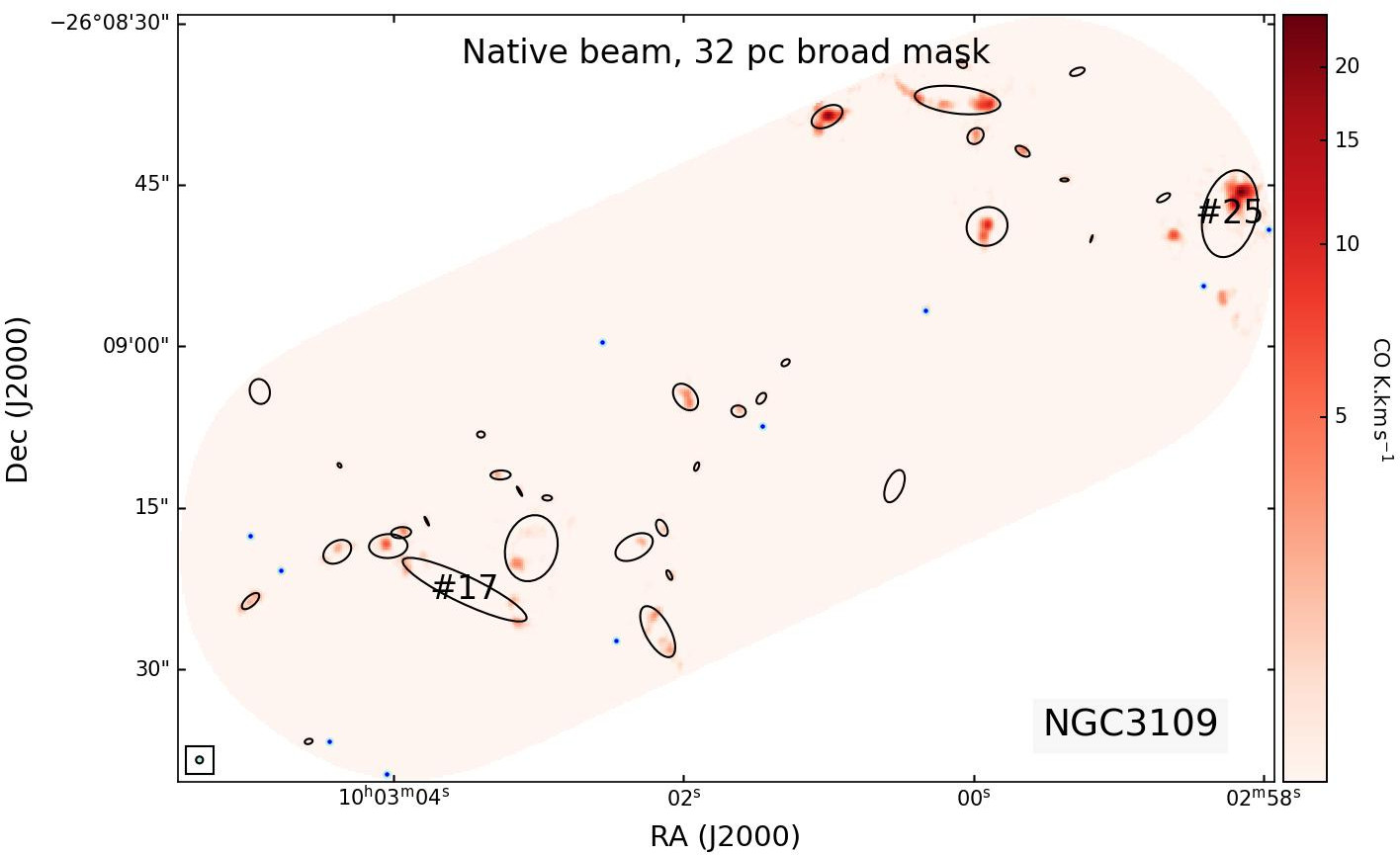} \\
\caption{As Fig. \ref{fig:moment5152}, but for NGC\,3109 where Clouds \#17 and \#25 are labeled.
}
\label{fig:moment3109}
\end{figure*}

\begin{figure*}[t!]
\centering
\includegraphics[height=0.25\linewidth]{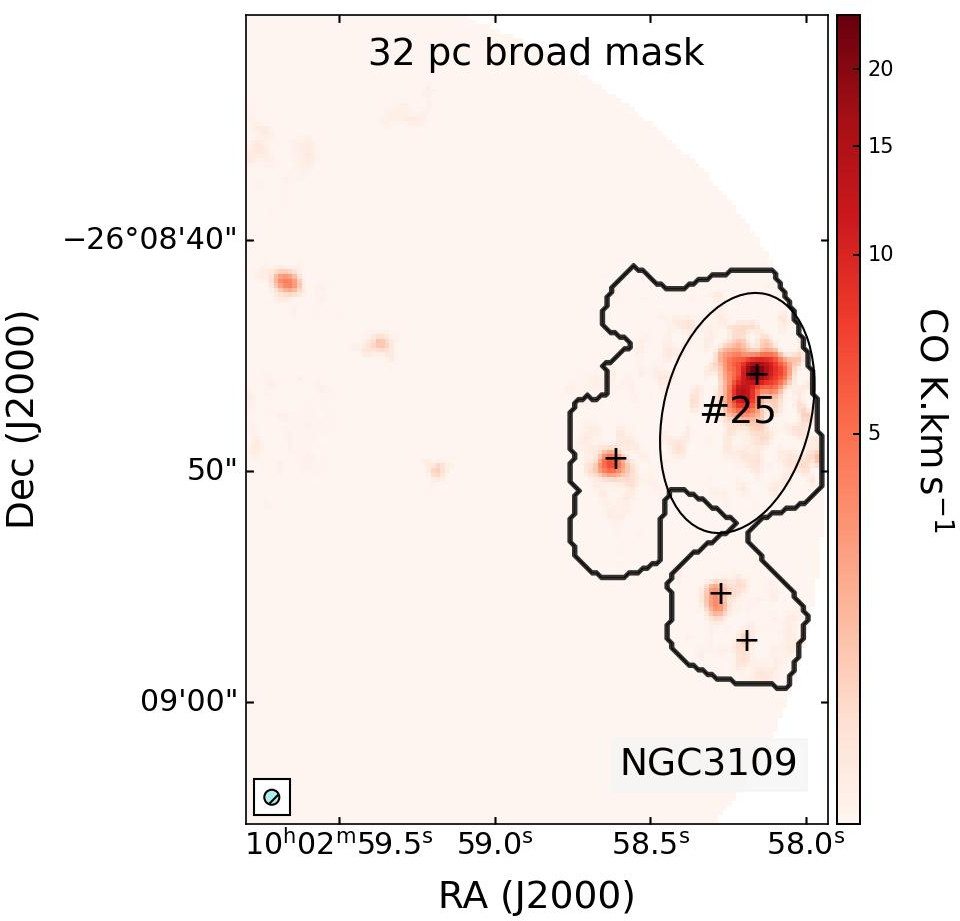}
\hspace{0.01\linewidth}
\includegraphics[height=0.24\linewidth]{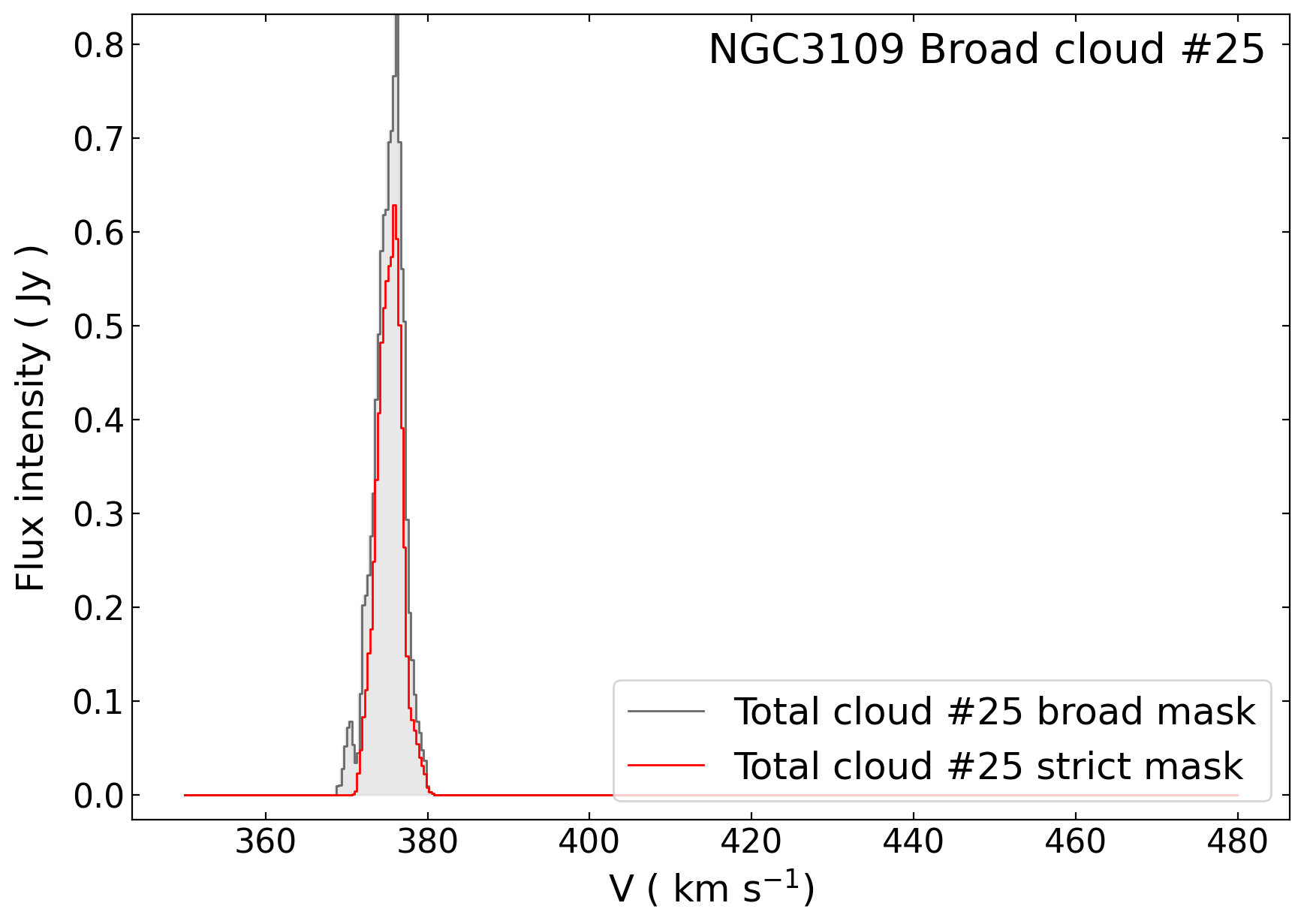} 
\hspace{0.005\linewidth}
\includegraphics[height=0.24\linewidth]{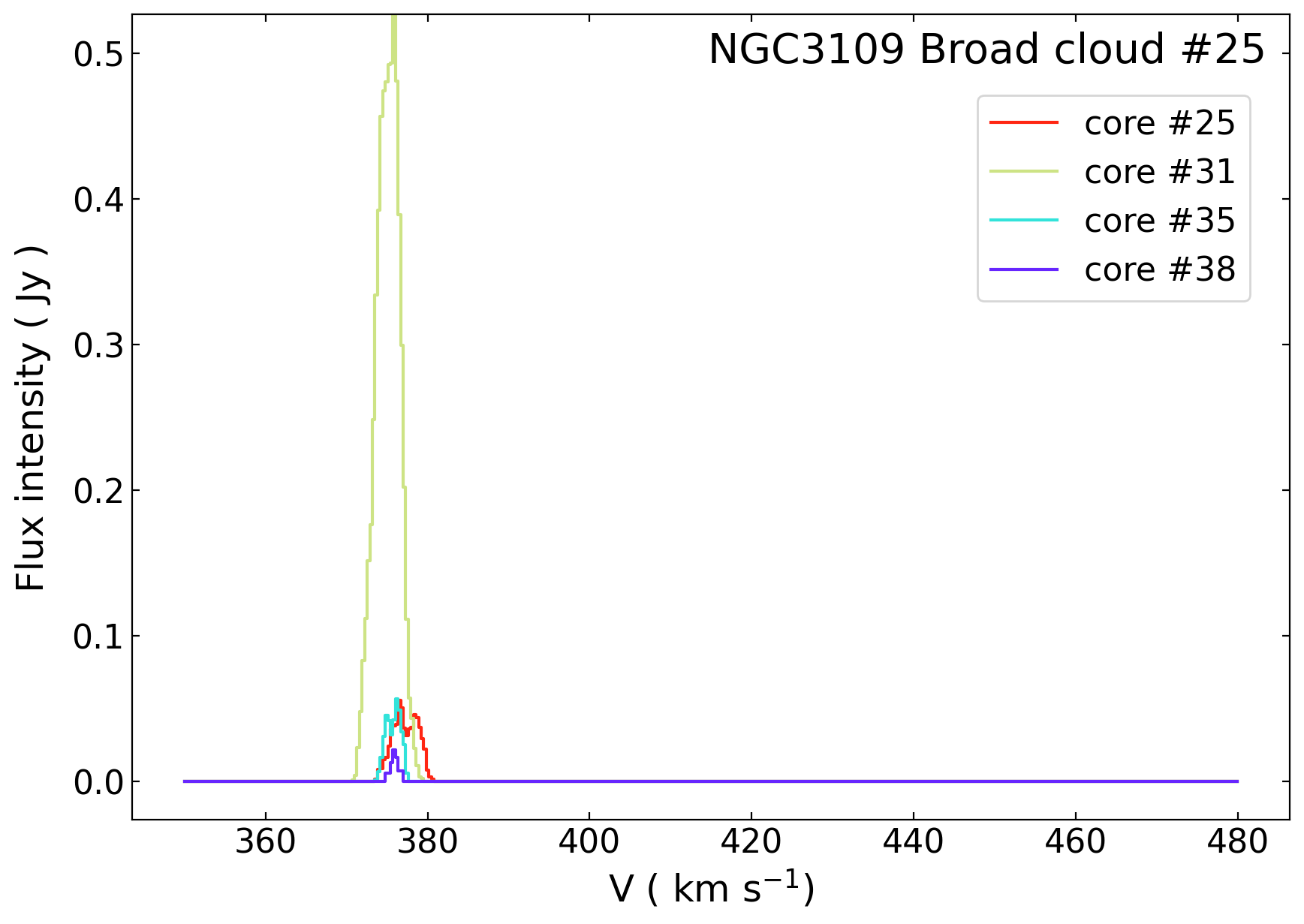} \\
\caption{As Fig. \ref{fig:ic5152_cloud5_39}, but for NGC\,3109, Cloud \#25 with 4 cores.
}
\vspace{5\baselineskip}
\label{fig:ngc3109_cloud25}
\end{figure*}

\begin{figure*}[t!]
\centering
\includegraphics[height=0.27\textwidth]{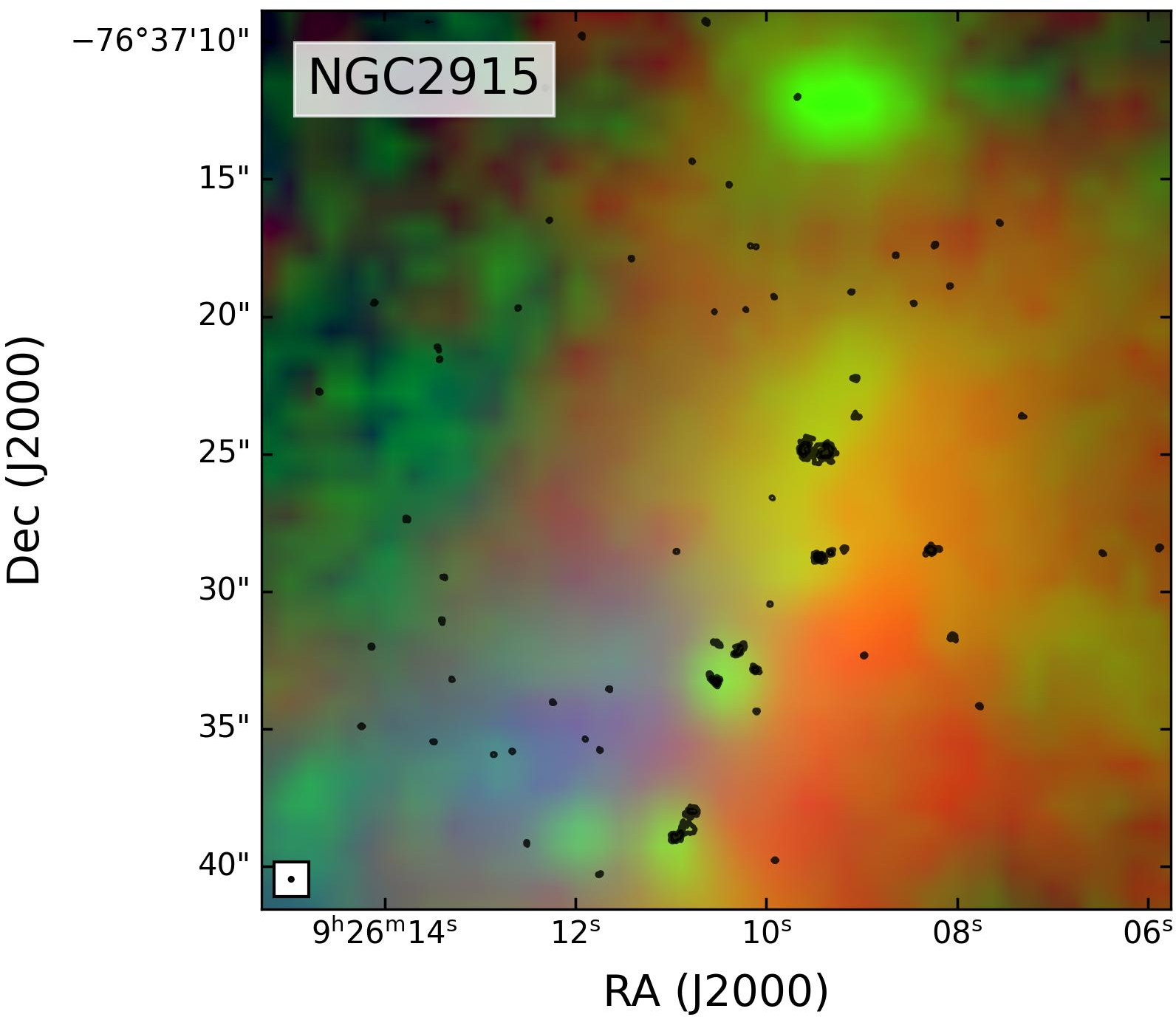}
\includegraphics[height=0.27\textwidth]{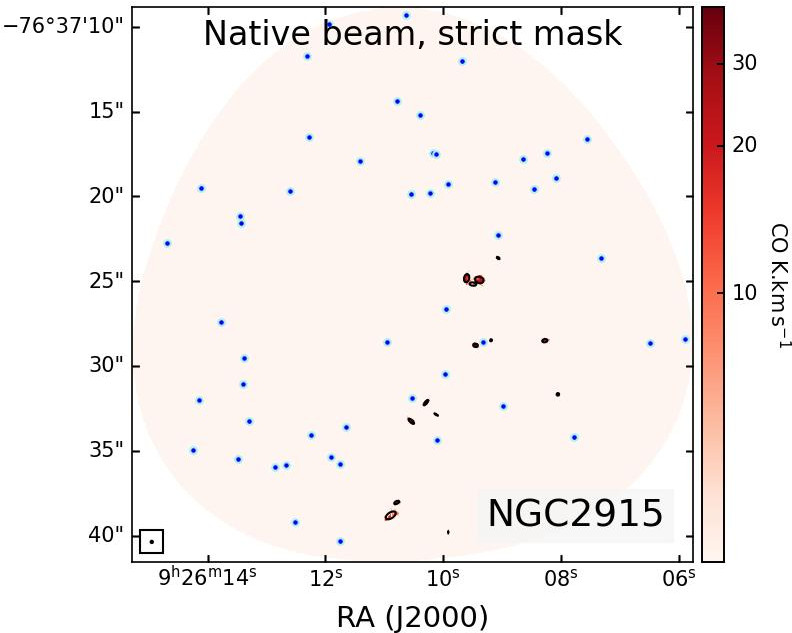} 
\includegraphics[height=0.27\textwidth]{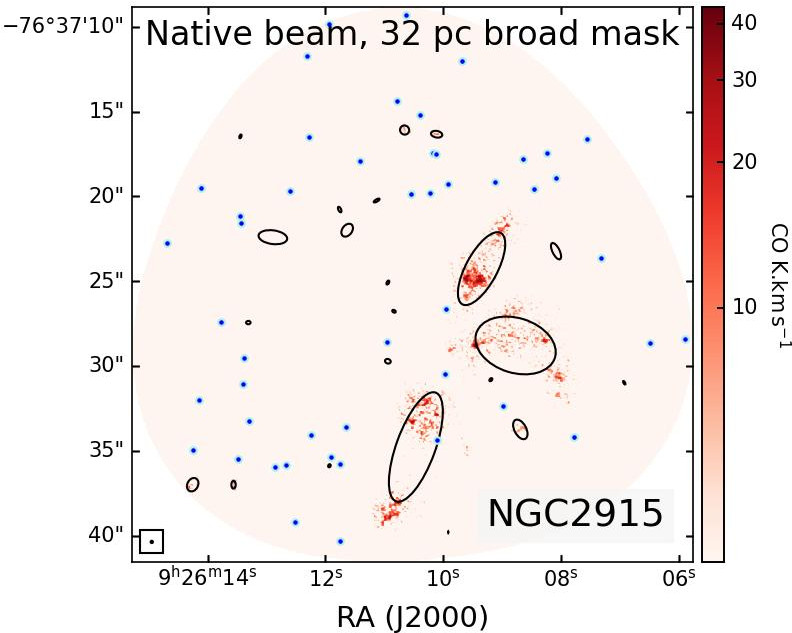} \\
\vspace{0.5\baselineskip}
\caption{As Fig. \ref{fig:moment5152}, but for NGC\,2915 with the RGB image shown
in the left panel, the strict native mask in the middle, and the broad 32\,pc (union) mask in the right.
Here the red in the left panel is \hers/PACS 70\,\micron.
}
\label{fig:moment2915}
\end{figure*}

\begin{figure*}[t!]
\centering
\includegraphics[height=0.265\textwidth]{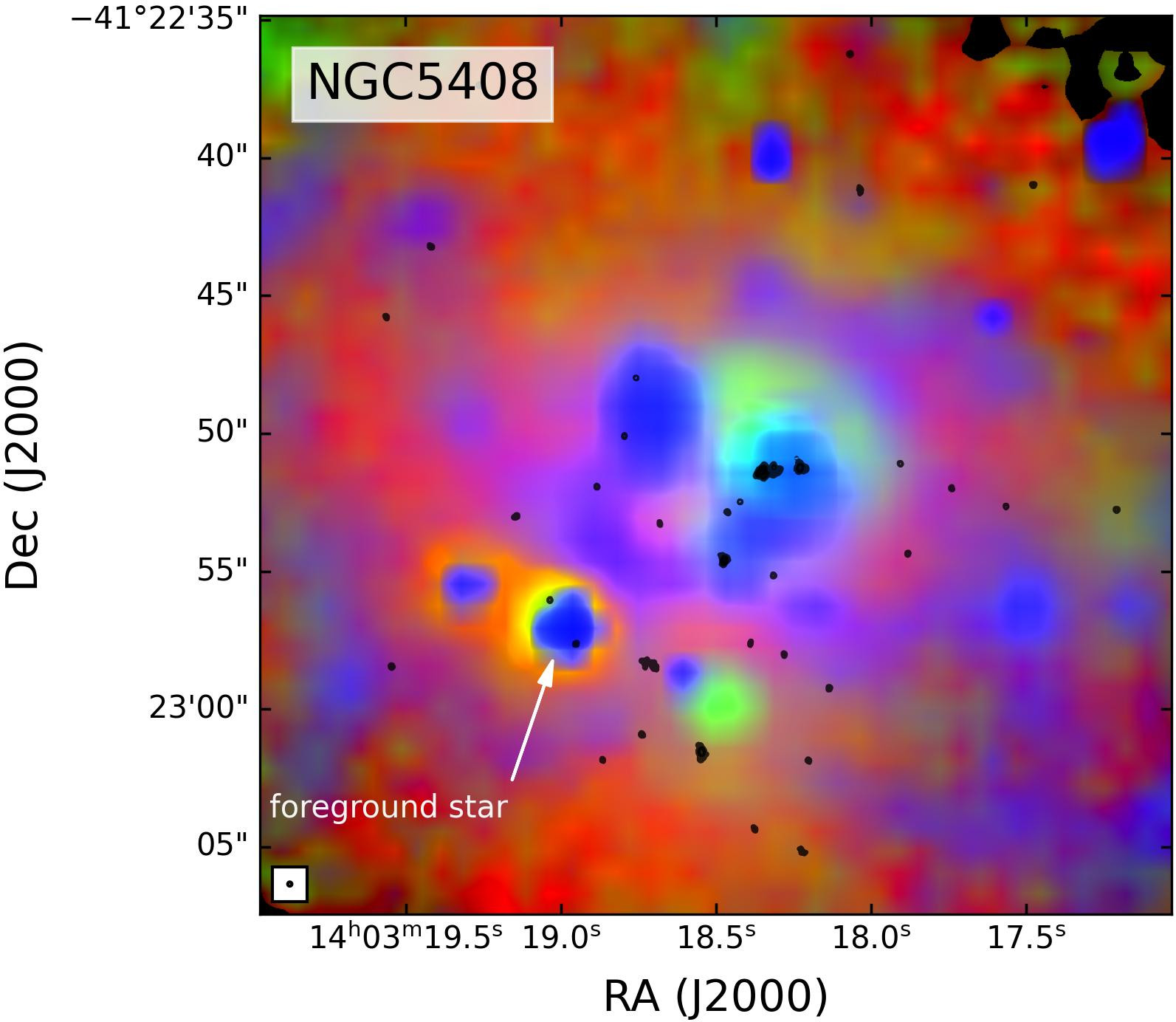}
\includegraphics[height=0.265\textwidth]{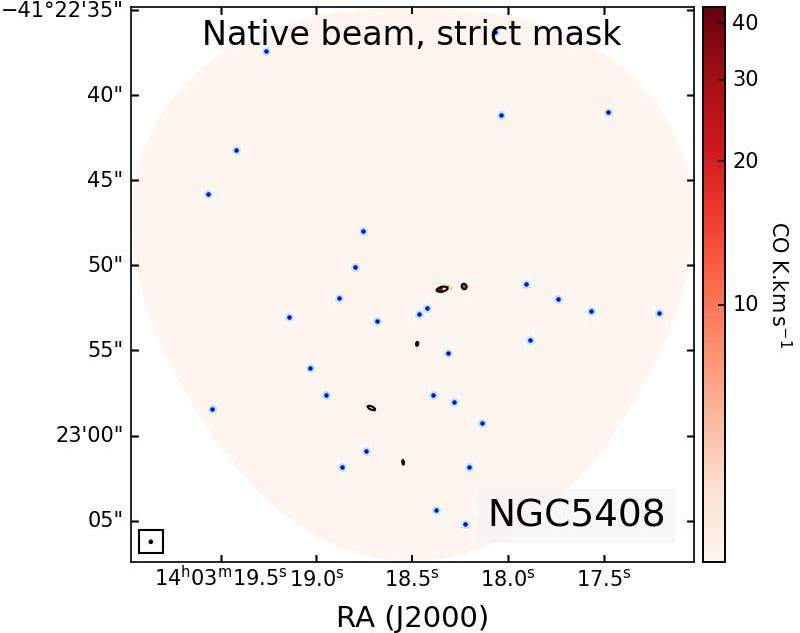} 
\includegraphics[height=0.265\textwidth]{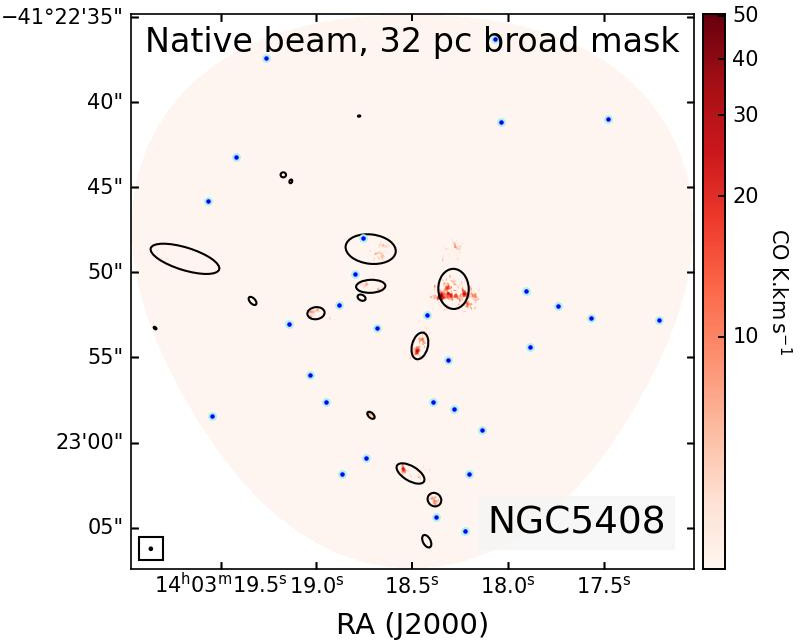} \\
\vspace{0.5\baselineskip}
\caption{As Fig. \ref{fig:moment5152}, but for NGC\,5408 with the RGB image shown
in the left panel, the strict native mask in the middle, and the broad 32\,pc (union) mask in the right.
The RGB image (with red as 70\,\micron\ \hers/PACS, green as 8\,micron, \spit/IRAC channel 4),
differs from previous ones in the blue channel,
with blue corresponding to continuum-subtracted
\ha\ from the SINGS data compilation \citep{dale07},
because there is no GALEX image available (\url{https://irsa.ipac.caltech.edu/data/SPITZER/SINGS/}).
}
\label{fig:moment5408}
\end{figure*}

\section{Properties of low-metallicity molecular clouds \label{app:proptables}}

The properties of the molecular clouds identified with the strict mask and the broad 32\,pc union mask described in Sect. \ref{sec:cloudprops}
are reported here.
The tables are snippets to guide content, and the integral lists can be found in the on-line version of the paper.
The radii of clouds that are unresolved are given as $-99.9$, and the beam size can be considered as an upper limit.

\begin{table}[h!]
\centering
\caption{Strict-mask properties of molecular cores in IC\,5152} 
\label{tab:IC5152-props-strictmask}
\resizebox{\linewidth}{!}{
\begin{tabular}{lclcccccccccc}
\hline
\hline
ID & S/N  & NPix & R.A. & Dec. &  V$_{\rm RAD}$ & $\sigma_{\rm v}$ & $\sigma(\sigma_{\rm v}$) & R & $\sigma$(R) & L$_{\rm CO}$ & $\sigma$(L$_{\rm CO}$) & M$_{\rm vir}$ \\
 & & & (deg) & (deg) & (kms$^{-1}$) & (kms$^{-1}$) & (kms$^{-1}$) & (pc) & (pc) & (Kkms$^{-1}$pc$^2$) & (Kkms$^{-1}$pc$^2$) & (M$_\odot$) \\
\hline
  1 &   6.7 &   256 &  330.6667 &  -51.2994 &   149.4  &    1.1 &   0.2 &  -99.9 &  -99.9 &  182 &  44  &  -99.9\\ 
  2 &  6.7 &   537  &  330.6669 & -51.2967 &   149.5 &     1.0 &   0.2 &   4.0 &  0.61 &   398 &  384 &  4056 \\ 
  3 &  6.0 &   512  &  330.6676 & -51.2993 &   147.8 &     1.8 &   0.3 &   2.5 &  0.60 &   391 &  489 &  8129 \\ 
  4 &   5.2 &    51 &  330.6656 &  -51.2958 &   150.1  &    0.7 &   0.6 &  -99.9 &  -99.9 &  37 &  25  &  -99.9\\ 
  5 & 14.4 &   962  &  330.6680 & -51.3002 &   146.6 &     2.2 &   0.2 &   3.0 &  0.31 &   1328 &  335 &  15091 \\ 
\hline
\end{tabular}
}
\end{table}

\begin{table}
\centering
\caption{Broad 32\,pc-mask properties of molecular clouds in IC\,5152} 
\label{tab:IC5152-props-Broad32}
\resizebox{\linewidth}{!}{
\begin{tabular}{lclcccccccccc}
\hline
\hline
ID & S/N  & NPix & R.A. & Dec. &  V$_{\rm RAD}$ & $\sigma_{\rm v}$ & $\sigma(\sigma_{\rm v}$) & R & $\sigma$(R) & L$_{\rm CO}$ & $\sigma$(L$_{\rm CO}$) & M$_{\rm vir}$ \\
 & & & (deg) & (deg) & (kms$^{-1}$) & (kms$^{-1}$) & (kms$^{-1}$) & (pc) & (pc) & (Kkms$^{-1}$pc$^2$) & (Kkms$^{-1}$pc$^2$) & (M$_\odot$) \\
\hline
  1 &  6.7 &  3947  &  330.6666 & -51.2966 &   150.3 &     1.2 &   0.1 &   7.6 &  0.30 &   854 &  53 &  11489 \\ 
  2 & 14.4 &  27326  &  330.6677 & -51.2999 &   147.7 &     1.9 &   0.1 &  17.1 &  0.33 &   4907 &  156 &  67303 \\ 
  3 &   5.2 &    51 &  330.6656 &  -51.2958 &   150.1  &    0.7 &   0.6 &  -99.9 &  -99.9 &  37 &  49  &  -99.9\\ 
  4 &  4.3 &   269  &  330.6661 & -51.2953 &   148.8 &     0.3 &   0.0 &   2.2 &  0.21 &   96 &  14 &  273 \\ 
  5 & 20.3 &  105975  &  330.6692 & -51.2963 &   135.0 &     2.6 &   0.1 &  33.7 &  0.49 &   13846 &  348 &  237629 \\ 

\hline
\end{tabular}
}
\end{table}

\begin{table}
\centering
\caption{Strict-mask properties of molecular cores in NGC\,2915} 
\label{tab:NGC2915-props-strictmask}
\resizebox{\linewidth}{!}{
\begin{tabular}{lclcccccccccc}
\hline
\hline
ID & S/N  & NPix & R.A. & Dec. &  V$_{\rm RAD}$ & $\sigma_{\rm v}$ & $\sigma(\sigma_{\rm v}$) & R & $\sigma$(R) & L$_{\rm CO}$ & $\sigma$(L$_{\rm CO}$) & M$_{\rm vir}$ \\
 & & & (deg) & (deg) & (kms$^{-1}$) & (kms$^{-1}$) & (kms$^{-1}$) & (pc) & (pc) & (Kkms$^{-1}$pc$^2$) & (Kkms$^{-1}$pc$^2$) & (M$_\odot$) \\
\hline
  1 &   4.1 &    14 &  141.5496 &  -76.6265 &   490.3  &    0.1 &   0.0 &  -99.9 &  -99.9 &  33 &  81  &  -99.9\\ 
  2 &   4.2 &    10 &  141.5440 &  -76.6222 &   489.6  &    0.1 &   0.0 &  -99.9 &  -99.9 &  19 &  44  &  -99.9\\ 
  3 &   4.6 &    27 &  141.5560 &  -76.6225 &   488.4  &    0.9 &   0.8 &  -99.9 &  -99.9 &  128 &  364  &  -99.9\\ 
  4 &   5.5 &    16 &  141.5490 &  -76.6266 &   487.1  &    0.1 &   0.0 &  -99.9 &  -99.9 &  22 &  8  &  -99.9\\ 
  5 &   4.3 &    16 &  141.5433 &  -76.6209 &   486.1  &    0.2 &   0.2 &  -99.9 &  -99.9 &  26 &  22  &  -99.9\\ 
\hline
\end{tabular}
}
\end{table}

\begin{table}
\centering
\caption{Broad 32\,pc-mask properties of molecular clouds in NGC\,2915} 
\label{tab:NGC2915-props-Broad32}
\resizebox{\linewidth}{!}{
\begin{tabular}{lclcccccccccc}
\hline
\hline
ID & S/N  & NPix & R.A. & Dec. &  V$_{\rm RAD}$ & $\sigma_{\rm v}$ & $\sigma(\sigma_{\rm v}$) & R & $\sigma$(R) & L$_{\rm CO}$ & $\sigma$(L$_{\rm CO}$) & M$_{\rm vir}$ \\
 & & & (deg) & (deg) & (kms$^{-1}$) & (kms$^{-1}$) & (kms$^{-1}$) & (pc) & (pc) & (Kkms$^{-1}$pc$^2$) & (Kkms$^{-1}$pc$^2$) & (M$_\odot$) \\
\hline
  1 &   4.1 &    14 &  141.5496 &  -76.6265 &   490.3  &    0.1 &   0.0 &  -99.9 &  -99.9 &  33 &  79  &  -99.9\\ 
  2 &   4.2 &    10 &  141.5440 &  -76.6222 &   489.6  &    0.1 &   -99.9 &  -99.9 &  -99.9 &  19 &  43  &  -99.9\\ 
  3 &   4.6 &    27 &  141.5560 &  -76.6225 &   488.4  &    0.9 &   0.8 &  -99.9 &  -99.9 &  128 &  323  &  -99.9\\ 
  4 &   5.5 &    16 &  141.5490 &  -76.6266 &   487.1  &    0.1 &   0.0 &  -99.9 &  -99.9 &  22 &  9  &  -99.9\\ 
  5 &   4.3 &    16 &  141.5433 &  -76.6209 &   486.1  &    0.2 &   0.2 &  -99.9 &  -99.9 &  26 &  38  &  -99.9\\ 
\hline
\end{tabular}
}
\end{table}

\begin{table}
\centering
\caption{Strict-mask properties of molecular cores in NGC\,3109} 
\label{tab:NGC3109-props-strictmask}
\resizebox{\linewidth}{!}{
\begin{tabular}{lclcccccccccc}
\hline
\hline
ID & S/N  & NPix & R.A. & Dec. &  V$_{\rm RAD}$ & $\sigma_{\rm v}$ & $\sigma(\sigma_{\rm v}$) & R & $\sigma$(R) & L$_{\rm CO}$ & $\sigma$(L$_{\rm CO}$) & M$_{\rm vir}$ \\
 & & & (deg) & (deg) & (kms$^{-1}$) & (kms$^{-1}$) & (kms$^{-1}$) & (pc) & (pc) & (Kkms$^{-1}$pc$^2$) & (Kkms$^{-1}$pc$^2$) & (M$_\odot$) \\
\hline
  1 &   4.9 &    14 &  150.7669 &  -26.1610 &   458.4  &    0.4 &   0.5 &  -99.9 &  -99.9 &  18 &  56  &  -99.9\\ 
  2 &   4.0 &    18 &  150.7434 &  -26.1484 &   423.8  &    0.2 &   0.3 &  -99.9 &  -99.9 &  35 &  98  &  -99.9\\ 
  3 &   4.3 &    21 &  150.7603 &  -26.1576 &   415.9  &    0.2 &   0.2 &  -99.9 &  -99.9 &  31 &  75  &  -99.9\\ 
  4 & 19.3 &   419  &  150.7669 & -26.1551 &   396.4 &     0.9 &   0.1 &   2.0 &  0.21 &   241 &  44 &  1568 \\ 
  5 &   4.4 &    49 &  150.7708 &  -26.1549 &   395.1  &    0.6 &   0.4 &  -99.9 &  -99.9 &  38 &  80  &  -99.9\\ 
\hline
\end{tabular}
}
\end{table}

\begin{table}
\centering
\caption{Broad 32\,pc-mask properties of molecular clouds in NGC\,3109} 
\label{tab:NGC3109-props-Broad32}
\resizebox{\linewidth}{!}{
\begin{tabular}{lclcccccccccc}
\hline
\hline
ID & S/N  & NPix & R.A. & Dec. &  V$_{\rm RAD}$ & $\sigma_{\rm v}$ & $\sigma(\sigma_{\rm v}$) & R & $\sigma$(R) & L$_{\rm CO}$ & $\sigma$(L$_{\rm CO}$) & M$_{\rm vir}$ \\
 & & & (deg) & (deg) & (kms$^{-1}$) & (kms$^{-1}$) & (kms$^{-1}$) & (pc) & (pc) & (Kkms$^{-1}$pc$^2$) & (Kkms$^{-1}$pc$^2$) & (M$_\odot$) \\
\hline
  1 &   4.9 &    14 &  150.7669 &  -26.1610 &   458.4  &    0.4 &   0.5 &  -99.9 &  -99.9 &  18 &  60  &  -99.9\\ 
  2 &  3.6 &   108  &  150.7554 & -26.1504 &   423.8 &     0.1 &   0.0 &   2.2 &  0.27 &   16 &  2 &  21 \\ 
  3 &   4.0 &    18 &  150.7434 &  -26.1484 &   423.8  &    0.2 &   0.2 &  -99.9 &  -99.9 &  35 &  107  &  -99.9\\ 
  4 &   4.3 &    21 &  150.7603 &  -26.1576 &   415.9  &    0.2 &   0.2 &  -99.9 &  -99.9 &  31 &  92  &  -99.9\\ 
  5 &  4.1 &   148  &  150.7445 & -26.1462 &   408.9 &     0.1 &   0.0 &   2.8 &  0.30 &   27 &  2 &  42 \\ 
\hline
\end{tabular}
}
\end{table}

\begin{table}
\centering
\caption{Strict-mask properties of molecular cores in NGC\,5408} 
\label{tab:NGC5408-props-strictmask}
\resizebox{\linewidth}{!}{
\begin{tabular}{lclcccccccccc}
\hline
\hline
ID & S/N  & NPix & R.A. & Dec. &  V$_{\rm RAD}$ & $\sigma_{\rm v}$ & $\sigma(\sigma_{\rm v}$) & R & $\sigma$(R) & L$_{\rm CO}$ & $\sigma$(L$_{\rm CO}$) & M$_{\rm vir}$ \\
 & & & (deg) & (deg) & (kms$^{-1}$) & (kms$^{-1}$) & (kms$^{-1}$) & (pc) & (pc) & (Kkms$^{-1}$pc$^2$) & (Kkms$^{-1}$pc$^2$) & (M$_\odot$) \\
\hline
  1 &   5.0 &    34 &  210.8287 &  -41.3811 &   528.8  &    0.2 &   0.2 &  -99.9 &  -99.9 &  63 &  160  &  -99.9\\ 
  2 &   4.7 &    31 &  210.8239 &  -41.3811 &   528.5  &    0.3 &   0.3 &  -99.9 &  -99.9 &  28 &  18  &  -99.9\\ 
  3 &   4.4 &    36 &  210.8266 &  -41.3845 &   528.1  &    0.2 &   0.3 &  -99.9 &  -99.9 &  76 &  269  &  -99.9\\ 
  4 &   4.6 &    51 &  210.8314 &  -41.3829 &   527.0  &    0.6 &   0.6 &  -99.9 &  -99.9 &  395 &  780  &  -99.9\\ 
  5 &   4.5 &    40 &  210.8309 &  -41.3787 &   525.9  &    0.1 &   0.0 &  -99.9 &  -99.9 &  54 &  148  &  -99.9\\ 
\hline
\end{tabular}
}
\end{table}

\begin{table}
\centering
\caption{Broad 32\,pc-mask properties of molecular clouds in NGC\,5408} 
\label{tab:NGC5408-props-Broad32}
\resizebox{\linewidth}{!}{
\begin{tabular}{lclcccccccccc}
\hline
\hline
ID & S/N  & NPix & R.A. & Dec. &  V$_{\rm RAD}$ & $\sigma_{\rm v}$ & $\sigma(\sigma_{\rm v}$) & R & $\sigma$(R) & L$_{\rm CO}$ & $\sigma$(L$_{\rm CO}$) & M$_{\rm vir}$ \\
 & & & (deg) & (deg) & (kms$^{-1}$) & (kms$^{-1}$) & (kms$^{-1}$) & (pc) & (pc) & (Kkms$^{-1}$pc$^2$) & (Kkms$^{-1}$pc$^2$) & (M$_\odot$) \\
\hline
  1 &   5.0 &    34 &  210.8287 &  -41.3811 &   528.8  &    0.2 &   0.2 &  -99.9 &  -99.9 &  63 &  179  &  -99.9\\ 
  2 &  4.1 &   253  &  210.8327 & -41.3815 &   528.5 &     0.2 &   0.0 &   1.9 &  0.24 &   208 &  29 &  46 \\ 
  3 &   4.7 &    31 &  210.8239 &  -41.3811 &   528.5  &    0.3 &   0.3 &  -99.9 &  -99.9 &  28 &  12  &  -99.9\\ 
  4 &   4.4 &    36 &  210.8266 &  -41.3845 &   528.1  &    0.2 &   0.3 &  -99.9 &  -99.9 &  76 &  230  &  -99.9\\ 
  5 &   4.6 &    51 &  210.8314 &  -41.3829 &   527.0  &    0.6 &   0.5 &  -99.9 &  -99.9 &  395 &  797  &  -99.9\\ 

\hline
\end{tabular}
}
\end{table}

\section{Additional scaling relations for clouds/cores \label{app:scaling}}

\begin{figure*}[h!]
\centering
\includegraphics[width=\linewidth]{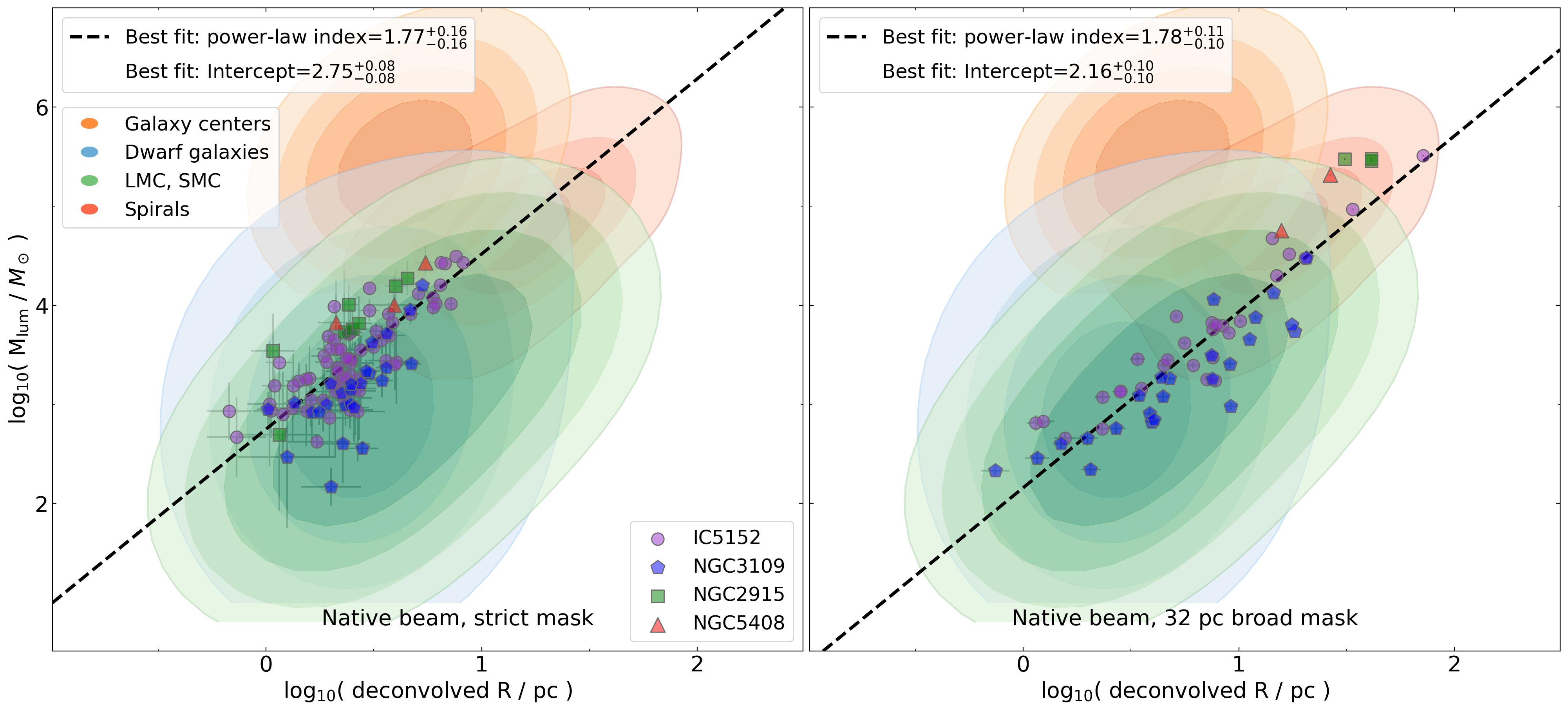}
\caption{Mass inferred from CO luminosity \mlum\ vs. cloud radius R (deconvolved)  for the clouds in our sample selected 
with two masks: the native-resolution strict mask (left panel), and 
the 32-pc broad mask (right).
The groupings for giant molecular clouds from other nearby galaxies, including dwarf galaxies and the Milky Way, 
are represented as colored contours that describe their 2D distributions;
see text for more details.
The best-fit \texttt{linmix} regression is shown as a dashed line.
Only cores/clouds with S/N$\geq$5 on the peak temperature are considered.
}
\label{fig:cloudprops_2}
\vspace{-5\baselineskip}
\end{figure*}

Figure \ref{fig:cloudprops_2} adds to Sect. \ref{sec:cloudscaling},
showing the mass inferred from CO luminosity \mlum\ plotted against deconvolved cloud radius R.

\end{appendix}

\end{document}